\documentclass[review]{elsarticle}

\usepackage{natbib}
\usepackage{caption}
\usepackage{subcaption}
\usepackage{graphicx}
\usepackage{booktabs}
\usepackage{amsmath}
\usepackage{bm}
\usepackage{color}
\usepackage{colortbl}
\usepackage[version=4]{mhchem}
\usepackage{siunitx}
\usepackage{longtable,tabularx}
\usepackage{makecell}
\usepackage{hyperref} 
\usepackage{mathrsfs} 
\usepackage{comment}
\usepackage{tikz}
\usepackage{rotating}
\usepackage{pgfplots}
\pgfplotsset{compat=newest}
\usetikzlibrary{calc}
\usetikzlibrary{shapes.arrows, decorations.markings, arrows.meta, calc, positioning}
\usetikzlibrary{arrows.meta, decorations.markings, patterns, calc, shapes.geometric}

\hypersetup{
    colorlinks = true,
    urlcolor   = blue,
    citecolor  = blue,
}

\newcommand{\pcr}{p_\mathrm{cr}} 
\newcommand{\Tcr}{T_\mathrm{cr}} 
\newcommand{\rhor}{\rho_\mathrm{r}} 
\newcommand{\pr}{p_\mathrm{r}} 
\newcommand{\Tr}{T_\mathrm{r}} 

\newcommand{\Tpb}{T_\mathrm{pb}}

\newcommand{\dd}{\;\mathrm{d}}

\makeatletter
\def\ps@pprintTitle{%
  \let\@oddhead\@empty
  \let\@evenhead\@empty
  \def\@oddfoot{\reset@font\hfil\thepage\hfil}
  \let\@evenfoot\@oddfoot
}
\makeatother

\usepackage{amssymb}
\usepackage{nomencl}
\makenomenclature
\usepackage{etoolbox}
\renewcommand\nomgroup[1]{%
  \item[\bfseries
  \ifstrequal{#1}{A}{Latin letters}{%
  \ifstrequal{#1}{B}{Greek symbols}{%
  \ifstrequal{#1}{C}{subscripts}{%
  \ifstrequal{#1}{E}{Abbreviations}{%
  \ifstrequal{#1}{F}{Dimensionless number}{%
  \ifstrequal{#1}{D}{superscripts}
  {}}}}}}%
]}

\DeclareTextFontCommand\textsfi{\usefont{OT1}{cmss}{m}{sl}}
\DeclareMathAlphabet\mathsfi            {OT1}{cmss}{m}{sl}
\DeclareTextFontCommand\textsfb{\usefont{OT1}{cmss}{bx}{n}}
\DeclareMathAlphabet\mathsfb            {OT1}{cmss}{bx}{n}
\DeclareTextFontCommand\textsfbi{\usefont{OT1}{cmss}{m}{sl}}
\DeclareMathAlphabet\mathsfbi            {OT1}{cmss}{m}{sl}
\IfFileExists{t1phv.fd}
{\DeclareTextFontCommand\textsfbi{\usefont{T1}{phv}{b}{it}}
	\DeclareMathAlphabet\mathsfbi            {T1}{phv}{b}{it}
}{}
\IfFileExists{ot1phv.fd}
{\DeclareTextFontCommand\textsfbi{\usefont{OT1}{phv}{b}{it}}
	\DeclareMathAlphabet\mathsfbi            {OT1}{phv}{b}{it}
}{}

\begin{document}

\begin{frontmatter}
\title{Microscale physics and macroscale convective \\ heat transfer in supercritical fluids}


\author[inst1]{Zhouhang Li$^\dagger$}
\author[inst2]{Daniel T. Banuti$^\dagger$}
\author[inst32,inst3]{Jie Ren}
\author[inst4]{Junfu Lyu} 
\author[inst1]{Hua Wang}
\author[inst5,inst6]{Xu Chu$^\ast$}
\ead{x.chu@exeter.ac.uk}
\cortext[cor1]{Corresponding author}

\affiliation[inst1]{School of Metallurgical and Energy Engineering, Kunming University of Science and Technology, Yunnan 650093, China}
\affiliation[inst2]{Institute for Thermal Energy Technology and Safety, Karlsruhe Institute of Technology, 76131 Karlsruhe, Germany}
\affiliation[inst3]{Institute of Aerodynamics and Gas Dynamics, University of Stuttgart, 70569 Stuttgart, Germany}
\affiliation[inst32]{School of Mechatronical Engineering, Beijing Institute of Technology, Beijing 100081, China}
\affiliation[inst4]{Department of Energy and Power Engineering, Tsinghua University, Beijing 100084, China}
\affiliation[inst5]{Department of Engineering, University of Exeter, EX4 4PY Exeter, UK}

\affiliation[inst6]{Cluster of Excellence SimTech, University of Stuttgart, 70569 Stuttgart, Germany}

\begin{abstract}
The worldwide push towards energy efficiency, use of alternative/waste energy, and power plant simplifications continues to drive the development of energy conversion and utilization systems, now covering a wide range of industrial applications including nuclear reactors, solar/fossil-fired power generation, geothermal systems, aerospace engineering, refrigeration/heat pumps, and waste heat utilization. Driven by fundamental thermodynamic efficiency considerations, an emerging trend in these systems is that the working fluid operates at a pressure above the critical pressure (i.e. supercritical pressure) in certain components or the whole system. Energy transport is thus accompanied by dramatic and strongly nonlinear variations of fluid thermophysical properties, which cause abnormal heat transfer behavior and non-ideal gas effects. This situation raises a crucial challenge for the heat exchanger and turbomachinery design, overall energy and exergy efficiency, as well as the safe operation of these thermal systems. The proposed review aims to provide a multi-scale overview of the flow and thermal behavior of fluid above the critical point: microscopic physics and macroscopic transport.
Microscopic physics, i.e. near-critical thermophysical properties, phase transition and fluid dynamics, are introduced based on the most recent findings through fundamental thermodynamics analysis, molecular dynamics simulations, and in situ neutron imaging measurements. A particular focus will be on the supercritical ‘pseudo boiling’ process, which is a generalization of classical liquid-vapor phase transitions to non-equilibrium supercritical states. Pseudo boiling was found to introduce a new, thermal, jet break-up mechanism, and to be intricately linked to supercritical heat transfer deterioration. These new perspectives lead to a revised view of the state space. Further, recent results demonstrated the possibility of stable supercritical fluid interfaces without surface tension.
On the macroscale, recent progress on the physical understanding and modeling of turbulent flow and convective heat transfer of supercritical fluids are summarized. Direct numerical simulation is able to fully resolve the entire turbulence spectrum of flows at supercritical conditions and to offer insights into the physics of thermal fluids. We start with a description of fundamental fluid mechanics problems related to supercritical fluids, such as velocity and temperature transformations and boundary-layer stability. It turns out that pseudo boiling is a sufficient physical mechanism to cause supercritical heat transfer deterioration, strongly resembling the subcritical boiling crisis. In addition, the heat transfer deterioration in supercritical fluids is found to be closely connected to the flow relaminarization by the non-uniform body-force. Finally, various modeling approaches such as recently developed advanced Reynolds-averaged turbulence modeling as well as machine-learning methods are summarized. 
\end{abstract}


\end{frontmatter}


\def\thefootnote{$\dagger$}\footnotetext{These authors contributed equally to this work}\def\thefootnote{\arabic{footnote}}

\newpage

\tableofcontents

\newpage



\nomenclature[A921]{\(T\)}{Temperature}
\nomenclature[A94]{\(U\)}{Velocity}
\nomenclature[A90]{\(p\)}{pressure}
\nomenclature[A5]{\(g\)}{Gravitational acceleration}
\nomenclature[A920]{\(R\)}{gas constant}
\nomenclature[A6]{\(h\)}{enthalpy}
\nomenclature[A4]{\(D\)}{Diameter}
\nomenclature[A3]{\(c_p\)}{isobaric specific heat capacity}
\nomenclature[A922]{\bm{\mathrm{T}}}{Transfer entropy}
\nomenclature[A61]{\bm{\mathrm{H}}}{Shannon entropy}
\nomenclature[A95]{$v_m$}{molar volume}
\nomenclature[A2]{$b$}{molecular repulsion}
\nomenclature[A1]{$a$}{intermolecular attraction}
\nomenclature[A11]{$A_s$}{Riedel parameter}
\nomenclature[A12]{$A_\mathrm{SRK}$}{The slope of pseudo boiling line in Soave-Redlich-Kwong equation of state}
\nomenclature[A91]{$q$}{heat flux}
\nomenclature[A912]{$\boldsymbol{q}$}{perturbation vector}
\nomenclature[A71]{$\ell_\mathrm{th}$}{thermal length} 
\nomenclature[A72]{$\ell_\mathrm{VLE}$}{interfacial thickness} 
\nomenclature[A8]{$\dot{m}$}{mass flow rate}

\nomenclature[B]{\(\alpha\)}{thermal conductivity}
\nomenclature[B]{\(\rho\)}{density}
\nomenclature[B]{\(\mu\)}{dynamic viscosity}
\nomenclature[B]{\(\nu\)}{kinematic viscosity}
\nomenclature[B]{\(\tau\)}{wall shear stress}
\nomenclature[B]{\(\kappa\)}{von Kármán constant}
\nomenclature[B]{\(\theta\)}{temperature difference}
\nomenclature[B]{\(\Delta\)}{Mesh size in the simulation}
\nomenclature[B]{\(\eta\)}{Kolmogorov scale}
\nomenclature[B]{\(\eta_\theta\)}{Batchelor scale}

\nomenclature[C1]{$\mathrm{cr}$}{critical}
\nomenclature[C21]{$\mathrm{r}$}{reduced properties, non-dimensional}
\nomenclature[C51]{$\mathrm{w}$}{wall}
\nomenclature[C50]{$\mathrm{VD}$}{van Driest transformation}
\nomenclature[C40]{$\mathrm{t}$}{turbulent}
\nomenclature[C31]{$\mathrm{st}$}{structural component}
\nomenclature[C41]{$\mathrm{th}$}{thermal}
\nomenclature[C20]{$\mathrm{pb}$}{pseudoboiling}
\nomenclature[C6]{$\infty$}{free stream quantity}

\nomenclature[D]{\(+\)}{wall-unit scaling}
\nomenclature[D]{\(\star\)}{semi-local wall-normal scaling}
\nomenclature[D]{$\prime$}{fluctuation/perturbation quantity}

\nomenclature[E]{CFD}{Computational Fluid Dynamics}
\nomenclature[E]{DNS}{Direct Numerical Simulation}
\nomenclature[E]{ANN}{Artificial Neural Network}
\nomenclature[E]{DNN}{Deep Neural Networks}
\nomenclature[E]{ML}{Machine Learning}
\nomenclature[E]{LES}{Large-Eddy Simulation}
\nomenclature[E]{RANS}{Reynolds-Averaged Navier-Stokes}
\nomenclature[E]{SCWR}{SuperCritical Water-cooled Reactors}
\nomenclature[E]{LCOE}{Levelized Cost Of Electricity}
\nomenclature[E]{CSP}{Concentrating Solar Power}
\nomenclature[E]{HTD}{Heat Transfer Deterioration} 
\nomenclature[E]{CHF}{Critical Heat Flux} 
\nomenclature[E]{CNT}{Classical Nucleation Theory} 
\nomenclature[E]{VLE}{Vapor-Liquid Equilibria}
\nomenclature[E]{AIM}{Anisotropy Invariant Map}
\nomenclature[E]{PCHE}{Printed Circuit Heat Exchanger}
\nomenclature[E]{CF}{CrossFlow}
\nomenclature[E]{TS}{Tollmien–Schlichting}
\nomenclature[E]{AHFM}{Algebraic Heat Flux Model}
\nomenclature[E]{GGDH}{Gradient Diffusion Hypothesis}
\nomenclature[E]{SGDH}{Simple Gradient Diffusion Hypothesis}
\nomenclature[E]{SST}{Shear Stress Transport}
\nomenclature[E]{WALE}{Wall-Adapting Local Eddy-viscosity}
\nomenclature[E]{EOS}{Equation of State}
\nomenclature[E]{RMSE}{Root-Mean-Square Error}
\nomenclature[E]{MLP}{Multi-Layer-Perceptron}

\nomenclature[F5]{\(Re\)}{Reynolds number}
\nomenclature[F4]{\(Pr\)}{Prandtl number}
\nomenclature[F3]{\(Nu\)}{Nusselt number}
\nomenclature[F6]{SBO}{supercritical boiling number}
\nomenclature[F1]{\(B\)}{Boiling number}
\nomenclature[F2]{$Ec_{\infty}$}{ Eckert number }

\printnomenclature

\newpage
\section{Introduction}

\begin{figure}[hbt!]
\centering
\includegraphics[width=0.65\textwidth]{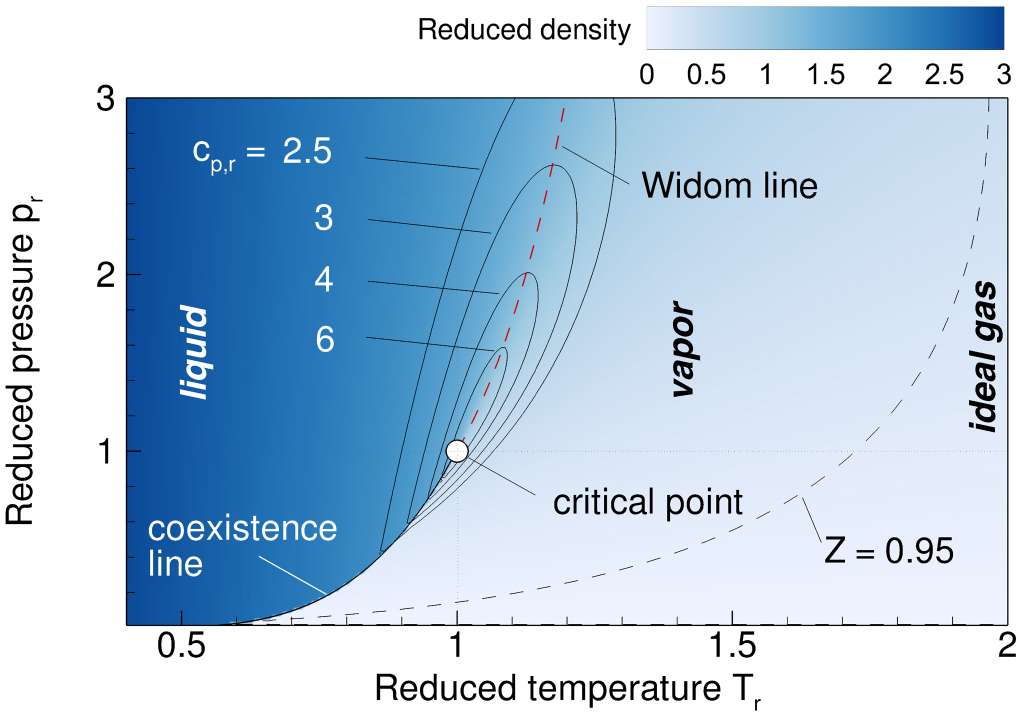}
\caption{Thermodynamic state plane and supercritical state structure. The Widom line is an extension to the coexistence line at supercritical pressure, defined as the locus of maximum isobaric heat capacity. It is a marker of the crossover between supercritical liquid-like and gas-like states. The contour represents the density distribution, showing the sharp transition at subcritical pressure and a smooth crossover at supercritical pressure. The dashed line at $Z=0.95$ denotes the transition to an ideal gas. The reduced density is $\rho_\mathrm{r}=\rho/\rho_\mathrm{cr}$, $R$ is the gas constant, and the isentropic exponent. Data for oxygen from NIST \citep{linstorm1998nist}, figure reproduced from \citep{BanutiPRE2017} with permission given by American Physical Society.}
\label{fig:RevPT}
\end{figure}

A supercritical fluid is a thermodynamic phase of a substance that exists at conditions exceeding its critical temperature and pressure. In the phase diagram Figure ~\ref{fig:RevPT}, the critical point marks the terminus of the liquid-vapor phase boundary, representing the state at which distinctive liquid and gas phases cease to coexist in equilibrium. The thermodynamic landscape at this juncture is highly unique and engenders a set of intriguing properties that warrant careful examination.
At the liquid-vapor critical point, the liquid and vapor densities become equal. Consequently, the surface tension, which usually acts as a barrier between these two distinct phases, diminishes to zero. The vanishing of the surface tension at the critical point means that the equilibrium meniscus or interface between the liquid and gas phases is eradicated. 
Along the coexistence line below the critical point, the transformation of a substance from liquid to vapor requires energy, quantified as the ``latent heat of vaporization''. This heat input corresponds to the energy necessary to overcome the intermolecular forces in the liquid state and to expand against the environmental pressure, allowing the molecules to move freely in the vapor state. As the critical point is approached, the differences between the coexisting equilibrium liquid and gas phases diminish, and the latent heat of vaporization gradually decreases until it becomes zero at the critical point. At supercritical states, fluid properties change continuously between gas-like (at high temperatures) and liquid-like (at lower temperatures) behavior. Thus, supercritical fluids can have the diffusive characteristics of a gas or the enhanced solvent capabilities of a liquid. This combination of properties allows supercritical fluids to behave in a manner that permits transformations that can occur without typical phase transitions.

The unique properties of supercritical fluids find applications in various fields such as analytical chemistry, material processing, and environmental science. For example, supercritical carbon dioxide is widely used in extraction processes, where it combines the penetrative power of a gas with the solvating power of a liquid, facilitating efficient extraction of desired compounds from complex matrices.

The behavior of supercritical fluids is governed by a complex interplay of thermodynamic and transport properties, making them a rich subject of study for researchers interested in fundamental physics and chemistry. Understanding the behavior of substances in the supercritical state not only contributes to the scientific comprehension of phase transitions and critical phenomena but also supports innovations in industrial processes that leverage the distinctive characteristics of supercritical fluids.
Supercritical fluids represent a fascinating state of matter that transcends the conventional boundaries of liquid and gas phases. Their existence above the liquid-vapor critical point allows them to undergo transformations without the traditional phase transitions and bestows them with a unique set of properties that are both scientifically intriguing and industrially desirable.


\subsection{Supercritical fluids for power generation}

\begin{figure}[hbt!]
\centering
\includegraphics[width=0.9\textwidth]{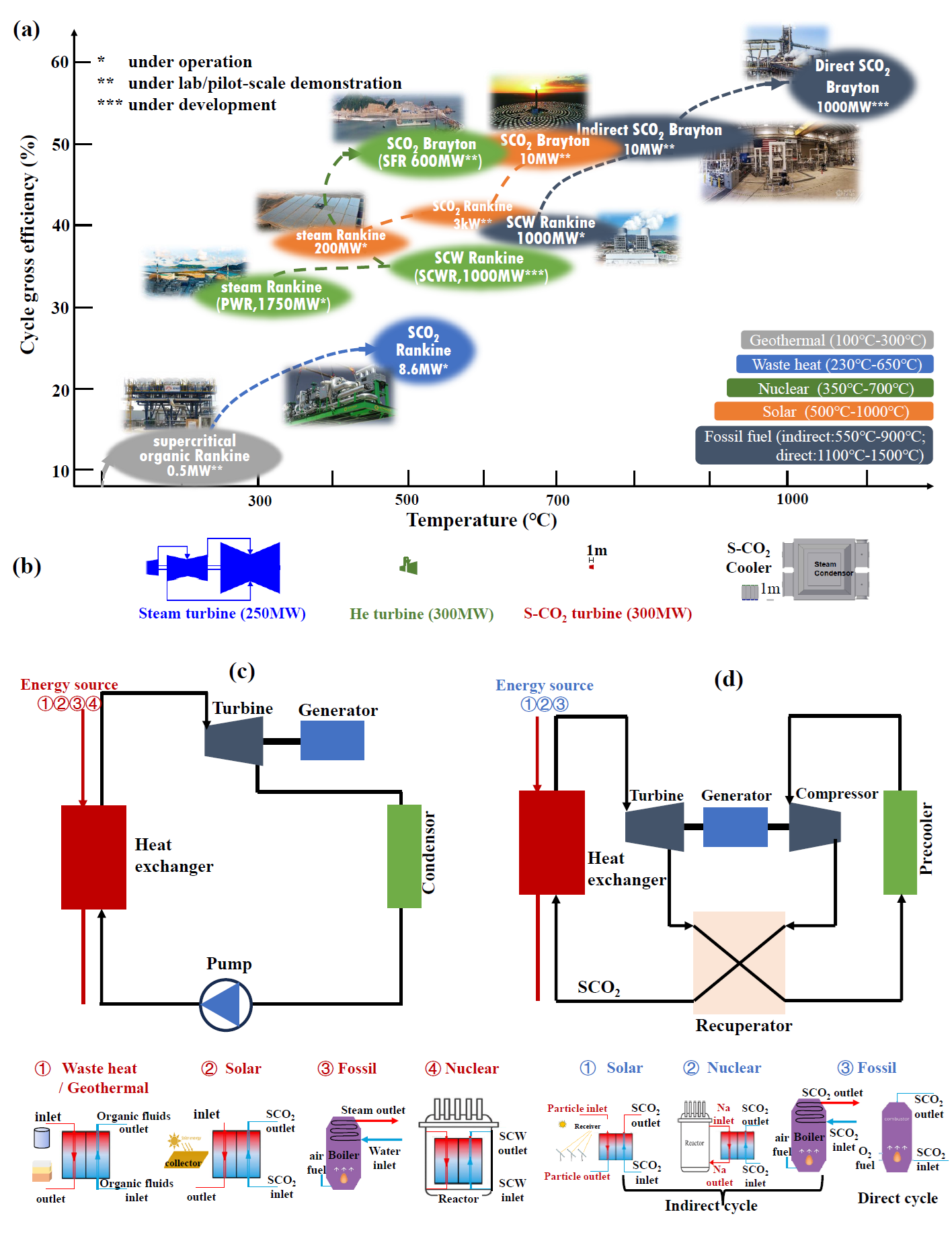}
\caption{\textbf{The developing trend of power cycles towards the supercritical state. a,} Supercritical power cycles and their efficiencies in various energy applications. \textbf{b,} Advantage of smaller turbine and cooler in sCO$_2$ Brayton cycle. \textbf{c,} Layout of a basic Rankine cycle with the heat exchanger adapting to various energy sources. \textbf{d,} Layout of a recuperated Brayton cycle with the heat exchanger adapting to various energy sources. Data in \textbf{a} from 
\citep{fig2_brun2017fundamentals,fig2_echogen,fig2_eDF,fig2_SCWR,fig2_SFR,fig2_sORCtesting,fig2_strakey2014technology,fig2_yamaguchi2008basic}. Part of the images in \textbf{a,b,c,d} reproduced or adapted from \citep{fig2_brun2017fundamentals,ahn2015review,fig2_image_eDF,Fig2_image_SFR,fig2b_image_wright2011supercritical,fig2c_zhu2017power,fig2c_zhang2007experimental,fig2d_fernandez2018dynamic,fig2d_pioro2017handbook}, with permission from Elsevier or license under CC BY 4.0.} 
\label{fig:trend}
\end{figure}

The developing trend of supercritical fluid based power cycles is summarized in Figure \ref{fig:trend}a in terms of source temperature and cycle efficiency. The inception of utilizing supercritical water as a working fluid for fossil-fuel-based steam generators can be traced to the mid-20th century, where it was lauded for its potential to augment thermal efficiency \citep{Starflinger2012,pioro2005sCO2,fan2022general,guo2020energy,schulenberg2013thermal,Schulenberg2014,Schulenberg2022,draskic2024steady}. In contemporary industrial contexts, supercritical water is ubiquitously employed as a working medium in the majority of newly constructed fossil-fuel power plants \citep{Pioro2004}, constituting the preeminent application of supercritical fluids in modern industry. Characteristically devoid of a liquid-vapor phase transition at supercritical pressure, phenomena such as critical heat flux or dry-out are obviated in this state.
The application of the supercritical-water-powered cycle extends to nuclear reactors, specifically in the form of supercritical water-cooled reactors (SCWR). With roots dating back to the 1950s and resurgence in the 1990s, the SCWR concept was endorsed by the Generation IV International Forum as one of six highly promising reactor designs \citep{schulenberg2013thermal,Schulenberg2014, Schulenberg2022}. SCWRs bear the capacity to elevate the thermal efficiency of contemporary nuclear power facilities from roughly 33–35$\%$ to proximate values of 45–50$\%$, constituting a substantial fiscal and environmental yield \citep{Pioro2004}.
Using carbon dioxide ($p_\mathrm{cr}=7.38$ MPa, $T_\mathrm{cr}$= 304.25 K) instead of water ($p_\mathrm{cr}$=22.06 MPa, $T_\mathrm{cr}$= 674.09 K) leads to reduced material and construction costs. Supercritical CO$_2$ Brayton cycles can operate more efficiently at lower temperatures, improving thermal efficiency and potentially lowering operational costs. 
The supercritical CO$_2$ Brayton cycle's compact size (see Figure \ref{fig:trend}b) and reduced complexity are a direct result of the fluid's superior thermophysical properties, enabling more efficient heat exchange and energy conversion processes.
The application domains of the S-CO$_2$ Rankine or Brayton cycle are multifarious, encompassing nuclear, fossil fuel, waste heat, and renewable energy sources such as solar thermal and fuel cells. One of the major differences in the cycles applying to various energy sources lies in the heat exchanger component as shown in Figure \ref{fig:trend}c, the performance of which highly depends on the transcritical phenomena of S-CO$_2$.

\begin{figure}[hbt!]
\centering
\includegraphics[width=1\textwidth]{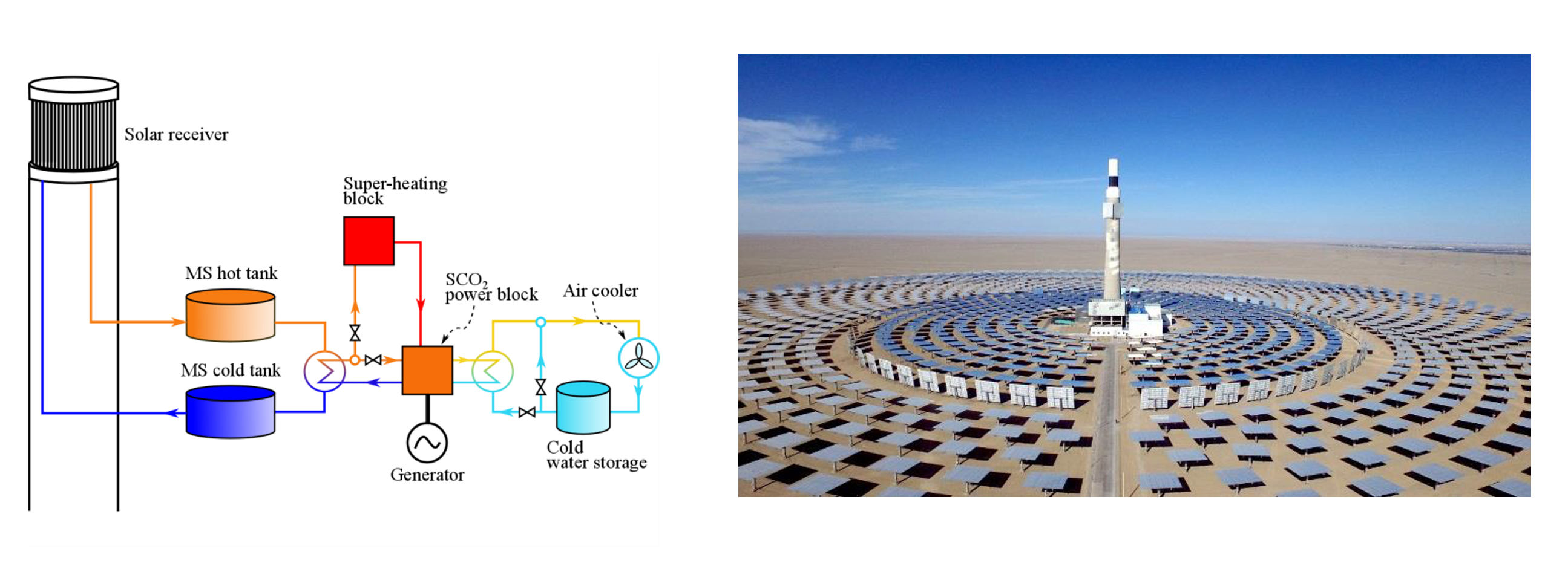}
\caption{10 MW CSP power plant combined with sCO$_2$ cycle, Shouhang-EDF. Reproduced from \citep{le2019shouhang},  licensed under CC BY 4.0.}
\label{fig:CSP}
\end{figure}

For example, concentrating solar power (CSP) systems are highly scalable and efficient, making them ideal for large-scale renewable energy production, particularly in regions with abundant sunlight \citep{dunham2014high,calderon2018high} (Figure \ref{fig:CSP}).
The SunShot project \citep{mehos2016path} anticipates that with the incorporation of advanced technologies such as the S-CO$_2$ power cycle, the Levelized Cost of Electricity (LCOE) for CSP will fall below \$\text{0.06/kWh} \citep{murphy2019potential}.
The future trend of CSP seems to lean towards higher operating temperatures. In the context of the third-generation CSP systems, it can be anticipated that the receiver's inlet temperature will surpass 750$^\circ$C. However, the capacity of the steam Rankine cycle to enhance thermal efficiency under these high-temperature conditions is restricted. Consequently, the adoption of the supercritical carbon dioxide cycle becomes more appealing, as it offers a thermal efficiency greater than 50\%.
\citet{wang2017integration} reviewed various S-CO$_2$ Brayton cycle layouts for potential integration into established molten salt solar power tower systems, employing a comprehensive modeling approach. They compared these cycles in terms of efficiency, specific work, and compatibility with thermal energy storage, revealing that while current S-CO$_2$ layouts enhance efficiency, challenges remain in optimizing specific work and temperature differential, necessitating further development of novel cycle layouts.
Furthermore, \citet{wang2017thermodynamic} presented an integrated molten salt solar power tower system combined with a S-CO$_2$ Brayton cycle, featuring a detailed model encompassing the heliostat field, solar receiver, thermal storage, and Brayton cycle with reheating. It focused on parametric analysis and optimization using a genetic algorithm to maximize exergy efficiency, exploring the impact of key thermodynamic parameters and component performance, and concludes that novel salts with higher allowable temperatures are crucial for enhancing system efficiency.
\citet{prieto2020carbonate} examined the potential of integrating S-CO$_2$ power cycles into CSP to enhance thermal-to-electric conversion efficiency beyond 50\%. It highlighted the need for delivering solar energy at temperatures around or above 700°C, a significant increase from the current 565°C in advanced plants. Their pioneering work in designing, building, and testing a high-temperature (700$^\circ$C) molten salt experimental pilot plant using carbonate salts is showcased, demonstrating the viability of large-scale operations within this temperature range. 
Research into the dynamic performance of CSP is also important due to their fluctuating and intermittent nature. These systems often deviate from their design points, presenting more instability compared to conventional power plants. Therefore, understanding the dynamic behavior of S-CO$_2$ solar power systems under off-design conditions and developing swift control strategies is crucial for ensuring their reliable and stable operation amidst varying parameters \citep{luu2018advanced,ma2020development}.

Another important application of S-H$_2$O/S-CO$_2$ cycles is in nuclear energy. One of the Global Generation IV nuclear reactors is the Supercritical-Water-Cooled Reactor (SCWR). 
Switching from water to \ce{CO2} as the coolant is attractive due to the lower critical temperature and pressure of \ce{CO2}. The conceptional idea started with the dissertation of \citet{Dostal2004}.
Recently, KAIST has led the design of the micro modular reactor (MMR) with a size of 3.8 m (height)×3.7 m (width)×7.0 m (length) with the 36.2MW core thermal power which can be contained in a small vessel and suitable for vehicle transportation \citep{kim2017concept}. In addition, S-CO$_2$ cycle can also be coupled with Sodium-cooled Fast Reactor (SFR)  to improve safety and economic \citep{jung2016investigation,eoh2013potential}.
Further, the application of S-CO$_2$ cycle on coal-fired power plants also attracts attention due to its potential high efficiency and compact design \citep{xu2018key,yang2020coupled,guo2020energy,le2013conceptual}.

\begin{figure}[hbt!]
\centering
\includegraphics[width=0.8\textwidth]{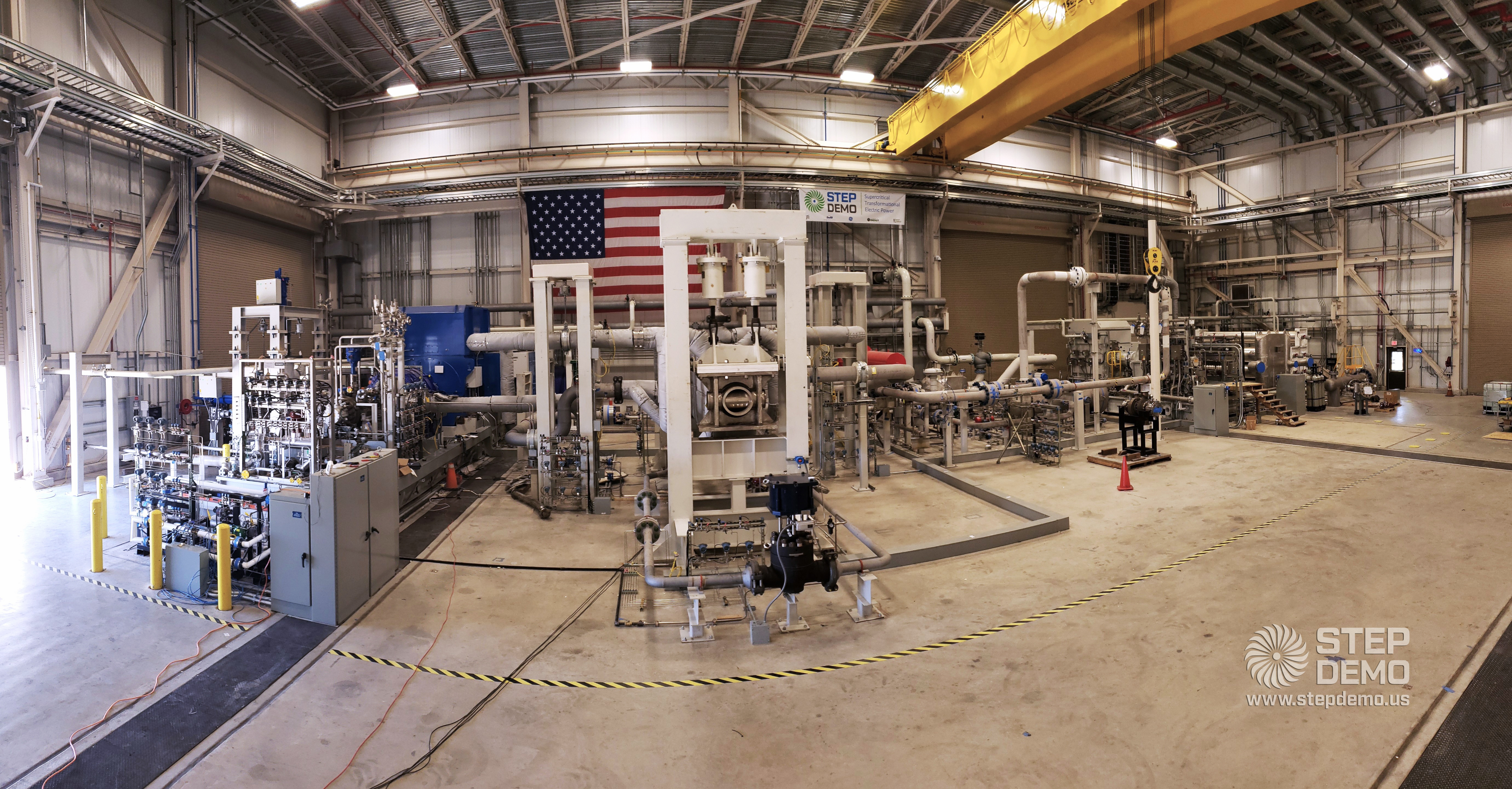}
\caption{10 MW Supercritical Transformational Electric Power (STEP) Demo pilot plant by Southwest Research Institute (SwRI), figure from \href{https://www.swri.org/press-release/step-demo-pilot-plant-makes-new-breakthrough-sco2-power-generation}{press release}, used for educational and informational purposes.}
\label{fig:STEP}
\end{figure}

More excitingly, in the last decade, the experimental loops and demonstration projects of the S-CO$_2$ cycle were set up worldwide including the United States, Korea, Japan and China. 
A team led by Gas Technology Institute (GTI), Southwest Research Institute (SwRI) and General Electric Global Research (GE) has initiated a project to design, construct, commission, and operate a versatile and reconfigurable 10 MWe Supercritical Carbon Dioxide (S-CO$_2$) Pilot Plant Test Facility located at SwRI’s San Antonio, Texas campus (Figure \ref{fig:STEP}). The project called STEP Demo (Supercritical Transformational Electric Power) is one of the largest scale and most comprehensive in the world.
In China, the S-CO$_2$ cycle demonstration projects are being built for solar thermal power plants and fossil-fired power plants. A 10MW CSP plant was applied in Dunhuang (Figure \ref{fig:CSP}) \citep{moore2020development,li2019design,le2019shouhang}. 

In the S-CO$_2$ cycle, the performance of the components influences the thermodynamic performance and the commercial application such as turbomachinery \citep{ameli2019centrifugal,odabaee2016cfd,qi2017supercritical,pelton2017design}. The transcritical phenomena with the highly non-ideal gas require a reconsideration of the compressor design with a deeper understanding of the transcritical flow at the compressible range. 
Heat exchangers are also vital in S-CO$_2$ cycles due to their role in ensuring efficient heat transfer, which is crucial for maintaining the supercritical state of CO$_2$ under high temperatures and pressures. Their compact design contributes to the overall reduction in size and cost, while also enhancing thermal efficiency. Furthermore, the stability and safety of the S-CO$_2$ system heavily depend on the effective performance of these heat exchangers, given the extreme operational conditions they must endure \citep{kwon2018experimental,yoon2014assessment}. This can only be guaranteed with extensive knowledge of flow and convective heat transfer in the pseudo-critical region.

\begin{figure}[hbt!]
\centering
\includegraphics[height=0.335\textwidth]{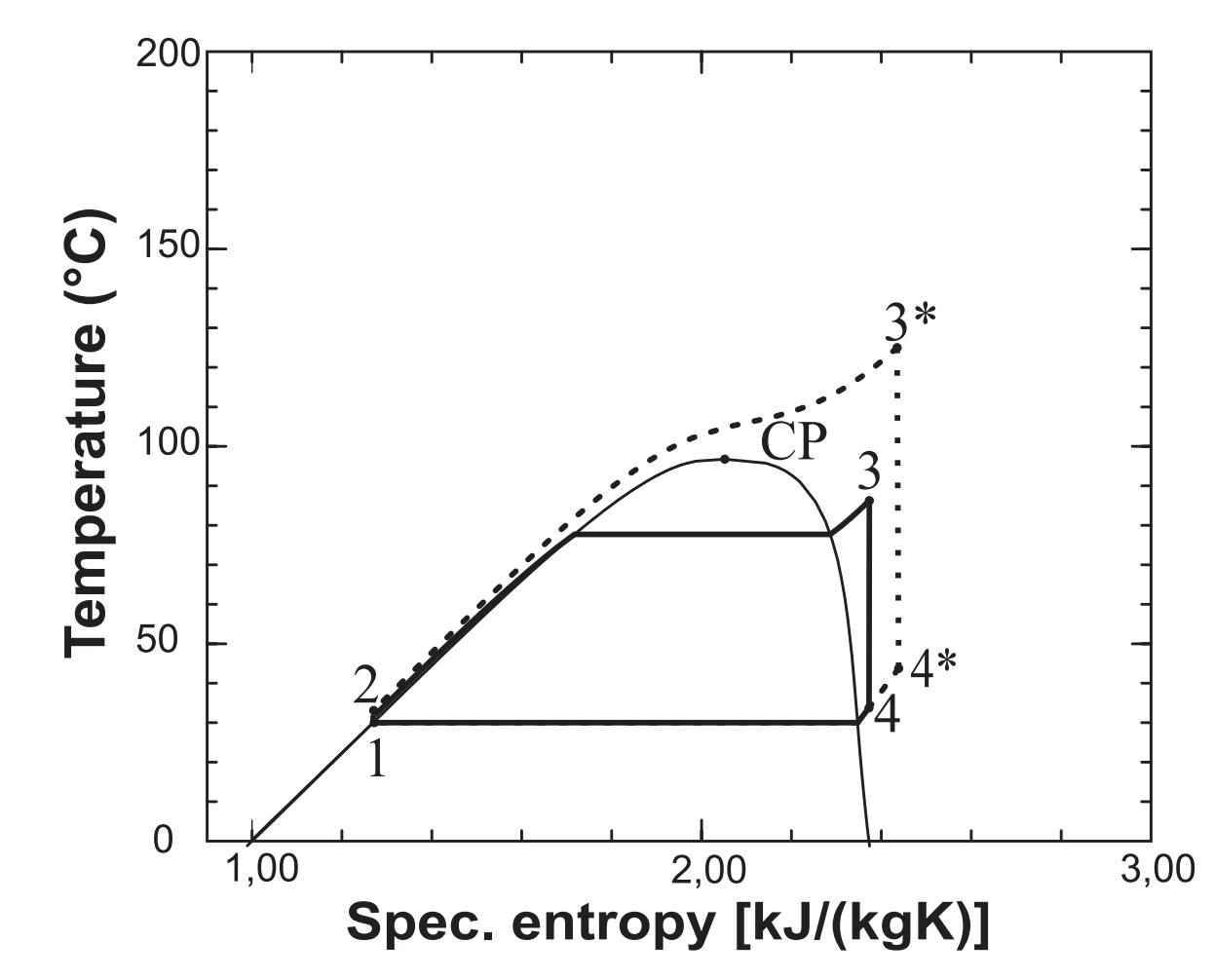}
\includegraphics[height=0.32\textwidth]{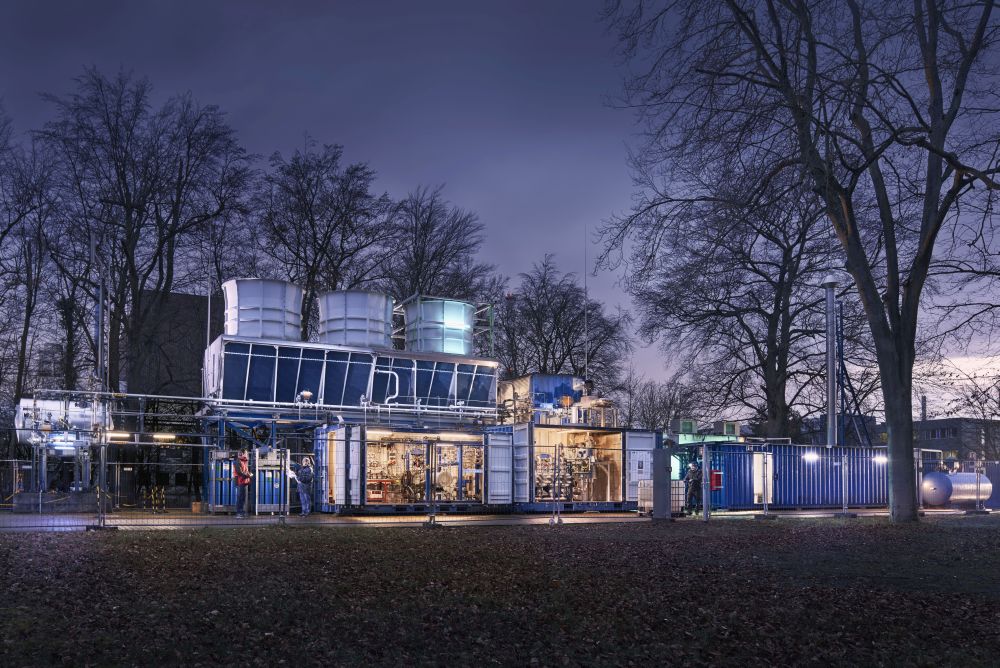}
\caption{The supercritical organic Rankine cycle MoNiKa (Modular Low Temperature Cycle Karlsruhe) at the Karlsruhe Institute for Technology (KIT). Left: Comparison of sub- and supercritical Rankine cycle, reproduced from \citep{Vetter2017} with permission given by Elsevier. Right: Photo of the facility, a 1~MW$_\mathrm{thermal}$ supercritical ORC using propane as working fluid.}
\label{fig:MoNiKa}
\end{figure}

Supercritical organic Rankine cycles (ORC), on the other hand, are well suited for low temperature heat sources, like geothermal or process waste heat. Figure~\ref{fig:MoNiKa}(left) compares sub- and supercritical cycle processes in a $T$-$s$ diagram. It is apparent how, by increasing the pressure of heat addition to supercritical values, the cycle power is increased substantially now that heat is no longer added isothermally during a phase transition. Further, heat exchangers can be operated more efficiently when an approximately constant temperature difference can be sustained. Due to the low overall temperatures, the increased turbine inlet temperature caused by the increased pressure is of no concern here.  Figure~\ref{fig:MoNiKa} (right) exemplarily shows the Modular Low Temperature Cycle Karlsruhe (MoNiKa) of the Karlsruhe Institute for Technology (KIT), a 1~MW$_\mathrm{thermal}$ supercritical ORC using propane as working fluid \citep{Vetter2017}.
Ongoing challanges in supercritical ORC technologies are transients, compressor, expander/turbine \citep{hsieh2017-TORC,kosmadakis2016-TORC,landelle2017-TORC}.

\subsection{Supercritical fluids in aerospace engineering and transportation}

High pressures are popular in a wide range of transport applications for road, air, and space. It can improve the efficiency of burning fuel and increase the thrust produced by engines, which is crucial for designing and improving propulsion systems \citep{gerber2021fluid,madana2021unpicking,BanutiCnF2016,BanutiCnF2018,BanutiDiss,Banuti2010,Banuti2013,Banuti2008,Banuti2009, Banuti2014, pizzarelli2009numerical, pizzarelli2010numerical, pizzarelli2015heat,jofre2015diffuse,jofre2017interface}.
Specifically, in the realm of rocket engines, these pressures range from 40 to 560 bar, reflecting the stringent demands associated with space exploration and satellite deployment \citep{Jofre2021}. This high-pressure environment facilitates the generation of substantial thrust, rendering it a cornerstone of modern rocketry. In contrast, gas-turbine jet engines employed in commercial and military aviation typically operate within a pressure window of 20 to 60 bar during takeoff \citep{BanutiCnF2016}. 
The critical point of China RP-3 aviation kerosene is approximately $p_\mathrm{cr}=24$ bar and $T_\mathrm{cr}=640$ K.
Diesel engines, widely utilized in various ground-based applications such as transportation and industrial machinery, operate at  modest to high pressure levels. In high performance engines, a compression ratio from 20 to 23 is expected, corresponding to pressures of 66-80 bar \citep{BanutiPoF2016}.

Even within this pressure regime, the optimization of combustion pressure remains a critical factor, impacting not only fuel efficiency and power output but also influencing parameters such as noise, vibration, and emissions.
The dynamic interaction of pressure, temperature, and fuel mixture within the combustion process poses intricate challenges and offers fertile ground for further research and development. Advances in computational modeling, materials science, and sensor technology are contributing to a more profound understanding of high-pressure combustion dynamics, opening avenues for innovations that promise enhanced performance, reliability, and sustainability across a wide range of propulsion applications.
The optimal control of pressure within combustion systems represents a critical aspect of propulsion engineering, with implications extending across efficiency, environmental compliance, and operational robustness. The continued exploration of high-pressure combustion dynamics, facilitated by technological advancements and interdisciplinary collaboration, is poised to yield transformative impacts on the future landscape of mobility and energy utilization, aligning with broader societal and environmental imperatives.

\begin{figure}[hbt!]
\centering
\includegraphics[width=1\textwidth]{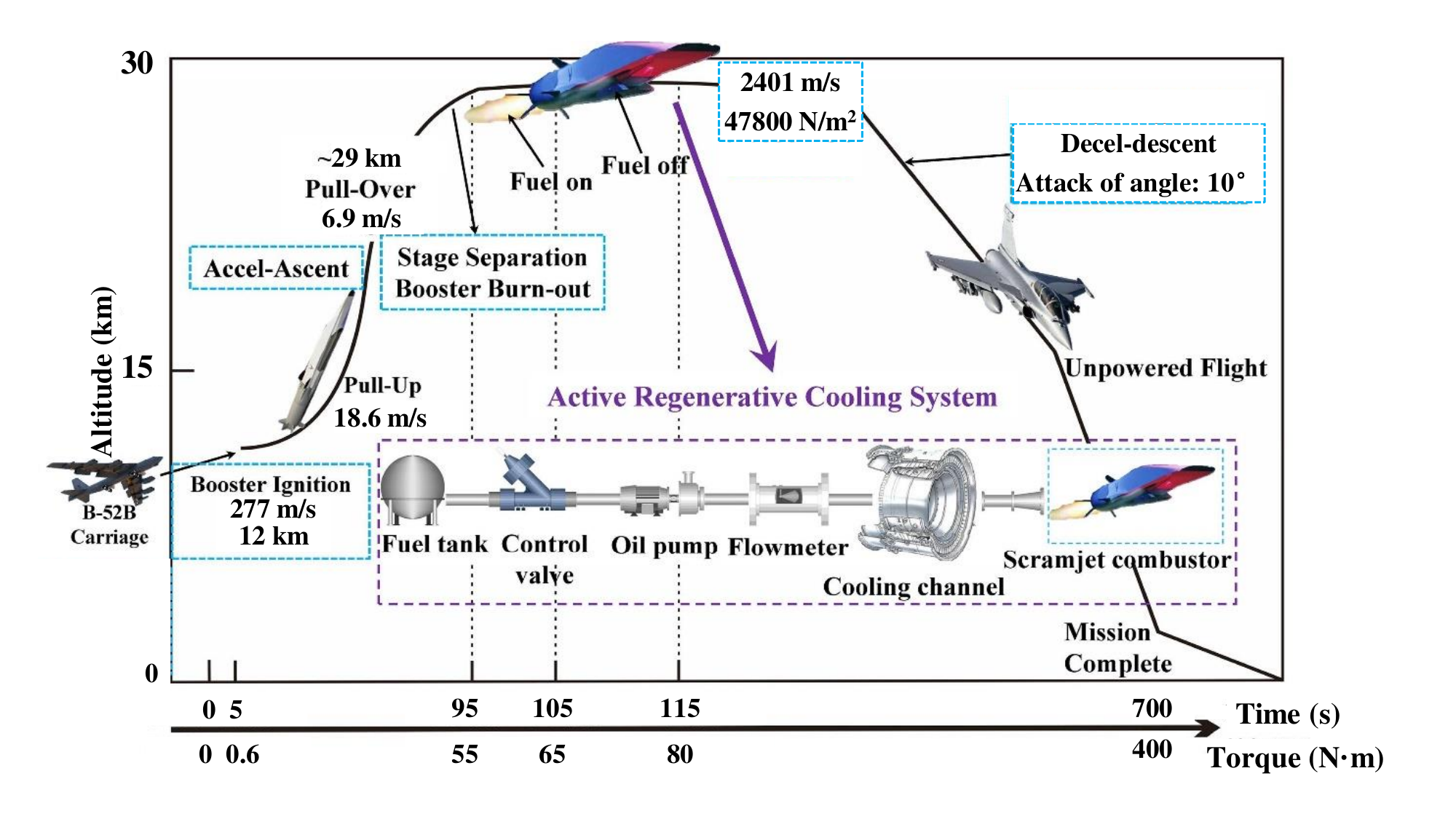}
\caption{Active regenerative cooling, reproduced from \citep{chen2020regenerativecooling} with permission given by Elsevier}
\label{fig:rege}
\end{figure}

In scramjet combustors, supercritical aviation kerosene, which also serves as fuel, is used for active regenerative cooling \citep{tsujikawa1996effects}, as shown in Figure \ref{fig:rege}. This technique is essential for managing the heat generated during high Mach number flights. The process begins with the compression of aviation kerosene using a pump, preparing it for combustion. The kerosene is then channeled around the combustor, where it absorbs the heat generated by combustion. After this heat absorption, the now-heated aviation kerosene is directed into the combustion process. Additionally, for aviation aircraft equipped with turbines, aviation kerosene can be utilized in a kerosene-air heat exchange system.





\subsection{Objective and outline}

In conjunction with the original research being published, there has been an effort to collate these findings into comprehensive reviews from various perspectives \citep{fan2022general,guo2020energy,ahn2015review,Jofre2021,Ehsan2018,Zhang2020,Yoo2013,Liu2019,crespi2017supercritical, pizzarelli2018status}. A recent review by \citet{guo2020energy} systematically examined the supercritical carbon dioxide power cycle for energy industries, exploring state-of-the-art technologies, key issues, and future prospects. \citet{Jofre2021} discussed transcritical diffuse-interface hydrodynamics of propellants in high-pressure chemical propulsion systems for aerospace, with an emphasis on the theoretical foundations of transcritical multi-phase and multi-component flow. The review by \citet{Ehsan2018} focused on heat transfer and pressure drop characteristics with supercritical CO$_2$ under heating and cooling conditions, particularly on experimental observations and the creation of heat transfer correlations. Meanwhile, \citet{Zhang2020} explored the application and enhancement of heat transfer of supercritical CO$_2$ in low-grade heat conversion. About a decade ago, \citet{Yoo2013} offered a review on the deteriorated heat transfer phenomena based on newly developed direct numerical simulation (DNS) from his group, shedding light on fundamental fluid mechanics within a limited range of conditions. Since then, many exciting research developments have brought new insights into fluid physics. Therefore, there is a compelling need for a review that encompasses the state-of-the-art understanding of the fundamental thermodynamics and fluid mechanics of single-phase and mono-component supercritical fluids, an area currently unexplored. We aspire for our review to impart valuable information for research and design across various levels and contribute to the advancement of renewable energy technology and aerospace engineering.
Recently, \citet{guardone2024nonideal} summarized the progress on non-ideal compressible fluid dynamics of dense vapors and supercritical fluids. The behavior of single-phase, non-reacting fluids near vapor-liquid saturation, near the vapor-liquid critical point, or under supercritical conditions, both quantitatively and qualitatively, deviates from the classical gas dynamics described for dilute, ideal gases in textbooks.

Our review article summarizes inspiring new fundamental perspectives of mostly single-phase and single-component supercritical fluids and shows the potential to use them in engineering research and development. 

Section \ref{thermodynamics} introduces the relevant microscopic physics of near-critical thermophysical properties, phase transitions, and fluid dynamics. These are explored based on fundamental thermodynamic analysis, molecular dynamics simulations, and in situ neutron imaging measurements. Special emphasis is placed on the concept of supercritical `pseudo boiling', a phenomenon that extends the classical understanding of liquid-vapor phase transitions to non-equilibrium supercritical processes. Since its introduction by \citet{Banuti2015}, the quantitative view of pseudo boiling has been successfully applied to (i) identify and explain a new, thermal, jet break-up mechanism (\citet{BanutiPoF2016}) for injection at supercritical pressure; (ii) reveal that supercritical heat transfer deterioration (HTD) can be solely explained as a thermodynamic phenomenon (\citet{longmire2022onset}); (iii) realize that a latent heat of vaporization can be identified for the supercritical liquid-gas (and vice versa) non-equilibrium phase transition (\citet{BanutiPPCE2019,BanutiJSF2020}); (iv) demonstrate the existence of supercritical stable and gradient-steepening interfaces between liquid and gaseous states in absence of surface tension (\citet{LongmireNatComms2023}). These unexpected insights prompt a reevaluation of the existing understanding of the thermodynamic state space and its processes.

Section \ref{fluid} provides a review of the research on convective heat transfer and fluid flow in the past decade. Initially, the progress on the scaling law of the velocity and temperature is discussed. The development of the scaling laws with the consideration of supercritical fluids lays down a foundational framework for forthcoming advancements in turbulence modeling. Then, the section offers a comprehensive benchmark of supercritical heat transfer in pipe flow with DNS from various research groups. The observed quantitative discrepancies in predicting wall temperature indicates the difference introduced by different numerical solvers. Additionally, the section introduces new DNS results in other conditions such as  horizontal flows, non-circular pipes, and non-uniform heat fluxes. These extensions significantly broaden the scope of flow and heat transfer conditions, going beyond previous works \citep{Bae2005,Bae2008}, and furnish a more thorough comprehension of heat transfer phenomena.


Section \ref{stability} presents the latest progress in analyzing flow stability, focusing on non-ideal fluids. This part is crucial for the ongoing endeavors to understand the transition from laminar to turbulent flows. Present insights predominantly cover the linear domain, yet grasping the full scope of the transition necessitates considering nonlinear interactions. These are significant where the sharp gradients of fluid properties can markedly influence the process.

Section \ref{model} provides a summary of recent developments in turbulence modeling, including both RANS and LES methods to predict the heat transfer of supercritical fluids. Advanced RANS models move beyond traditional two-equation $k-\varepsilon$ and $k-\omega$ models and encompass recent advances in physics-informed machine learning. The LES section details progress in both wall-resolved and wall-modeled LES for supercritical fluids. Additionally, the section highlights the application of machine learning, specifically Artificial Neural Networks (ANN), for the rapid prediction of heat transfer in supercritical fluids. According to the latest research, ANN is capable of predictions as fast as conventional heat transfer correlations but with a significant improvement in accuracy. This novel approach has strong potential for contributing to system-level modeling of heat exchangers.

\section{Physics and thermodynamics of supercritical fluids}
\label{thermodynamics}

The coexistence region, terminating at the critical point where the pressure $p$, temperature $T$, and density $\rho$ reach their respective critical values $\pcr$, $\Tcr$, $\rho_\mathrm{cr}$, is the most striking feature in the fluid state space. At subcritical pressures $p<\pcr$, this coexistence region has a large impact on the nature of heat transfer, where condensation and vaporization can release or absorb amounts of energy to an extent not observed in single phase fluids, enhancing or inhibiting heat transfer beyond simple Fourier conduction \cite{Carey2008}: heat flux from a wall to a pool is enhanced up to a maximum value (critical heat flux--CHF) when rising vapor bubbles absorb energy while limiting temperature increase, simultaneously inducing a heat-flux enhancing flow. Upon further wall temperature increase, heat flux from a wall to a pool is inhibited towards the Leidenfrost point (boiling crisis), when the bubbles coalesce into an insulating vapor film at the wall. In addition, phase change phenomena cause large variations in density, making convective heat transfer susceptible to acceleration and orientation; vaporization causes rapid expansion of the fluid, which in turn may affect boundary layer flow or turbulence.

This begs the question: Is there something akin to phase transitions at supercritical pressures, beyond the coexistence region, that may affect heat transfer in a manner similar to subcritical processes? We will address this question in this section.

\subsection{Introduction to supercritical fluids}
For this to answer, we need to discuss characteristics of supercritical fluids, a topic which has seen much advancement over the last decade.

\subsubsection{High pressure fluid properties}
\begin{figure}[hbt!]
\centering
\includegraphics[height=.24\textwidth]{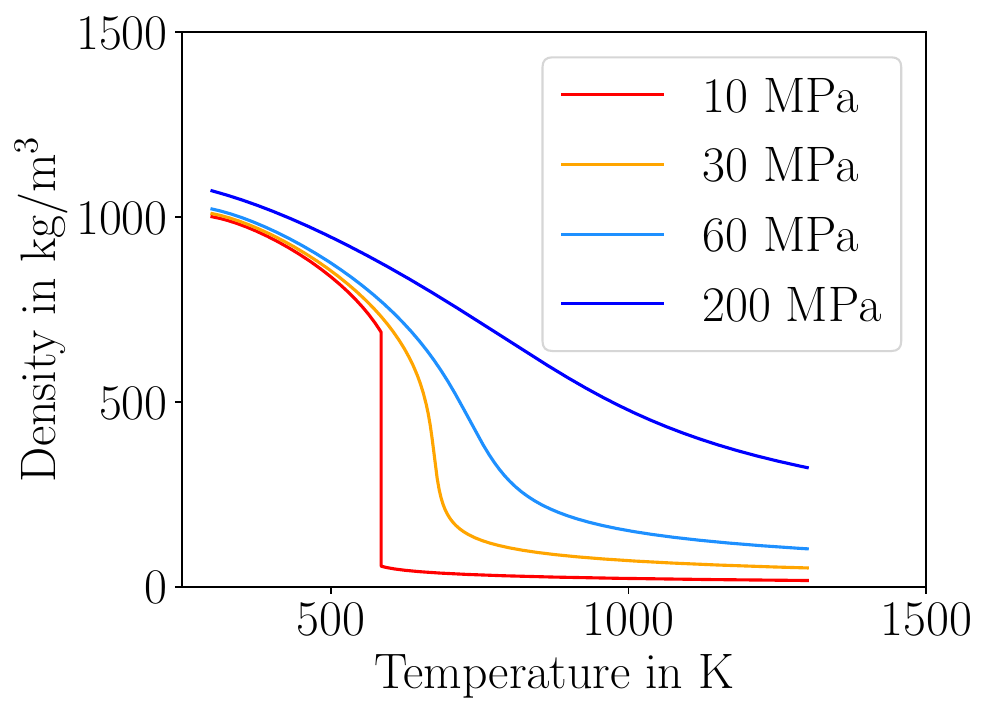}
\includegraphics[height=.24\textwidth]{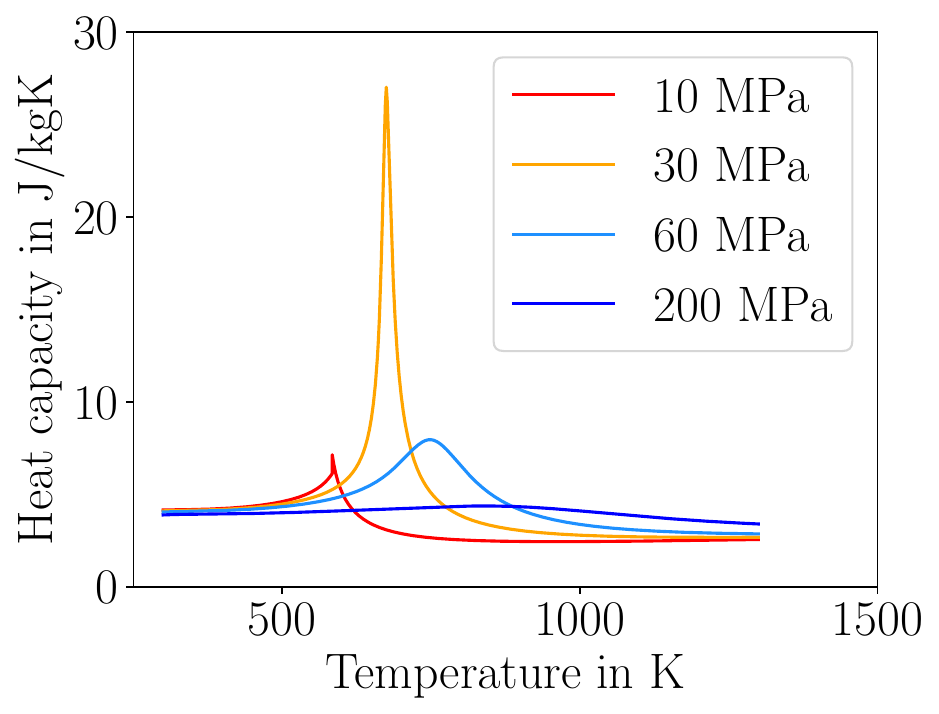}
\includegraphics[height=.24\textwidth]{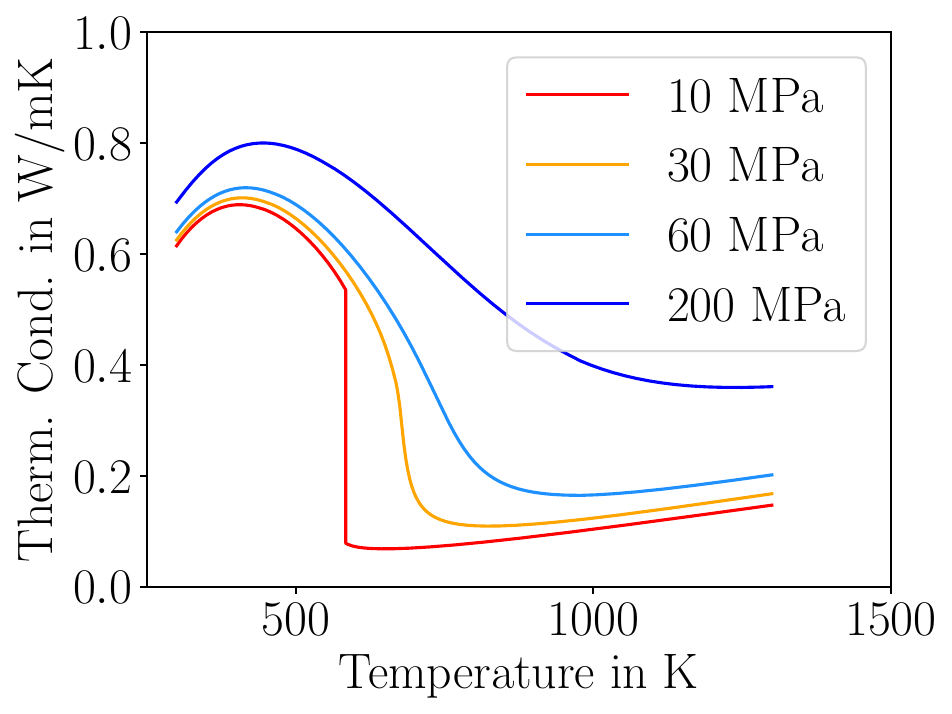}
\includegraphics[height=.24\textwidth]{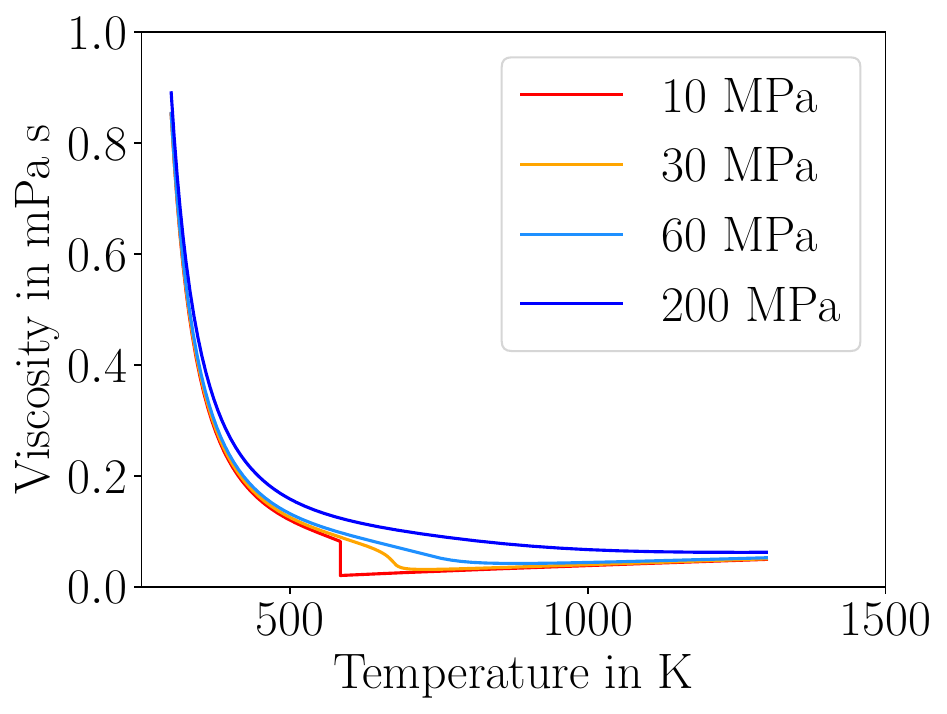}
\includegraphics[height=.24\textwidth]{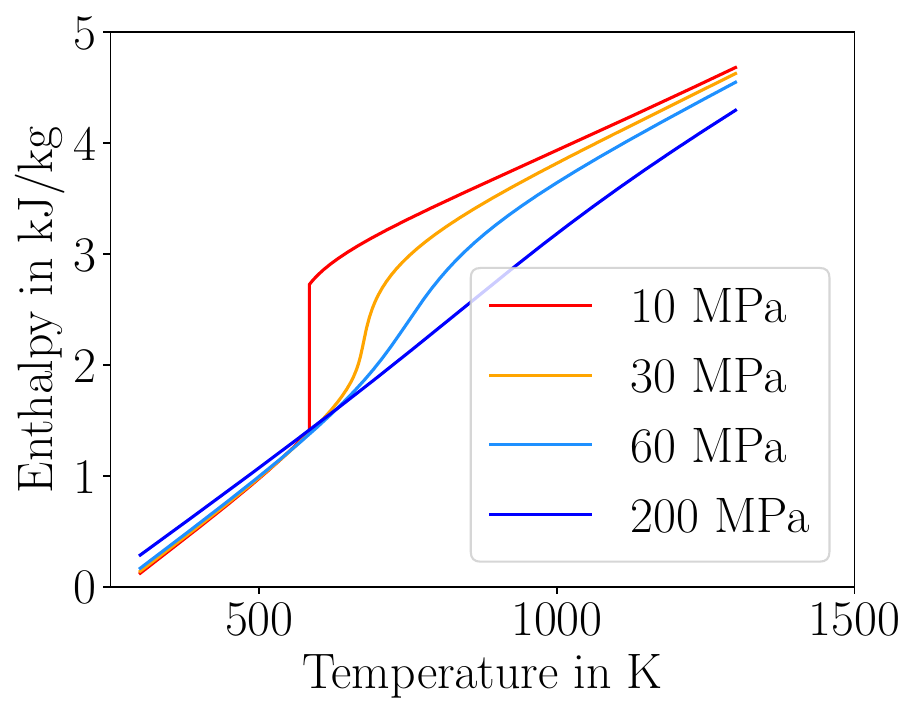}
\includegraphics[height=.24\textwidth]{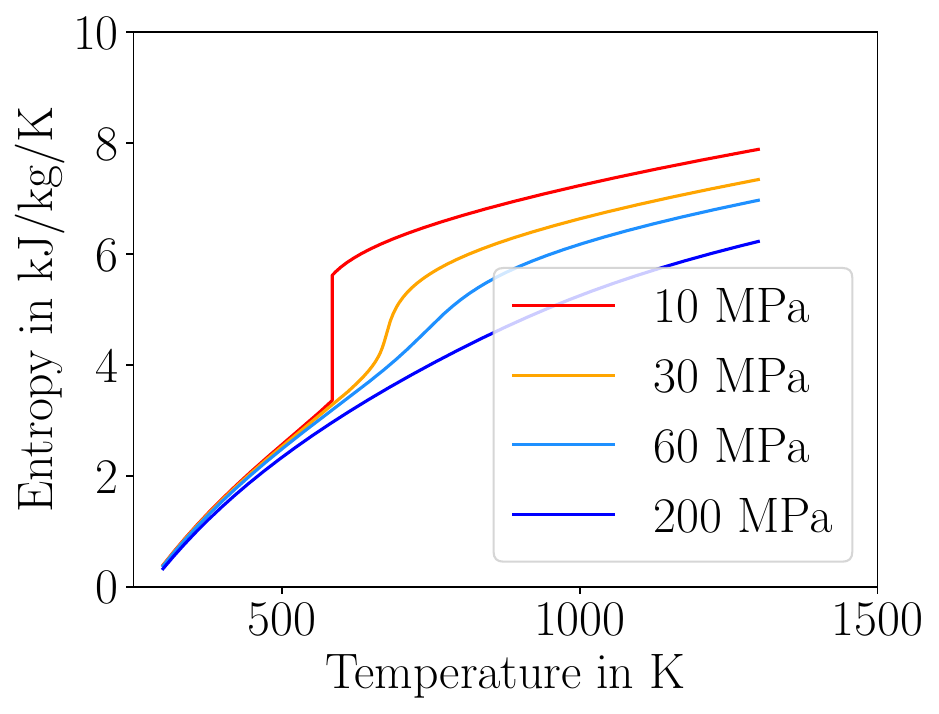}
\caption{Fluid properties for water at subcritical (10~MPa), and supercritical (30, 60, 200~MPa) pressures.}
\label{fig:PB:waterprops}
\end{figure}

\begin{figure}[hbt!]
\centering
\includegraphics[height=.24\textwidth]{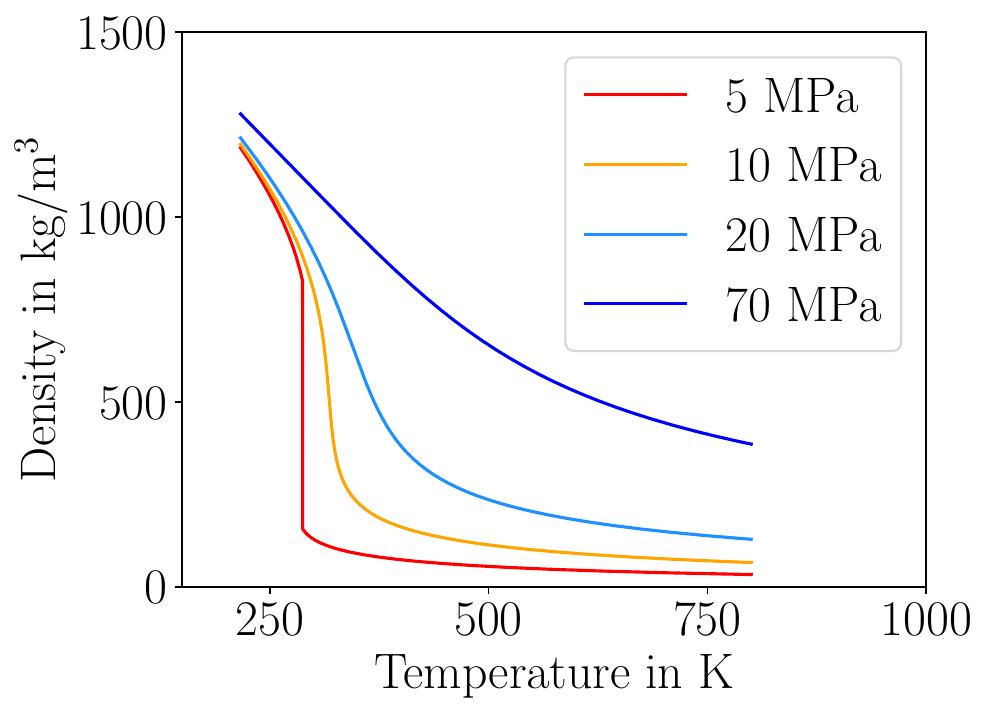}
\includegraphics[height=.24\textwidth]{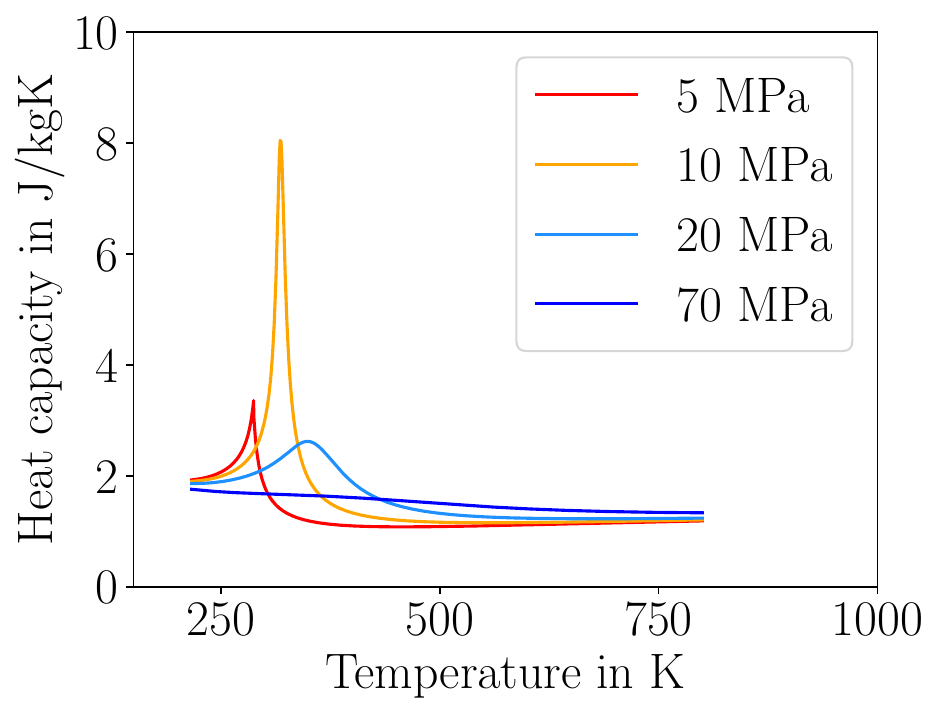}
\includegraphics[height=.24\textwidth]{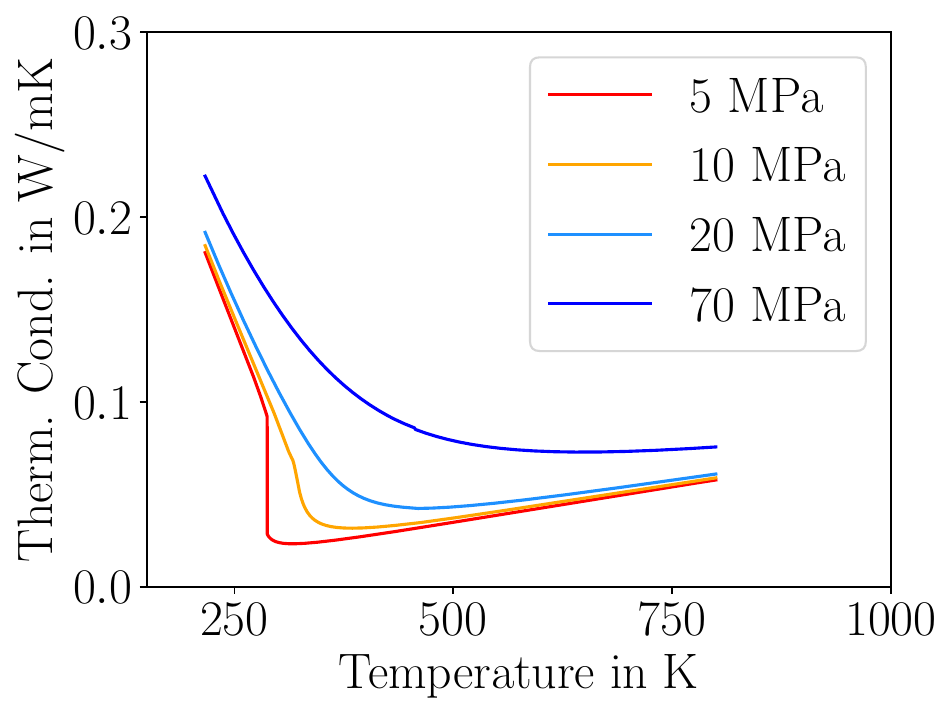}
\includegraphics[height=.24\textwidth]{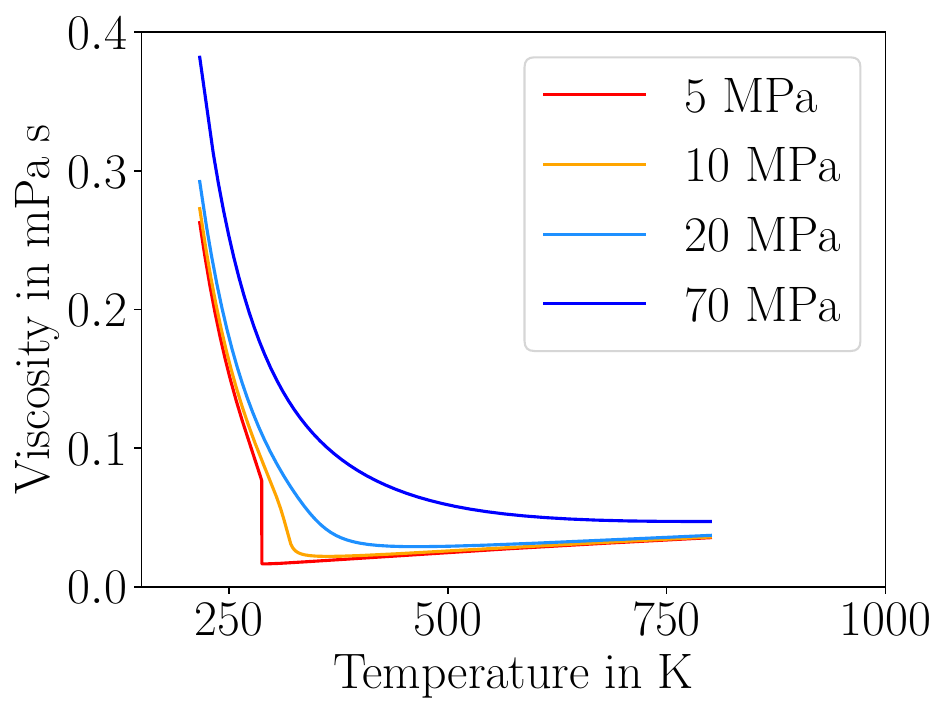}
\includegraphics[height=.24\textwidth]{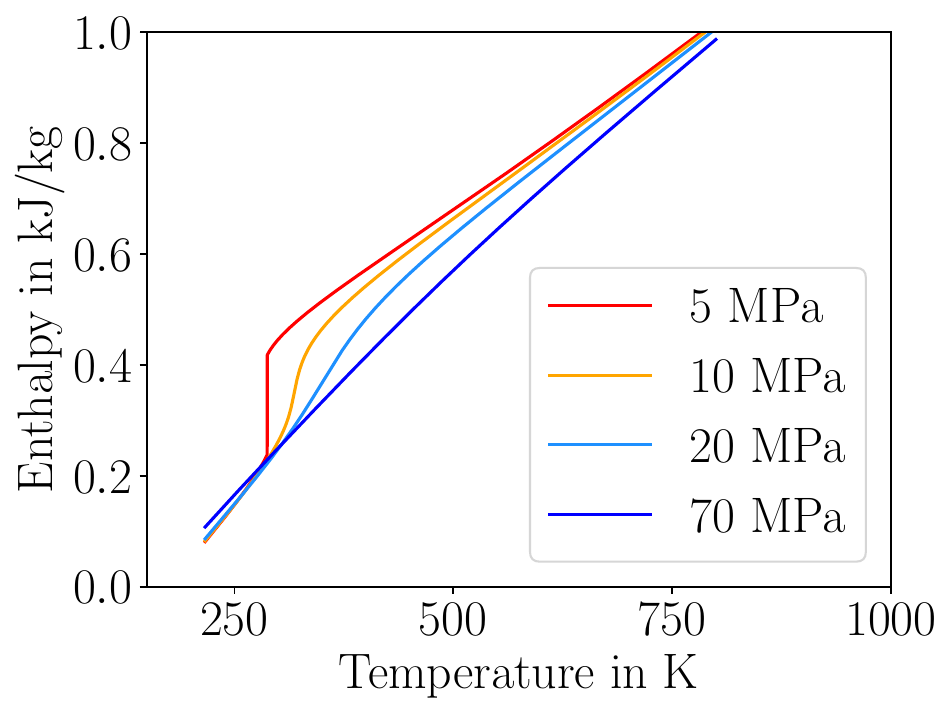}
\includegraphics[height=.24\textwidth]{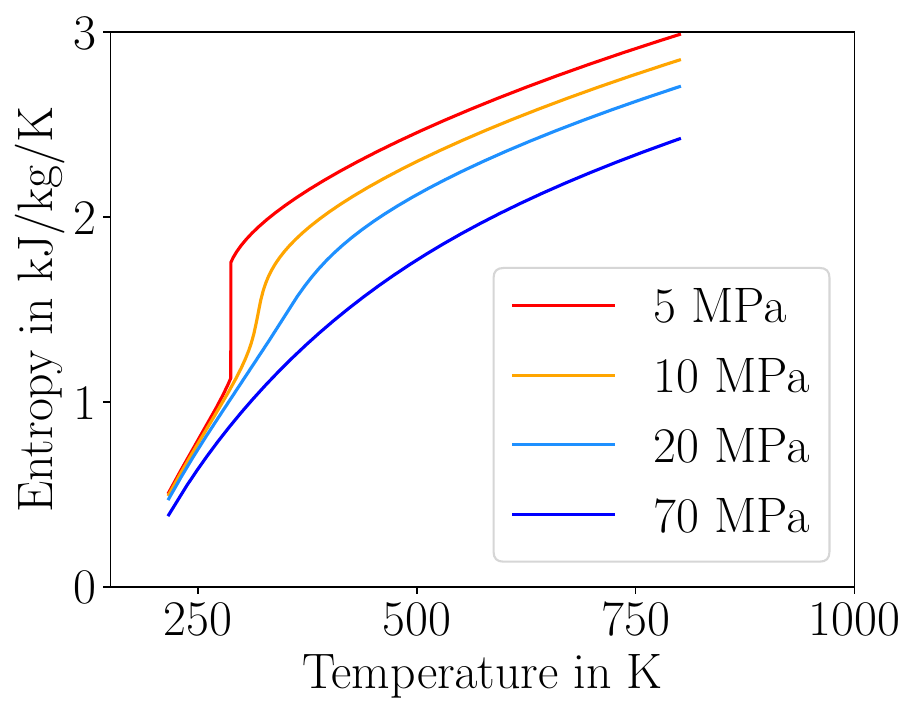}
\caption{Fluid properties for carbon dioxide at subcritical (5~MPa), and supercritical (10, 20, 70~MPa) pressures.}
\label{fig:PB:co2props}
\end{figure}

The particular and extraordinary properties of fluids at near-critical and supercritical pressures have been known since at least the 1970s \cite{Schmidt1960,Hendricks1970} and remain an important point to this date \cite{Yoo2013,Yoon2018}. Figures~\ref{fig:PB:waterprops} and \ref{fig:PB:co2props} show various fluid properties for water and carbon dioxide, respectively. Two things are of particular note: first, fluids exhibit an extreme variation of properties in a narrow temperature range in extension of the coexistence line; second, this is a general property seen across different fluids.

\subsubsection{Discovery and concept of supercritical fluids}
Two centuries ago, in 1822, Cagniard de la Tour sealed fluids in in glass tubes and discovered that the liquid-vapor meniscus vanished when sufficiently heated \cite{Andrews1869}. 50 years later, Andrews explained this phenomenon in terms of the critical point beyond which phase coexistence was no longer possible \cite{Andrews1869}, and van der Waals developed his equation of state \cite{Waals1873}
\begin{equation}
    p = \frac{RT}{v_m\underbrace{-b}_{\text{repulsion}}} - \underbrace{\frac{a}{v_m^2}}_{\text{attraction}} \; .
\label{eq:vdW}
\end{equation}
This equation finally captured relations between pressure $p$, temperature $T$, and molar volume $v_m$ across gaseous, liquid, and supercritical fluid states, by accounting for molecular repulsion (or displacement) $b$, and intermolecular attraction $a$, using the gas constant $R$.

\begin{figure}[hbt!]
\centering
\includegraphics[height=.38\textwidth]{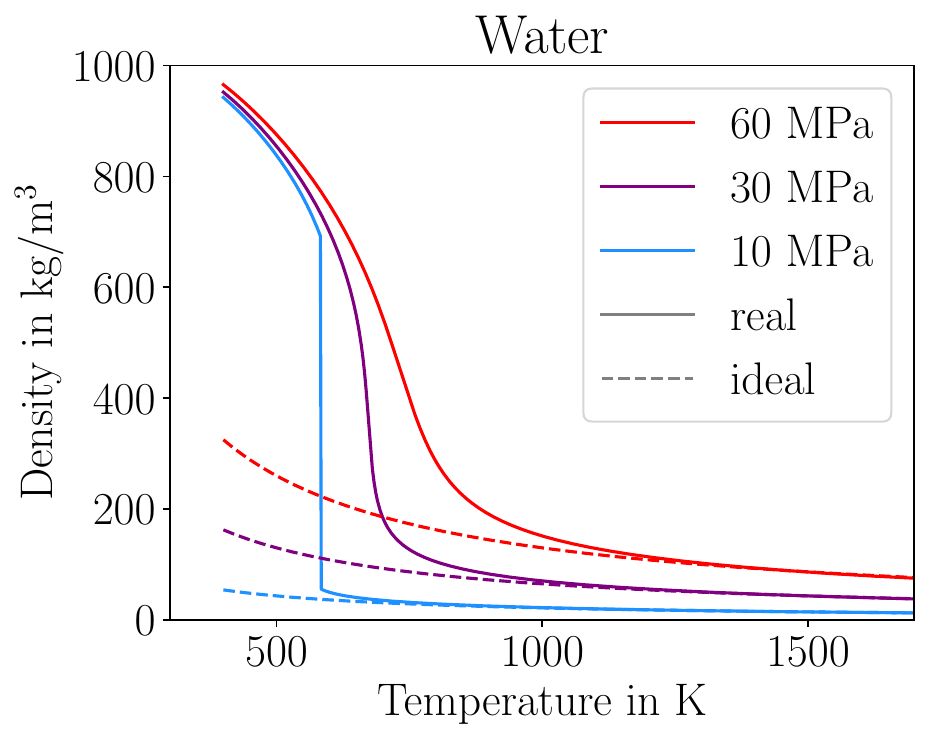}
\includegraphics[height=.38\textwidth]{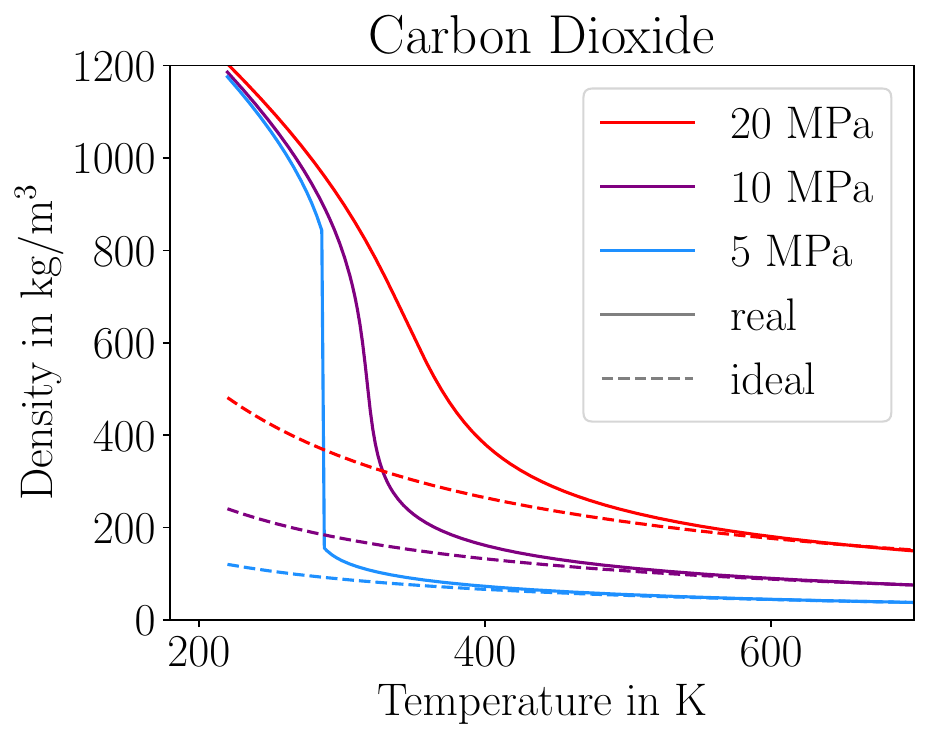}
\caption{Densities of water and carbon dioxide for accurate real fluid data (solid lines) and the ideal gas equation (dashed). The blue lines represent a subcritical pressure, purple and red correspond to supercritical pressures. Across all pressures, all isobar approach the same incompressible liquid state towards low temperatures; towards high temperatures, the isobar approach ideal gas behavior. In between, a transitional region can be found.}
\label{fig:PB:densities}
\end{figure}

Andrews realized that even at supercritical temperatures the fluid still exhibited limiting behavior of ideal gases at low pressures, and liquid behavior at high pressures. Similarly, Figure~\ref{fig:PB:densities} shows how an incompressible liquid limit is reached towards low temperatures, and an ideal gas limit towards higher temperatures, if the pressures is not too high.

\begin{figure}[hbt!]
\centering
\includegraphics[height=.38\textwidth]{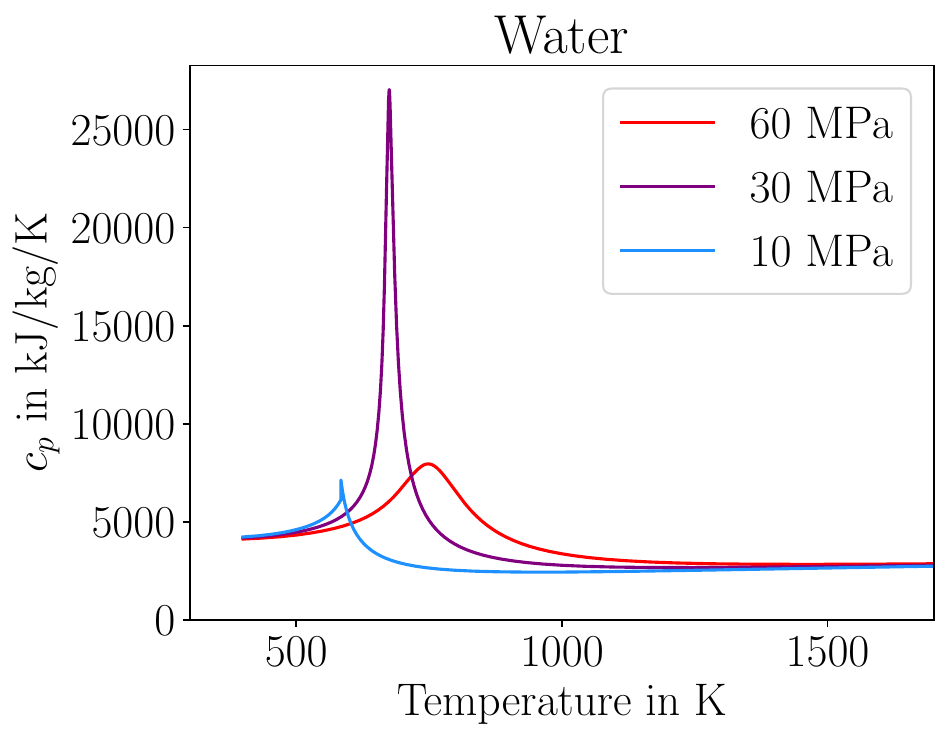}
\includegraphics[height=.38\textwidth]{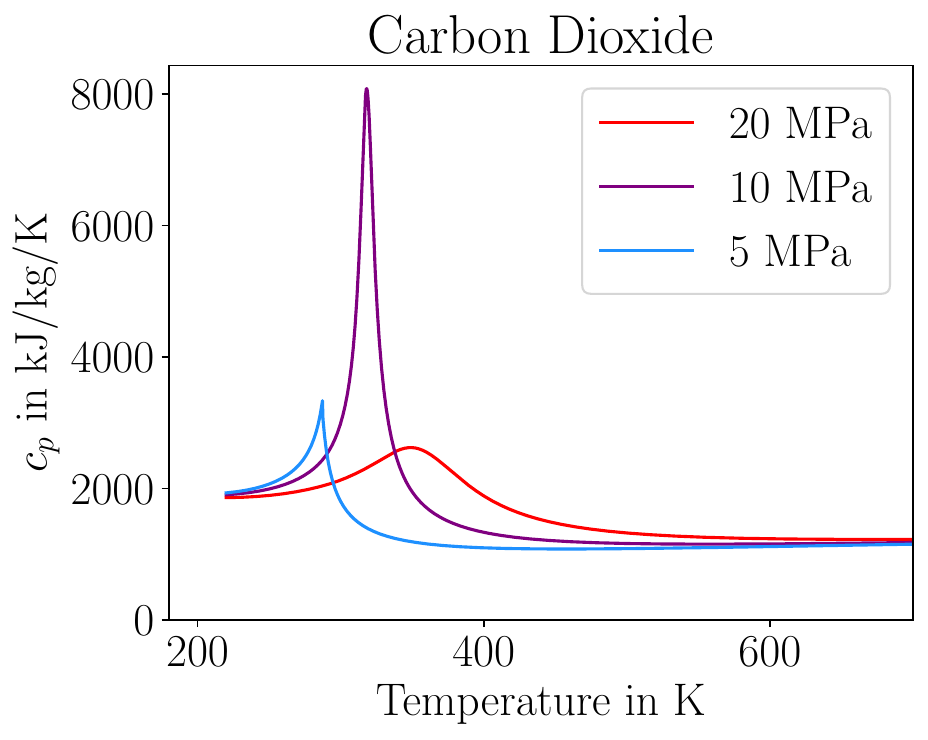}
\caption{Isobaric specific heat capacities of water and carbon dioxide at subcritical (blue) and supercritical (purple and red) pressures. Like with densities, all isobar approach the same incompressible liquid state towards low temperatures; towards high temperatures, the isobar approach ideal gas behavior. In between, a transitional region can be found.}
\label{fig:PB:cp}
\end{figure}

\subsubsection{Supercritical liquids and gases}

I.e., the absence of an \emph{equilibrium coexistence} of distinct supercritical liquid and gaseous states does not preclude the existence of distinguishable supercritical liquid and gaseous states. This is important enough to discuss further: the classical way to evaluate whether a liquid is present in an experimental vessel is to check for the meniscus that separates liquid and gas \cite{Andrews1869,Hendricks1970}. I.e., the liquid in that sense is defined by its \emph{equilibrium existence next to a gas}. But, again, this does not prevent the existence of fluid states that can molecularly be identified as either liquids or gases at supercritical conditions. 

And indeed, molecular dynamics simulations \cite{BanutiARB2017,RajuNSR2017, Maxim2019, Maxim2020, BanutiJSF2020} have shown that supercritical liquid-like fluids are identical to subcritical liquids, and that supercritical gas-like fluids are identical to gases.


Thus, regions of different temperatures in flows of supercritical fluids can cause the simultaneous existence of supercritical liquids and gases, c.f. Figure~\ref{fig:PB:densities}---a situation that is regularly found in heat transfer problems.

\subsection{Transitional states}

The existence of distinct liquid and gaseous states at supercritical conditions suggests to study where exactly a transition between the states takes place. This is not an obvious step, as it is still textbook knowledge \cite{Cengel2005thermodynamics} that \emph{"above the critical state, there is no line that separates the compressed liquid region and the superheated vapor region.''} There is a variety of definitions and concepts, some of which are contradictory \cite{BanutiJSF2020}.

\subsubsection{Pseudocritical line}
\begin{figure}[hbt!]
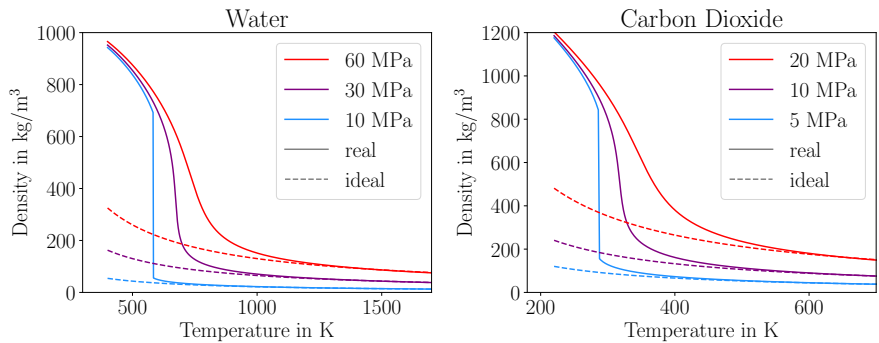

\centering
\includegraphics[height=.38\textwidth]{figs/H2O_densities.pdf}
\includegraphics[height=.38\textwidth]{figs/CO2_densities.pdf}
\caption{Densities of water and carbon dioxide for accurate real fluid data (solid lines) and the ideal gas equation (dashed). The blue lines represent a subcritical pressure, purple and red correspond to supercritical pressures. Across all pressures, all isobar approach the same incompressible liquid state towards low temperatures; towards high temperatures, the isobar approach ideal gas behavior. In between, a transitional region can be found.}
\label{fig:PB:latentheat}
\end{figure}

The \emph{pseudocritical} temperature $T^*$, defined as the temperature at which the isobaric specific heat capacity $c_p$ exhibits a distinct peak along an isobar, was suggested as a transition point between the supercritical `pseudoliquid' and `pseudovapor' \cite{Hendricks1970}. \citet{Hendricks1970} go so far as to associate a latent heat with the supercritical liquid--gas transition, through extrapolation of the liquid and gaseous branches of the enthalpy isobar towards $T^*$, c.f. Figure~\ref{fig:PB:latentheat}. The transition takes place at $T^*$ and can be mapped out in extension to the subcritical coexistence line as the locus of states with a maximum isobaric heat capacity.

\subsubsection{Widom line and 'the ridge': response functions}
Nishikawa and coworkers \cite{Nishikawa1995,Nishikawa1997,Morita1997} identified what they called a `ridge' in fluid properties in extension of the coexistence line, which they interpreted as a higher order liquid--gas phase transition. More recently, the \emph{Widom line} has come into focus as a generalized term for supercritical maxima of different thermodynamic response functions. Definitions based on the correlation length \cite{Simeoni2010,May2012}, isothermal compressibility \cite{Sciortino1997,Abascal2010}, or the isobaric specific heat capacity \cite{Xu2005, Santoro2008, Gorelli2006, Ruppeiner2012} are used. Originally introduced as a line separating two different liquid states in subcooled water \cite{BU2003, Poole1992}, it has since been applied to the supercritical liquid--gas transition. In that sense, the Widom line can be interpreted as a generalization of the pseudocritical line. In mixtures, one or more Widom lines can be identified, depending on the mixture characteristics \cite{RajuNSR2017}.

\subsubsection{Fisher-Widom and Frenkel lines}
Fisher and Widom \cite{Fisher1969} analyzed a supercritical liquid--gas transition based on the molecular structure of the fluid states and suggested a dynamic transition between the different states. The Frenkel line \cite{Brazhkin2012,Bolmatov2013} is a related idea which predicts a structural transition at $c_v=2k_\mathrm{B}$. A thorough analysis \cite{BanutiJSF2020} shows that some results previously attributed to the Widom line \cite{Gorelli2006,Simeoni2010} can be attributed to the Frenkel line instead. These data match the $c_v=2k_\mathrm{B}$ criterion fairly well up to $\pr=100$ and $\Tr = 4.5$, far beyond the region where extrema in response functions can be found. A simple correlation for the Frenkel line \cite{BanutiJSF2020} is\footnote{The correlation is a based on a least squares fit of NIST \cite{NIST} data for argon of the $c_v=2k_\mathrm{B}$ criterion. The fit line \emph{almost} passes through the critical point, and maybe, physically, it should.}
\begin{equation}
    \pr = 34.981 \Tr - 34.498 \; .
    \label{eq:PB:Frenkel}
\end{equation}
Supporting experimental data exist for $10\pcr<p<100\pcr$; it is, however, unclear whether the Frenkel line extends beyond these limits.

\subsubsection{Pseudo boiling line}
There are a few difficulties with both the pseudocritical line and the Widom lines. First, one may ask why the heat-capacity-based line is chosen over, say, the thermal expansion -- both expansion and absorption of energy are phenomena associated with subcritical boiling. Which, then, is the `correct' Widom line? Second, the $c_p$ maxima in isobaric scans move to higher temperatures as the pressure increases only for $\pr<6$; beyond, the associated transition temperature reduces as the pressure increases \cite{BanutiJSF2020}, with the exception of the quantum fluids hydrogen \cite{Hendricks1970} and helium. Which, then, is the `correct' end point? Third, even in a $\pr$-$\Tr$ plot, different species show different pseudocritical or Widom lines \cite{Hendricks1970}. Which is the `correct' one?

This is resolved by the \emph{pseudo boiling line}, introduced as a well-defined and thermodynamic meaningful marker of a supercritical phase transition. The term \emph{pseudo boiling} originated in the heat transfer community for supercritical boiling-like behavior \cite{Kafengauz1966,Kafengauz1968,Ackerman1970}, and was generalized to the pseudo boiling line as an extension of the coexistence line by Oschwald \cite{Oschwald2006}.

We are mainly interested in the thermodynamic similarity between subcritical boiling and supercritical pseudo boiling. While boiling is characterized by a discontinuous change in the slope of the Gibbs free enthalpy \cite{Hirschfelder1954}, it can be shown that this discontinuity is projected into the supercritical state space as a set of states with a \emph{maximum curvature} of the Gibbs free enthalpy \cite{BanutiJSF2020}. The locus of these states can be approximated as maxima of $c_p$ in the vicinity of the critical point, and a natural end point beyond which no maximum in the curvature is present anymore sets an upper pressure limit of $\approx 3 \pcr$. This answers our first two questions. For $\pcr <p < 3\pcr$, the pseudo boiling line is described by \cite{Banuti2015}
\begin{equation}
    \pr = \exp \left[ A_s (\Tr -1) \right] \; ,
    \label{eq:PBL}
\end{equation}
where $A_s$ is the slope of the coexistence line at the critical point, also known as the Riedel parameter \cite{Riedel1956}. It turns out that $A_s$ can be determined analytically from cubic equations of state \cite{BanutiJSF2020,BanutiASME2021}. For the Soave-Redlich-Kwong equation of state \cite{Soave1972}, we find 
\begin{equation}
    A_\mathrm{SRK} = 5.51934 + 4.80640 \omega - 0.537437 \omega^2 \; ,
    \label{eq:PB:As}
\end{equation}
i.e., the slope of the pseudo boiling line is a cubic function of the acentric factor. This answers our third question: the pseudo boiling line adheres to the extended corresponding states principle, rather than the corresponding states principle \cite{BanutiJSF2020}.

To reiterate, the pseudo boiling line is related to the pseudocritical line and the $c_p$-based Widom line, but has an exact thermodynamic meaning and a natural end point that is limited to the thermodynamic existence of the supercritical phase transition.


\subsubsection{Similarity of transition lines}
\begin{figure}[hbt!]
\centering
\includegraphics[height=.38\textwidth]{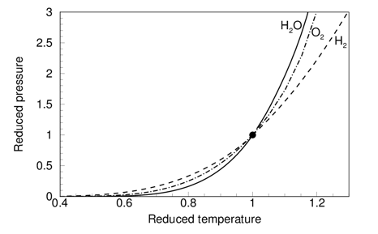}
\includegraphics[height=.38\textwidth]{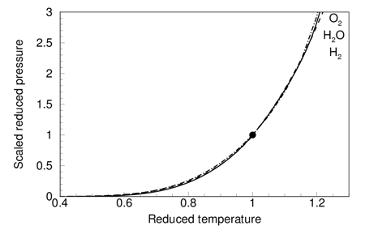}
\caption{Collapse of coexistence lines in reduced pressure and temperature coordinates. Following the corresponding states principle, the lines should collide. Left: However, due to the difference in species acentric factors, different slopes of the reduced coexistence lines can be expected.  Right: Using the scaled reduced pressure Eq.~(\ref{eq:prstar}), coexistence lines and pseudo boiling lines even of fluids with widely different acentric factors collapse. Reproduced from \citep{BanutiPRE2017} with permission given by American Physical Society.}
\label{fig:PB:CLprstar}
\end{figure}

The insight that different species have different pseudo boiling lines limits our ability to generalize findings obtained for a particular fluid species. A similarity parameter that collapses all fluids is thus desirable. We can first introduce a generalized equation for the pseudo boiling line and the coexistence line \cite{BanutiPRE2017},
\begin{equation}
    \pr= \exp\left[ \frac{A_s}{\min(\Tr, 1)}(\Tr-1)\right] \; .
    \label{eq:generalPBL}
\end{equation}
Raising both sides of Eq.~(\ref{eq:generalPBL}) to the power of $(A_0/A_s)$, where $A_0$ is a suitable reference value, eliminates all species-specific parameters, and we obtain the scaled reduced pressure $\pr^*$ as a truly species independent parameter \cite{BanutiPRE2017} within the range of applicability of Eq.~(\ref{eq:generalPBL}),
\begin{equation}
    \pr^* = \pr^{(A_0/A_s)} \; .
    \label{eq:prstar}
\end{equation}
The reference value $A_0$ is arbitrary, but it seems suitable to use a simple fluid as reference, i.e. a species adhering to the corresponding states principle \cite{Guggenheim1967}. We thus define \cite{BanutiPRE2017} $A_0 = A(\omega=0) \approx 5.52$. Figure~\ref{fig:PB:CLprstar} shows that using $\pr^*$ and $\Tr$ as nondimensional parameters does indeed make coexistence lines and pseudo boiling lines collapse.

\subsection{Transitional processes and pseudo boiling}

Figures~\ref{fig:PB:densities} and \ref{fig:PB:cp} clearly show that fluid properties at supercritical conditions reach liquid behavior towards low temperatures, indistinguishable from subcritical liquids, and ideal gas behavior towards high temperatures, indistinguishable from ideal gases. How then do we arrive at the conclusion that \emph{``two distinct phases do not exist at supercritical pressures''} \cite{Hall1971}? 

This view can only come from the method of identifying a liquid, as discussed by Andrews \cite{Andrews1869}, where a visible meniscus between a liquid phase and a gaseous phase marks the existence of a liquid. However, a meniscus is also not visible in a vessel filled purely with either a gas or a liquid -- or, indeed, a supercritical fluid. Thus, what does not exist at supercritical pressures is a macroscopic equilibrium coexistence between liquid and gaseous states. This, however, does not contradict the existence of the observed distinct liquid and gaseous supercritical states or the non-equilibrium coexistence of these supercritical states, e.g., in problems with strong temperature gradients such as injection or wall heat transfer.

Having thus established the existence and co-existence of supercritical liquids and gases, we need to understand the actual transition better. We already discussed that the pseudo boiling line is a marker of this transition, representing the locus of states with a maximum curvature in the Gibbs enthalpy. I.e., the textbook notions that  \emph{"above the critical state, there is no line that separates the compressed liquid region and the superheated vapor region''} and that \emph{``at pressures above the critical pressure, there is not a distinct phase-change process} \cite{Cengel2005thermodynamics} are not as clear is previously thought. 

Revisiting Figures.~\ref{fig:PB:densities} and \ref{fig:PB:cp}, we see that the end points of a liquid--gas transition are identical between the subcritical and the supercritical case. What is different is that at supercritical conditions, this transition spans a finite temperature interval and occurs through a continuous change in fluid states and properties \cite{Banuti2015}.

\subsubsection{Latent heat of supercritical fluids}
None other than Ernst Schmidt, namesake of the Schmidt number, observed that the latent heat is spread over a finite temperature interval at subcritical conditions as the critical point is approached \cite{Schmidt1960}\footnote{\emph{``Die als Verdampfungsw\"arme auf eine bestimmte Temperatur konzentrierte W\"armekapazit\"at wird in der Nähe der kritischen Temperatur sozusagen auf einen endlichen Temperaturbereich ausgebreitet.''} \cite{Schmidt1960}}. This is visible in the rising $c_p$ on either side of the coexistence line for both liquid and vapor states, unlike for very low pressures $p<0.1\pcr$, where the liquid and vapor heat capacities are constant as the coexistence line in approached. At supercritical pressures, the equilibrium part of the transition has vanished altogether, all of the latent heat, and indeed the transition, is now found distributed across the transitional region between liquid and gas \cite{Banuti2015,BanutiPPCE2019}.

\begin{figure}[hbt!]
\centering
\includegraphics[height=.38\textwidth]{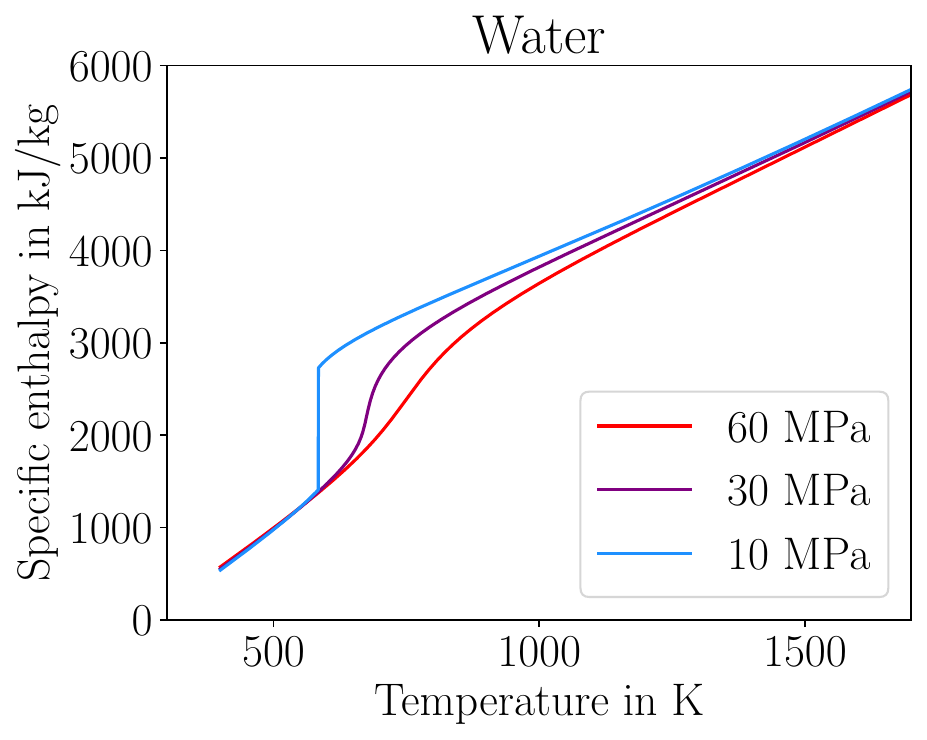}
\includegraphics[height=.38\textwidth]{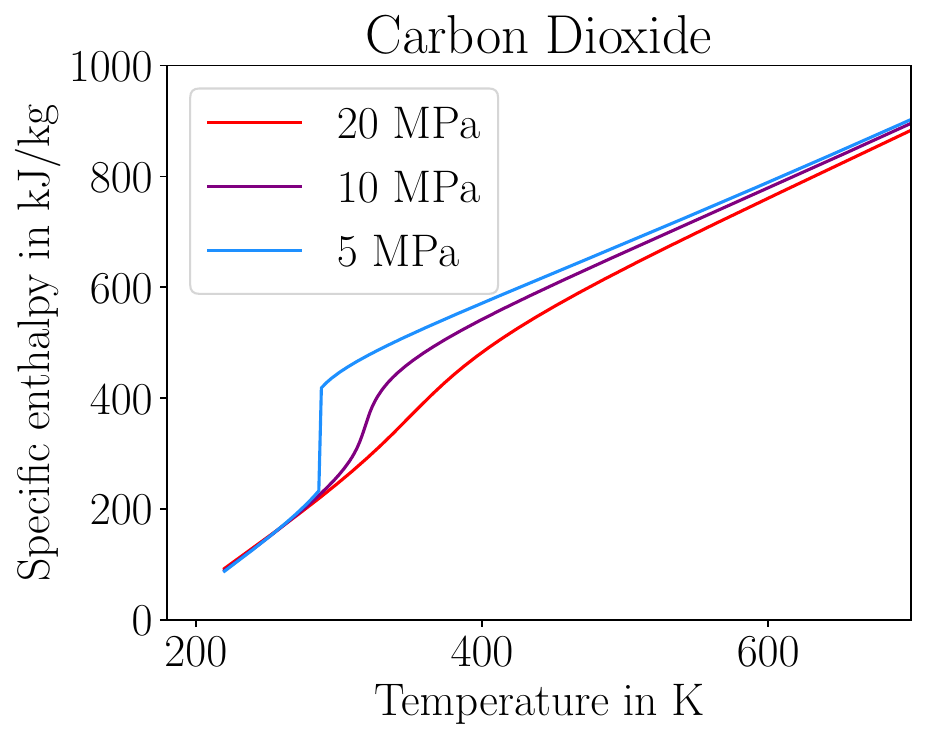}
\caption{Specific enthalpies of water and carbon dioxide at subcritical (blue) and supercritical (purple and red) pressures. All isobar approach the same incompressible liquid state towards low temperatures; towards high temperatures, the isobar approach ideal gas behavior. In between, a transitional region can be found. As the end points on liquid and gaseous limit states are pressure independent, the difference in enthalpy between the limit states is likewise pressure independent.}
\label{fig:PB:h}
\end{figure}

Figure~\ref{fig:PB:h} shows enthalpy isobar at subcritical and supercritical pressure. Much like in Figures~\ref{fig:PB:densities} and \ref{fig:PB:cp}, a common low temperature liquid limit and a common high temperature gaseous limit are approached. For enthalpy, this reveals something really interesting: The enthalpy required to heat the fluid across the transitional region from liquid to gas takes the same amount of energy, regardless of pressure, e.g. heating water from 500~K to 1500~K. I.e., the integral latent heat is invariant with respect to changes in pressure, even extending to supercritical pressures. Only the constant temperature coexistence part reduces with growing pressure and vanishes at the critical point. 

A different way to think about it is this: When the enthalpy of both ideal gases and incompressible liquids is pressure independent, then the transitional `difference' between both states must also be pressure independent \cite{Banuti2015,BanutiPPCE2019}.

\subsubsection{Pseudo boiling energy budget}
\begin{figure}[hbt!]
\centering
\includegraphics[height=.33\textwidth]{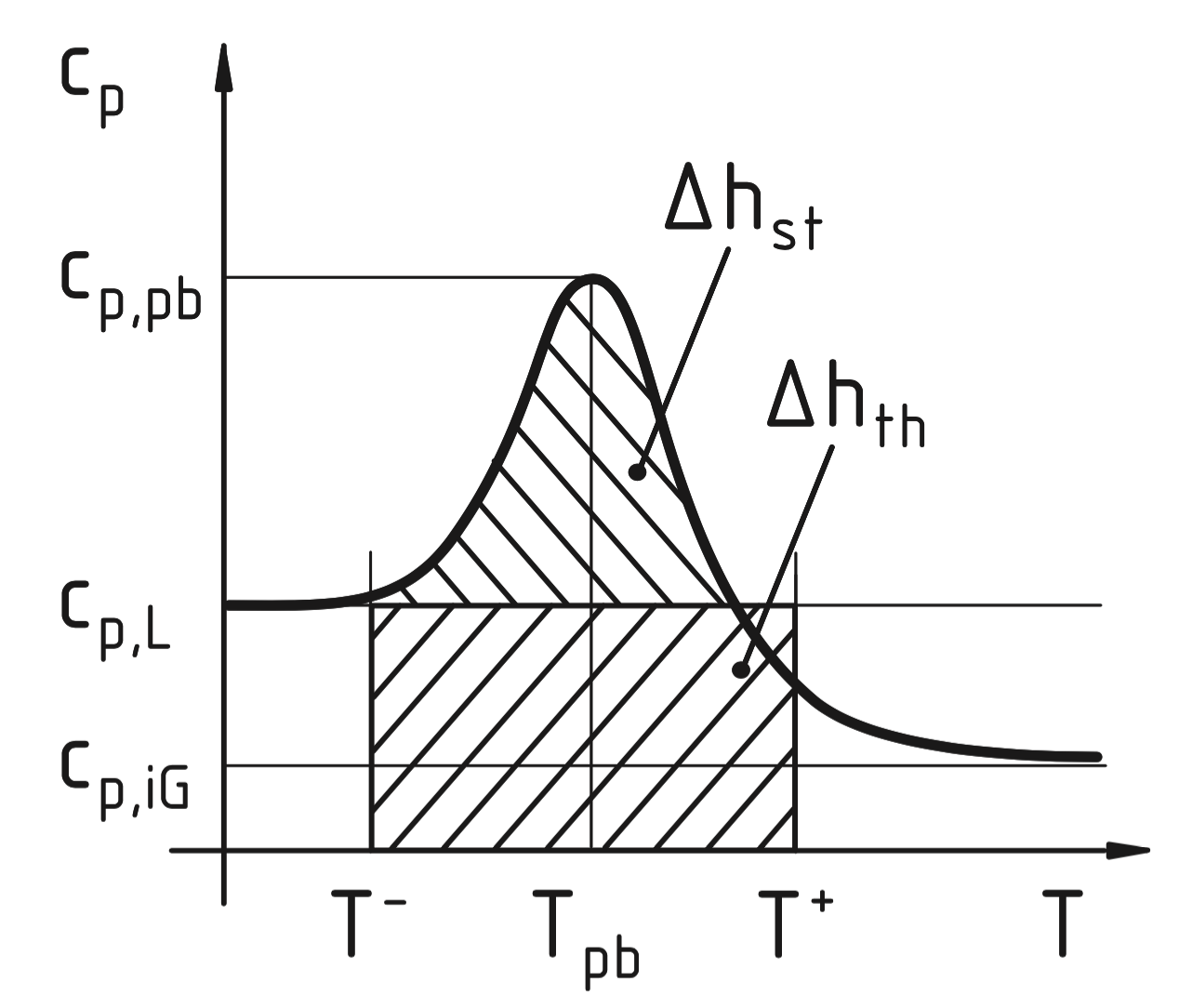}
\caption{Energy budget during pseudo boiling transition. The area under the curve is the change in enthalpy during the transition from T$^-$ to $T^+$. The total area can be subdivided into the thermal component $\Delta h_\mathrm{th}$ required to increase the temperature of the fluid, and the structural component $\Delta h_\mathrm{st}$ that accounts for molecular separation and expansion of the fluid. Reproduced from \citep{Banuti2015} with permission given by Elsevier.}
\label{fig:PB:cp-budget}
\end{figure}

But what happens during the pseudo boiling transition? Figure~\ref{fig:PB:cp-budget} gives a close-up of the isobaric specific heat capacity across an isobaric pseudo boiling process, spanning the finite temperature interval from $T^-$ to $T^+$ with a total enthalpy change during pseudo boiling of \cite{Banuti2015}
\begin{equation}
     h_\mathrm{pb} = \int_{T^-}^{T^+} c_p(T) \dd T = h(T^+) - h(T^-) \; .
\end{equation}

During an isobaric heating process, added heat contributes to two phenomena: First, energy is required to increase the temperature of the fluid from $T^-$ to $T^+$: considering as reference the fluid remaining a liquid with a heat capacity of $c_{p,\mathrm{L}}$ during that process, the required \emph{thermal} (th) component $\Delta h_\mathrm{th}$ is \cite{Banuti2015}
\begin{equation}
    h_\mathrm{th} = c_{p,\mathrm{L}}(T^+ - T^-) \; .
\end{equation}

Second, the fluid needs to undergo the structural changes that turn a densely packed liquid into a gas. This excess \emph{structural} component (st) $h_\mathrm{st}$ 
\begin{equation}
    h_\mathrm{st} = h_\mathrm{pb} - h_\mathrm{th}
    \label{eq:PBh}
\end{equation}
then accounts for these vaporization characteristics, i.e. $h_\mathrm{st}$ represents the amount of work needed to overcome intermolecular attraction and to expand the fluid against the background pressure.



\subsubsection{Microscale characterization of transitional fluid states}

The evidence for a supercritical phase transition from (macroscopic) thermodynamic properties suggests to investigate the molecular nature of pseudo boiling.

At subcritical pressures, classical nucleation theory (CNT) \cite{Fisher1967, Dillmann1991} predicts the formation of molecular clusters as precursors to macroscopic droplet formation and condensation. Under the capillarity assumption, the microscopic phases have the same properties as macroscopic liquid and vapor. With vanishing macroscopic surface tension as the critical pressure and temperature are exceeded, CNT predicts that nucleation should no longer happen.

However, the existence of molecular clusters at supercritical conditions is well documented from experiments and simulations \cite{Nishikawa1995, Nishikawa1997, Tucker1999, Skarmoutsos2009, Skarmoutsos2017, BanutiJSF2020} -- and indeed the reason why supercritical fluids are desired as solvents.

\begin{figure}[hbt!]
\centering
\includegraphics[height=.33\textwidth]{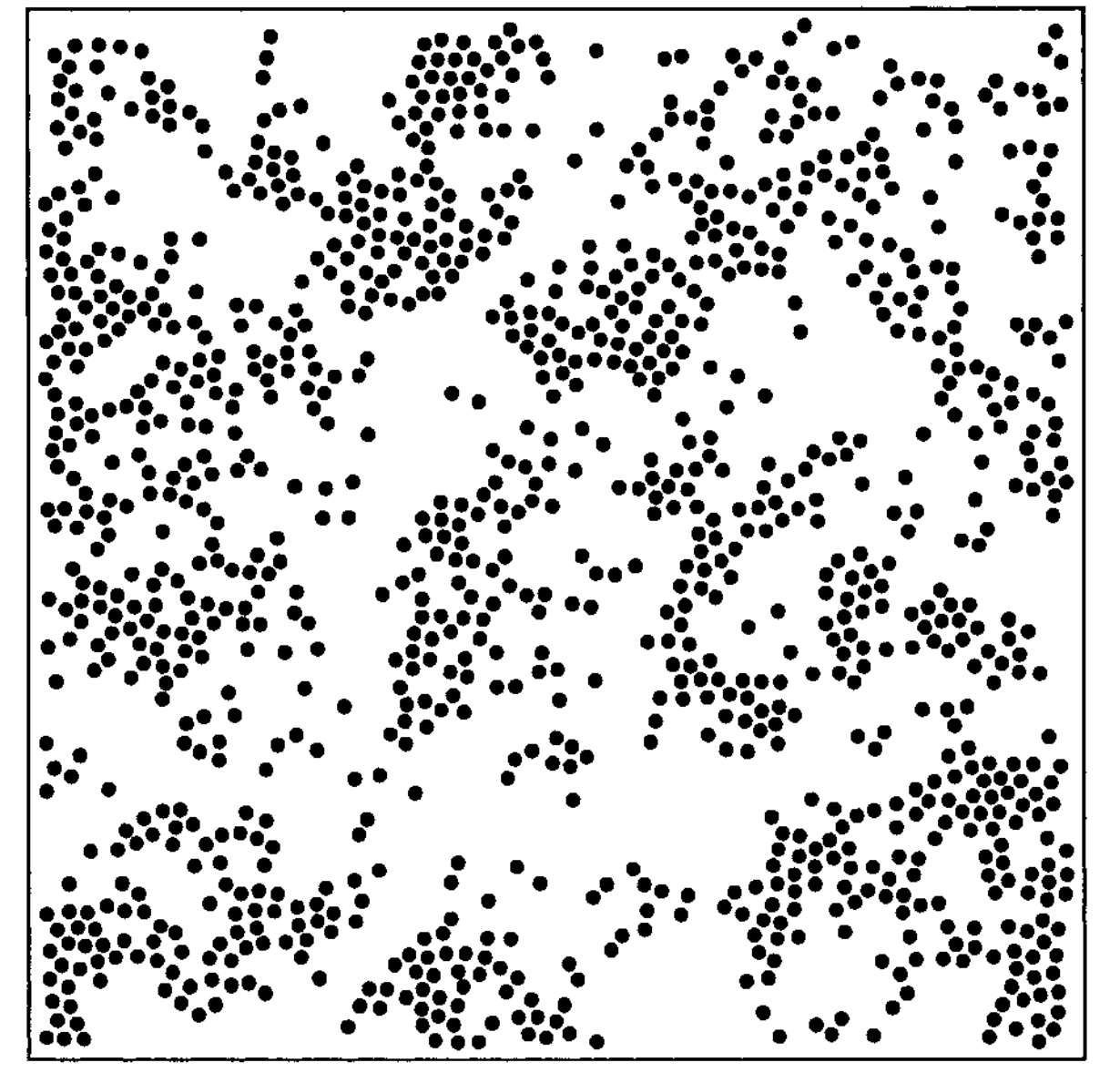}
\caption{Molecular snapshot showing cluster formation for model 2D Lennard-Jones fluid at $\Tr = 1.17$ and $\rhor = 0.86$. Adapted from \citep{Tucker1999} with permission given by American Chemical Society.}
\label{fig:PB:TuckerClusters}
\end{figure}

It turns out that pseudo boiling occurs as the mesoscale phase transition, among supercritical molecular clusters \cite{Maxim2019}. With increasing temperature, the compact liquid breaks up into smaller and smaller fragments; the required energy to separate particles is manifest in the structural pseudo boiling enthalpy, Eq.~(\ref{eq:PBh}). The critical point is characterized by the transition between a single large supercluster and its break-up into smaller clusters \cite{BanutiHELGa2022}. The reversed process of clusters consolidating into a bulk liquid is then pseudo condensation.

\begin{figure}[hbt!]
\centering
\includegraphics[height=.33\textwidth]{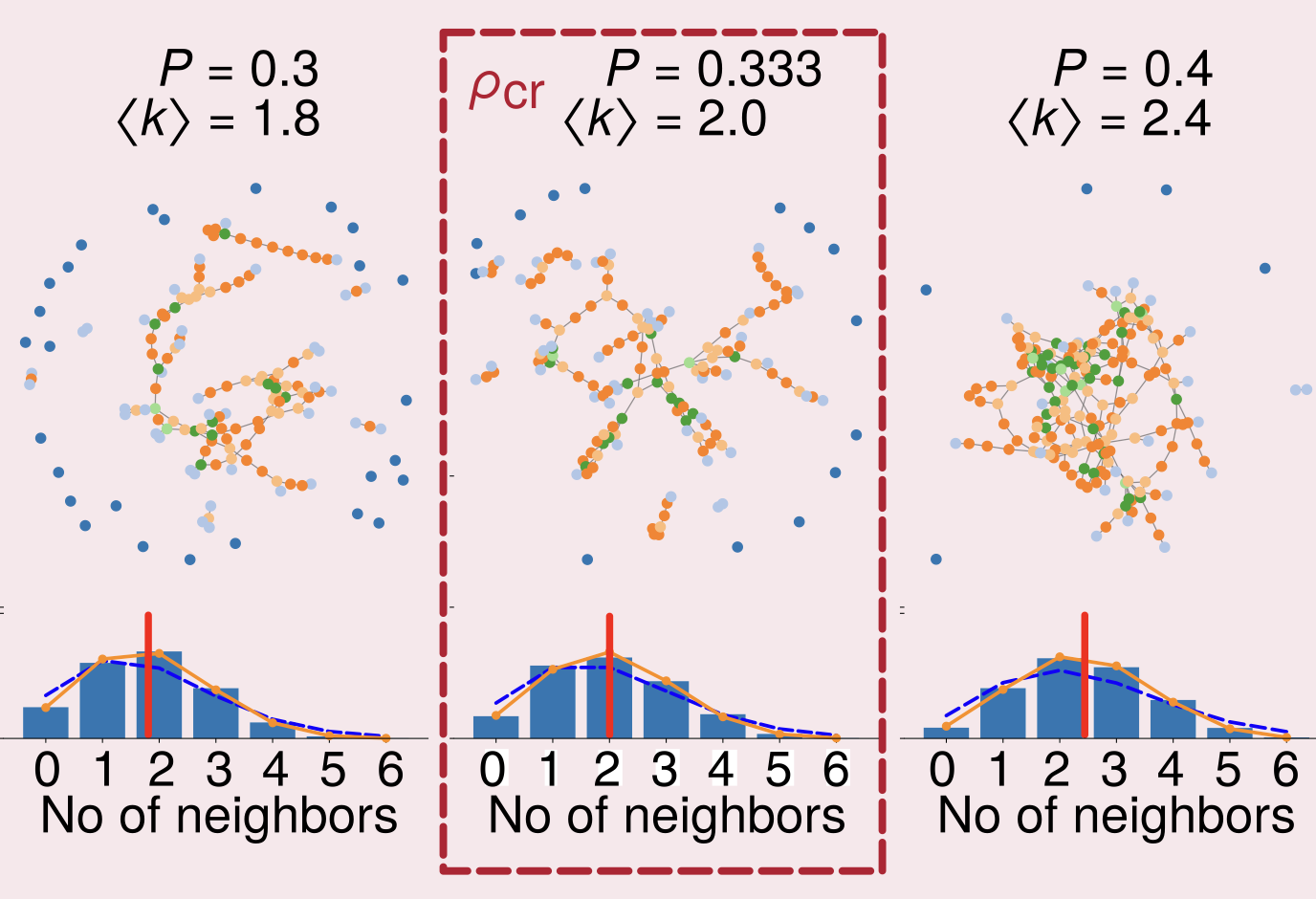}
\caption{Formation of molecular networks at near-critical conditions. In an isothermal scan from low to high densities at $\Tcr$, $\rhor=1$ marks the state at which each particle has on average two neighbors. $P$ is a nondimensional lattice density where 0 and 1 correspond to an empty and a fully occupied lattice, respectively. The number of neighbors follows a binomial distribution. Reproduced from \citep{BanutiJSF2020} with permission given by Elsevier.}
\label{fig:PB:BanutiNetwork}
\end{figure}

There is one more subtlety about the molecular clusters at critical conditions: rather than forming spheroidal formations, the particles associate in networks instead. Figure~\ref{fig:PB:BanutiNetwork} shows such a linearly branched structure at the critical density and temperature, with the distribution pdf of number of neighbors per particle. For an isothermal scan from low to high densities at $\Tcr$, $\rhor=1$ marks the state at which each particle has on average two neighbors.

\subsection{A revised fluid state space}
Rather than the traditional view of the fluid state space divided into four quadrants by the critical isotherm $T=\Tcr$ and the critical isobar $p=\pcr$, it is physically more appropriate to distinguish between liquid states at low temperatures and gaseous states at higher temperatures. Due to the thermodynamic similarity expressed in the corresponding states principle \cite{Elliott2012}, a single chart can be made to be (approximately) valid for general fluids when the reduced properties $\pr = p/\pcr$, $\Tr = T/\Tcr$, etc. are used instead of the dimensional pressure, temperature, etc., respectively.

\begin{figure}[hbt!]
\centering
\includegraphics[width=.7\textwidth]{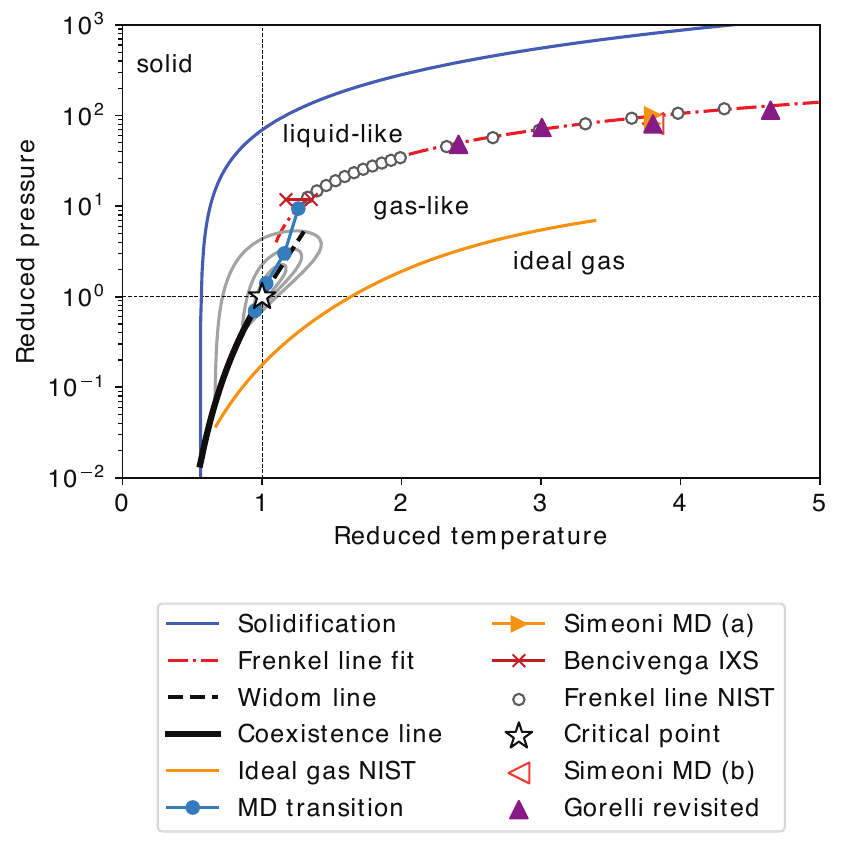}
\caption{A revised state space diagram. Rather than distinguishing the classical but arbitrary four quadrants around the critical point, one can identify distinct transition lines. Solid and liquid states are separated by the solidification line. At pressure exceeding the triple point, the liquid--gas transition occurs across different lines and means different things as the pressure increases: for $\pr<1$ the coexistence line marks liquid-vapor phase coexistence; for $1<\pr<3$ the pseudo boiling line (or $c_p$ based truncated Widom line) marks the thermodynamic pseudo boiling process; for $\pr>10$ the Frenkel line separates regions that dynamically behave like liquids and gases, e.g. a vanishing transversal sound propagation. The transition between pseudo boiling and Frenkel line is still an open question, as is a potential upper limit for the Frenkel line. Reproduced from \citep{BanutiJSF2020} with permission given by Elsevier.}
\label{fig:PB:statespace}
\end{figure}

Figure~\ref{fig:PB:statespace} illustrates this view, along with the transition lines. As an example, during an isobaric heating process at pressures above the triple pressure, substances undergo a transition solid--liquid--gaseous--ideal gas as they are heated. At low subcritical pressures, a distinct boiling process with a liquid-vapor equilibrium on the coexistence line is observed, where the vapor acts like an ideal gas. As the critical point is approached, real fluid effects get stronger: the vapor no longer behaves like an ideal gas, the liquid heat capacity exhibits a distinct increase as the temperature approaches the saturation temperature \cite{Schmidt1960}, until, at the critical point, it diverges \cite{Hendricks1970}. At supercritical pressures $1 <\pr<3$ the liquid-gas transition occurs across the pseudo boiling line, over a widening temperature interval \cite{Morita1997,Banuti2015}. A latent heat \cite{Banuti2015,BanutiPPCE2019} and distinct expansion can still be identified. The distinct pseudo boiling process loses its relevance for $\pr>3$; at $\pr>10$ neither distinct heat capacity peaks nor an inflection point in the density can be identified \cite{BanutiJSF2020}. For $\pr>10$, a transition can still be identified in terms of the cross-over in the transport coefficients, but it has lost its thermodynamic character \cite{Bolmatov2014}.

Figure~\ref{fig:PB:statespace} reveals something else: a gas can indeed be compressed to a liquid state at supercritical temperatures; it just does not reach an equilibrium phase coexistence on its way.

\subsection{Interfaces}
The absence of a pure fluid liquid-vapor phase equilibrium at supercritical pressures raises questions concerning the existence and possible nature of supercritical interfaces. This is particularly important for heat transfer applications, such as droplet evaporation or film boiling. In the following, we will briefly discuss the fundamental aspects of mixture and pure fluid interfaces at supercritical pressures; the main focus of the paper is in pure fluids.

\subsubsection{Equilibrium}
Mixture thermodynamics and liquid-vapor phase equilibria (vapor-liquid equilibria, VLE) were among the key applications of van der Waals' research and his equation of state \cite{Waals1873}, Eq.~(\ref{eq:vdW}). The exciting history and key concepts behind the theory of VLE were presented by Anneke Levelt Sengers in her wonderful book \cite{Sengers2002}.

The theory of VLE is a long-established textbook topic and was discussed in numerous books, c.f. \citet{Elliott2012} for a gentle but concise introduction and \citet{Prausnitz1985} for a thorough and deep discussion, including a molecular perspective. Further, see \citet{Reid1987} for a comprehensive compilation of property models and correlations. Given the focus of this paper on pure fluid heat transfer and the abundance of material available on this topic already available, we will provide only a very brief introduction, focused on recent advances.

Ultimately, the key insight in VLE is that all coexisting phases exhibit the same Gibbs free enthalpy $G=H-TH$ and generally the lowest possible value of $G$ is attained in spontaneous phenomena. From this, the exact conditions of the liquid to vapor transition (and vice versa) can be determined. As an example, using a cubic equation of state, such as van der Waals' Eq.~(\ref{eq:vdW}), we obtain three mathematical solutions from which two are physically realizable (thermodynamically stable) within the coexistence region for the fluid density when given temperature and pressure. The solution that will be realized in nature is then the one with the minimum $G$. In pure fluids, a VLE can no longer be reached for pressures at or above the critical pressure.

In mixtures, the same fundamental mechanism is at work--we still need identical Gibbs enthalpies across all coexisting phases where the minimum $G$ solution will be favored by nature. However, the difficulty now lies in determining the proper mixture conditions. In the ubiquitous cubic equations of state, such as van der Waals' or the more practical Peng-Robinson \cite{Peng1976} and Soave-Redlich-Kwong \cite{Soave1972} equations of state, representative mixture values of the parameters $a$ and $b$ of Eq.~(\ref{eq:vdW}) are evaluated and the mixture treated as if it was a pure fluid, but now with adapted values for $a=a_\mathrm{mix}$ and $b=b_\mathrm{mix}$. This formalism remains extremely successful in predicting phase envelopes of mixtures, e.g. in applications relevant for the oil industry \cite{Peng1976}.

\begin{figure}[hbt!]
\centering
\includegraphics[width=.7\textwidth]{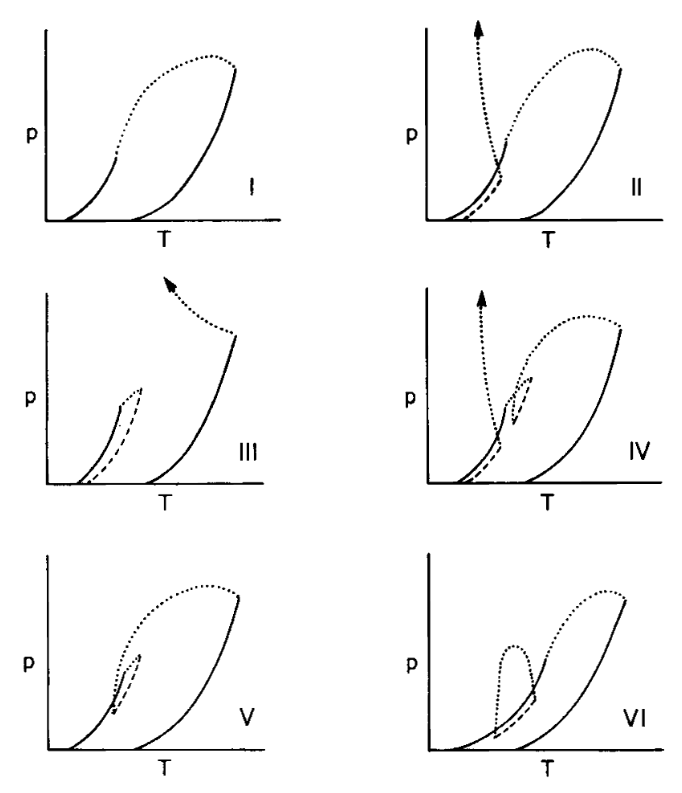}
\caption{Vapor liquid equilibrium (VLE) curves for binary mixtures types I to VI according to Scott and van Konynenburg \cite{Konynenburg1980}. Solid lines denote the pure component coexistence lines, dotted lines mark the VLE envelope. While fluid mixtures of similar fluids (e.g. hydrocarbons) exhibit an upper pressure limit to phase separation, Type I, many technical mixtures, such as between nitrogen and heptane, exhibit Type III behavior with no such upper limit. Adapted from \citep{Scott1987} with permission given by American Chemical Society.}
\label{fig:ScottTypes}
\end{figure}

An important phenomenon that is captured by this formalism is that of VLE at pressures exceeding the pure components' critical pressures, i.e. at nominally supercritical pressure. It is indeed common for mixtures at suitable compositions to exhibit a VLE at such nominally supercritical conditions. Interestingly, the character of these nominally supercritical VLE depends on the exact mixture at hand, but was formalized by Scott and van Konynenburg \cite{Konynenburg1980,Sengers2002}. All in all, six types have been identified, c.f. Figure~\ref{fig:ScottTypes}, five of which can be derived from van der Waals' equation. As examples, a Type I mixture requires its components to be `similar' and exhibits a maximum pressure at which a VLE can be observed. In a somewhat stretched analogy, one could argue this is similar to the pure fluid case. On the other hand, a Type III mixture does not appear to have such an upper pressure limit for VLE, and phase separation for many technically relevant mixtures such as hydrocarbon-nitrogen, hydrogen-oxygen, can occur at arbitrary pressures. 

This allows for the formation of distinct interfaces, e.g. around hydrocarbon droplets in air, even at high and nominally supercritical pressures \cite{Mayer1996,Manin2014,Crua2017}.

\subsubsection{Multi-component local equilibrium }

However, in the case of propellant injection (e.g. in rockets \cite{Mayer1996} or Diesel engines \cite{Manin2014,Crua2017}), we do not observe a global thermodynamic equilibrium. Instead, a temperature difference between the droplet and the carrier gas is present, allowing at best for a local interfacial equilibrium.

This complicates matters compared to VLE such as depicted in Figure~\ref{fig:ScottTypes}, because now, transport needs to be solved in addition to a thermodynamic equilibrium. Consider a quasi-isobaric mixing process between (liquid) $n-$dodecane at 363~K and (gaseous) nitrogen at 900~K, akin to the cases defined by \citet{Manin2014,Crua2017}. In fluid mechanical problems, we are not interested in the final global equilibrium state, such as if we put both components into a box and waited out the equilibration. Instead, the whole point are the dynamics of the local states. In the chosen example, we have the simultaneous transfer of mass and heat, a molecular diffusion into each component and thermal diffusion from the cooler to the warmer stream. 

\begin{figure}[hbt!]
\centering
\includegraphics[width=.8\textwidth]{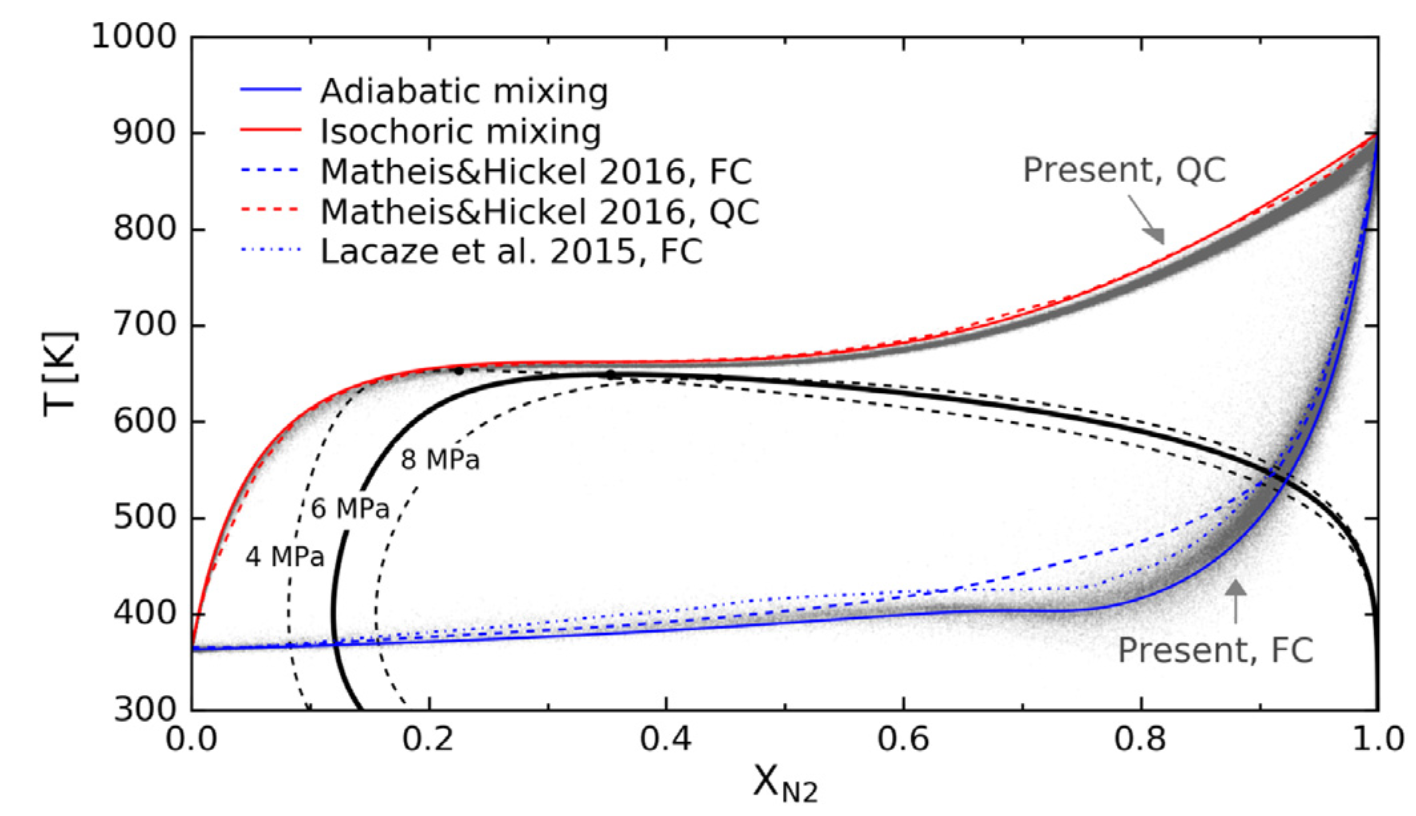}
\caption{Temperature-composition diagram for injection of a 363~K n-dodecane jet into a 900~K nitrogen environment. Grey scattered dots mark CFD results of the mixing process, connecting both pure boundary conditions. The black solid and dotted lines mark the vapor liquid equilibrium envelope at different pressures supercritical for both components, representative of a Type III mixture as shown in Figure~\ref{fig:ScottTypes}. Phase separation can be expected when a mixture state lies within the VLE envelope. Depending on the CFD solver, two main mixing trajectories emerge, isochoric (red) and adiabatic (blue), i.e. current state of the art solvers give ambiguous results as to whether the mixture would phase separate or not. Reproduced from \citep{Ma2019} with permission given by Elsevier.}
\label{fig:MaMixing}
\end{figure}

Figure~\ref{fig:MaMixing} illustrates this in a composition--temperature diagram \cite{Ma2019}, where the limiting conditions of cool pure $n-$dodecane can be found on the left for ($X_\mathrm{N2}=0$, T = 363~K) and the pure hot nitrogen can be found on the right at ($X_\mathrm{N2}=1$, T = 900~K). The detailed fluxes of both mass and heat now determine the exact mixing trajectory that will connect both limiting conditions through the composition--temperature space. In black, we find the VLE solution for the respective compositions. Assuming local equilibrium, if the mixing trajectory intersects the VLE dome, phase separation can be expected; of the mixing trajectory passes outside of the VLE dome, a single phase flow prevails. Unfortunately, the prediction of the mixing trajectory remains a challenge \cite{Ma2019}.

\subsubsection{Multi-component non-equilibrium}

\begin{figure}[hbt!]
\centering
\includegraphics[height=.4\textwidth]{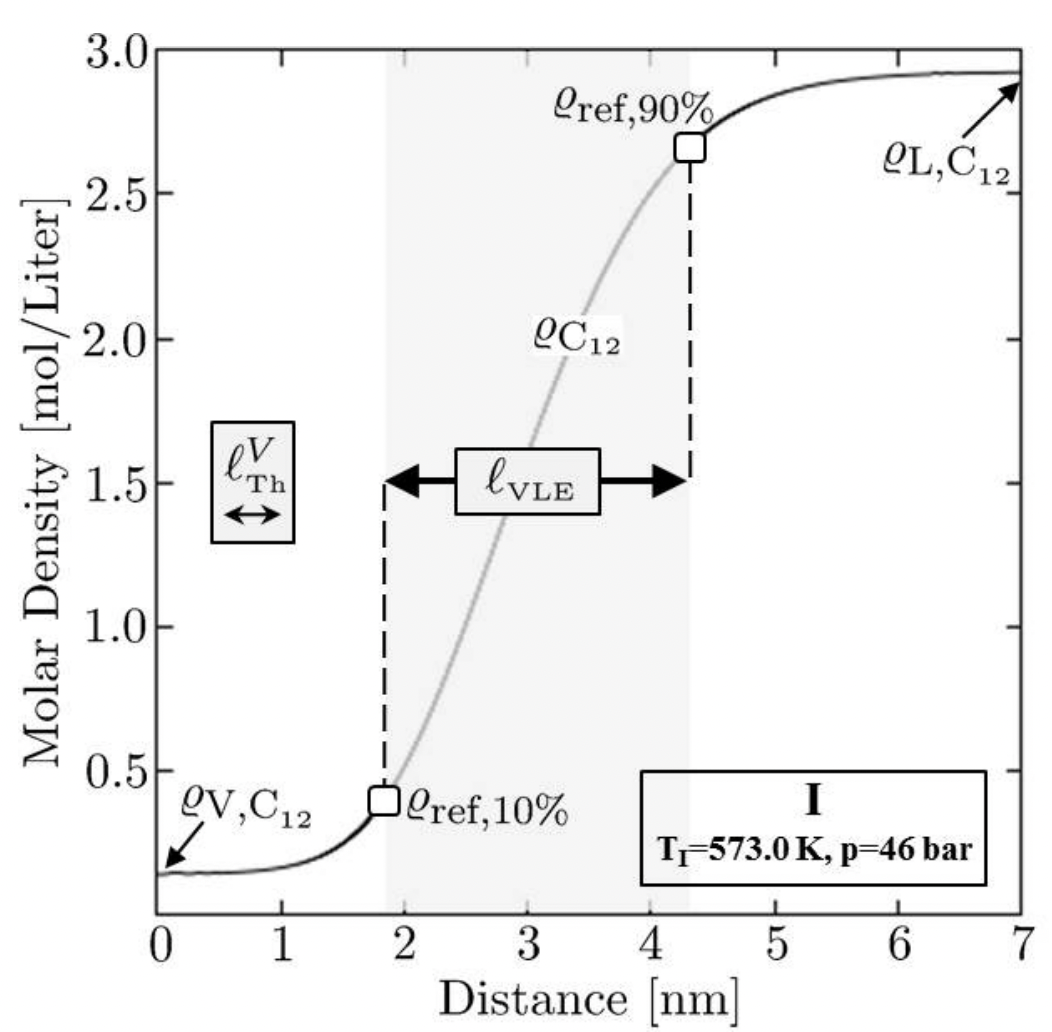}
\includegraphics[height=.395\textwidth]{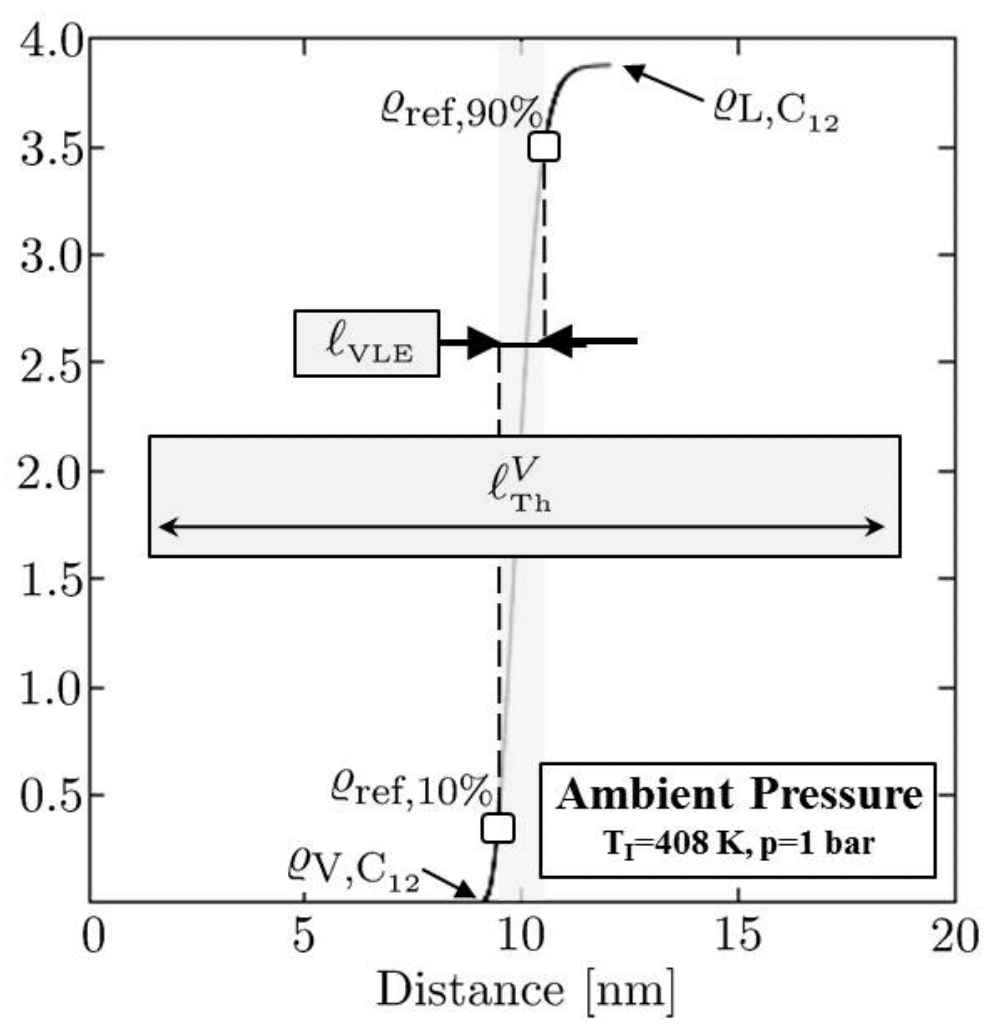}
\caption{Comparison of thermal length $\ell_\mathrm{th}$ to the interfacial thickness $\ell_\mathrm{VLE}$. A sharp interface forms for $\ell_\mathrm{th} \gg \ell_\mathrm{VLE}$, a diffuse interface occurs with $\ell_\mathrm{th} \ll \ell_\mathrm{VLE}$. Reproduced from \cite{DahmsPoF2016} with permission given by American Physical Society.}
\label{fig:DahmsTypes}
\end{figure}

A further analysis of such interfacial states was introduced by van der Waals in his linear gradient theory \cite{Waals1894}, which originally allowed for variation of one parameter, originally composition, across an phase interface. Dahms and Oefelein \cite{DahmsJPP2015,DahmsPCI2015} and Dahms \cite{DahmsPoF2016} applied and analyzed this framework to injection and introduced the concept of an interfacial Knudsen number to distinguish diffuse and sharp interfaces at nominally supercritical conditions. Originally hailing from the field of rarefied gas dynamics, the Knudsen number Kn is the nondimensional ratio of the mean free path of gas particles $\lambda$  and a geometrical length scale $\ell$ that allows to differentiate between the limits of continuum (many collisions and equilibrated fluid) and free molecular flows (few if any collisions) \cite{Hirschfelder1954}. Dahms and Oefelein instead relate a thermal length $\ell_\mathrm{th}$ to the interfacial thickness $\ell_\mathrm{VLE}$, Figure~\ref{fig:DahmsTypes} illustrates the formation of a sharp interface with $\ell_\mathrm{th} \gg \ell_\mathrm{VLE}$ and of a diffuse interface with $\ell_\mathrm{th} \ll \ell_\mathrm{VLE}$.

\subsubsection{Single component non-equilibrium}
We have discussed the well-established existence of phase equilibria with associated interfaces at subcritical pressures in pure fluids, and at nominally supercritical pressures in mixtures.

It turns out that there is another way interfaces can be stabilized at supercritical conditions, even in a pure fluid: in presence of a thermal gradient \citet{LongmireNatComms2023}. These `Thermal Gradient Induced Interfaces' (TGIIF) are characterized through a non-equilibrium heat transfer between the cold and the hot side, where a sufficient temperature difference needs to be established across the pseudo boiling regime.

\begin{figure}[hbt!]
\centering
\includegraphics[height=.25\textwidth]{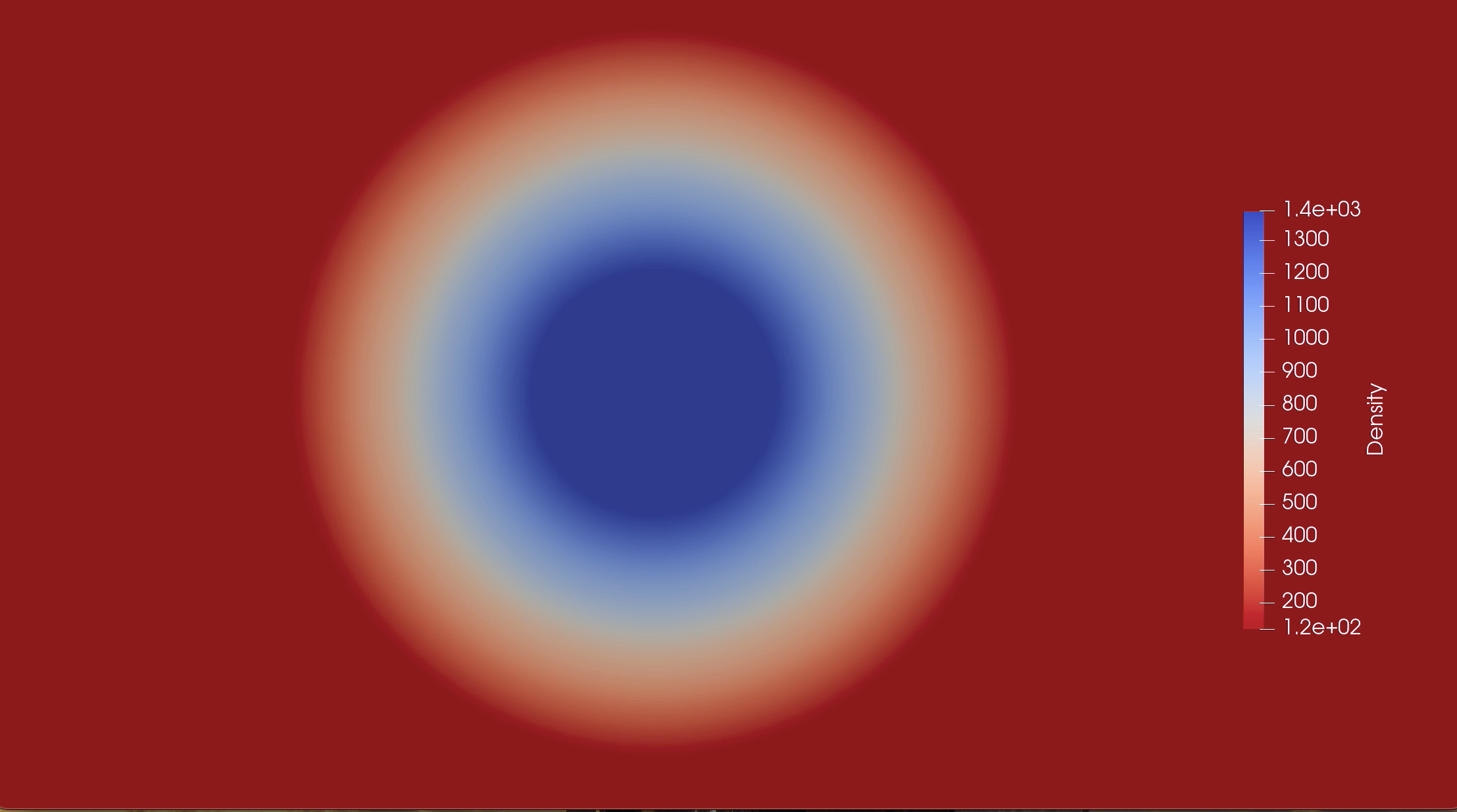}
\includegraphics[height=.25\textwidth]{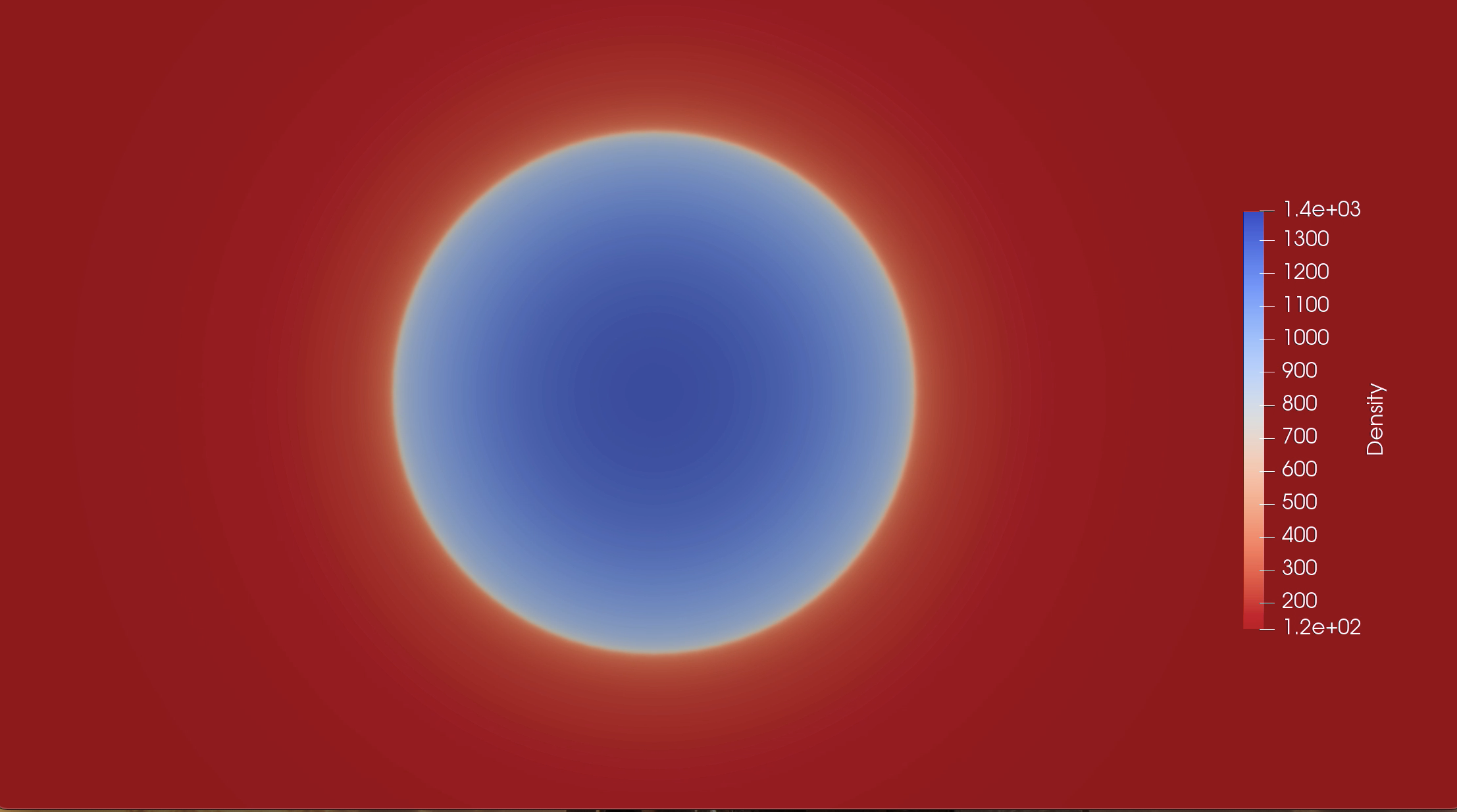}
\caption{Gradient steepening and thermal gradient induced interface (TGIIF) in a cold blob of oxygen embedded in a warm oxygen environment. Left: Initial condition with diffuse interface. Right: Sharpened interface with steepened density gradient at later point in time. The steepening of gradients is unexpected as diffusion tends to minimize the curvature of profiles, smearing them out. Reproduced from \cite{LongmireNatComms2023}, licensed under CC BY 4.0.}
\label{fig:LongmireScroplets}
\end{figure}

Figure~\ref{fig:LongmireScroplets} shows the density fields of a cool drop of oxygen in a warm oxygen environment \citep{LongmireNatComms2023} at supercritical pressure, at two instances in time. The counterintuitive aspect here is that the blurred blob with a low interfacial density gradient is the initial condition, the sharpened blob with a high interfacial density gradient is the state the field converges to before the blob eventually vaporizes. I.e., what we observe under certain conditions is the apparent diffusion against a density gradient without the presence of surface tension. The specific gradient the field is attracted to is case-specific, when the initial density gradient is lower, a density steepening process can be observed.

In fact, \citet{LongmireNatComms2023} found analytical conditions that identify thermodynamic regimes where such an interface is possible even as a steady state. Figure~\ref{fig:DahmsTypes} illustrated that even the sharp interface of a phase equilibrium is truly a smooth, continuous transition from a dense to a light region in the field. The initial ansatz of \citet{LongmireNatComms2023} hence was to look for the general physical possibility of a sustained, steady state density inflection point, i.e. solutions of 
\begin{equation}
\rho_{xx} = \rho_{TT}T_x^2 + \rho_T T_{xx} = 0 \; ,
    \label{eq:TGIIF_ansatz}
\end{equation}
and of Fourier's law of heat conduction in 1D 
\begin{equation}
q'' = k T_x \; ,
\label{eq:q}
\end{equation}
where subscripts denote partial differentiation, e.g., $\rho_T = (\partial \rho / \partial T)$ and $\rho_{TT} = (\partial^2 \rho / \partial T^2)$. Rather than directly solving Eqs.~(\ref{eq:TGIIF_ansatz},\ref{eq:q}), \citet{LongmireNatComms2023} analyze sign compatibility to find 
\begin{equation}
\mathrm{sgn}(\rho_{TT}) = 
- \mathrm{sgn}(k_T)
\label{eq:TGIIF}
\end{equation}
as necessary condition for the existence of a density inflection point, i.e. a steady interface. Comparing Eqs.~(\ref{eq:TGIIF_ansatz}) and (\ref{eq:TGIIF}), we see that the latter no longer features any case boundary conditions, such as spatial gradients. Instead, Eq.~(\ref{eq:TGIIF}) is purely a fluid specific criterion! The existence of a thermal gradient induced interface is thus fluid-inherent.

\begin{figure}[hbt!]
\centering
\includegraphics[height=.35\textwidth]{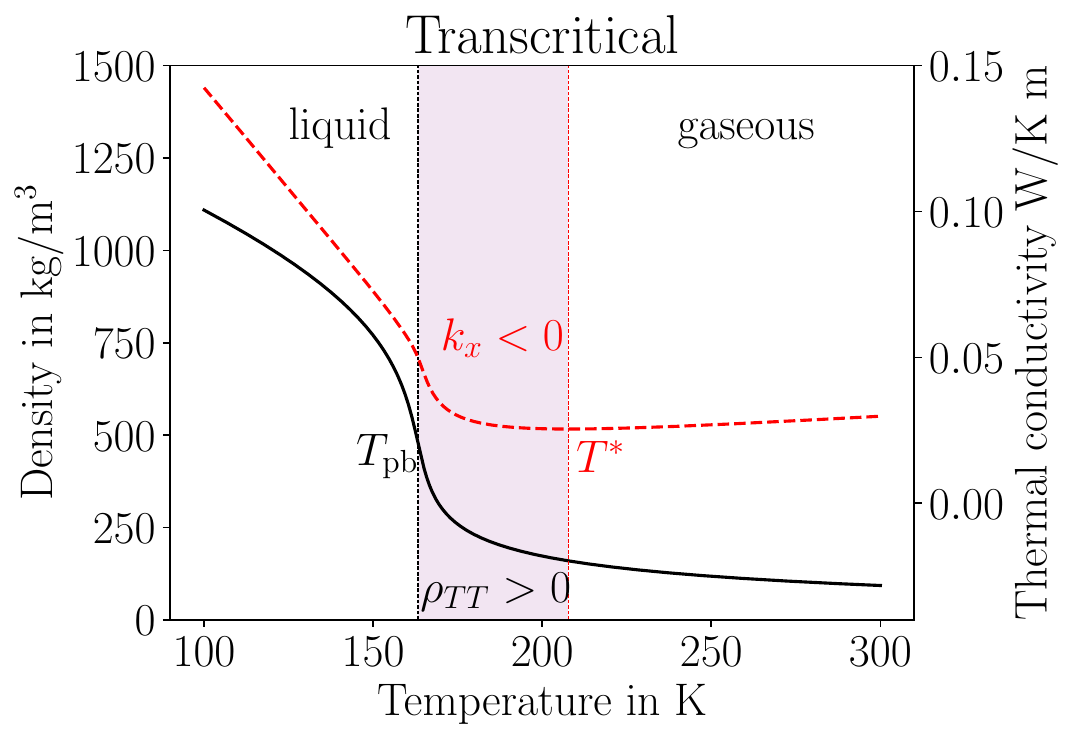}
\includegraphics[height=.35\textwidth]{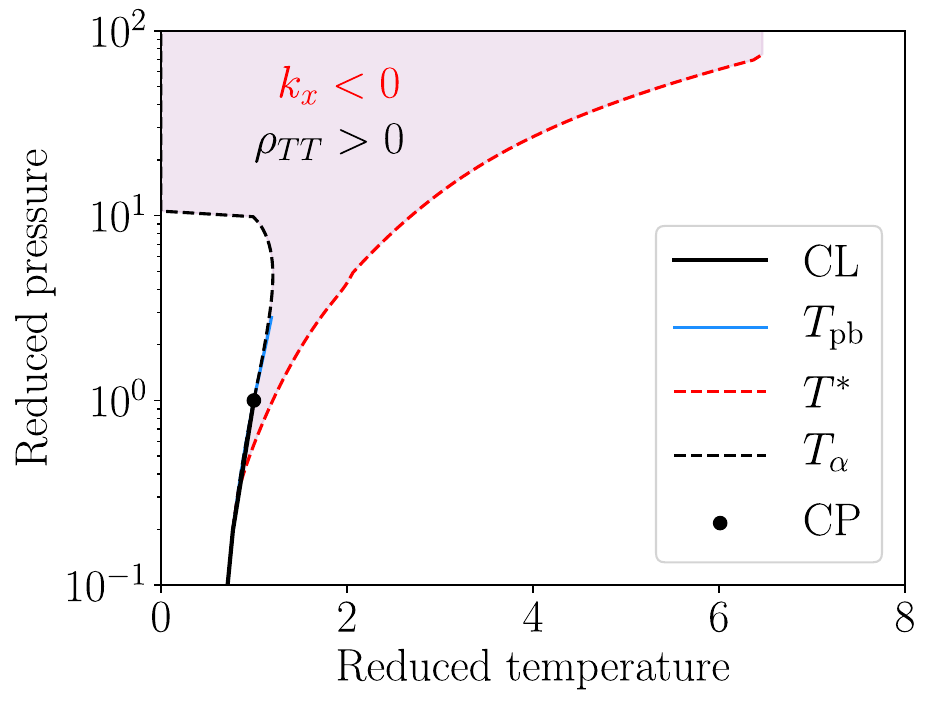}
\caption{A steady persistent interface in presence of a temperature gradient, characterized by a density inflection point, is a fluid property and possible if $\mathrm{sgn}(\rho_{TT}) = - \mathrm{sgn}(k_T)$. This condition is not fulfilled in liquids or ideal gases but is observed in the vicinity and beyond the critical point (shaded red region). Reproduced From \cite{LongmireNatComms2023}, licensed under CC BY 4.0.}
\label{fig:LongmireCriteria}
\end{figure}

So where do we find a fluid that fulfills Eq.~(\ref{eq:TGIIF})? What we are looking for is a fluid where the sign of the thermal density dependence curvature is the opposite of the thermal conductivity gradient sign. Figure~\ref{fig:LongmireCriteria} (left) shows density and thermal conductivity along an isobaric trace at slightly supercritical pressure for oxygen. At low temperatures $\lessapprox 160$~K the fluid behaves like a liquid, with $\rho_{TT}>0$ and $k_T<0$; at high temperature $T \gtrapprox 210$~K it behaves like a gas, with $\rho_{TT}<0$ and $k_T>0$. Neither case fulfills Eq.~(\ref{eq:TGIIF}). 

However, a transcritical fluid exhibits the desired behavior for $\Tpb<T<T^*$, where $T^*$ is the temperature at which the thermal conductivity reaches its minimum. Figure~\ref{fig:LongmireCriteria}(right) maps out the regions where Eq.~(\ref{eq:TGIIF}) is fulfilled; we see that it even extends to subcritical pressures close to the coexistence line.

According to the fundamental corresponding states principle \cite{Prausnitz1985} this can be expected to be general fluid behavior.

\subsection{Characteristic parameters}
Nondimensional characteristic parameters play a fundamental role in our understanding of heat transfer and fluid mechanics. In this section, we discuss parameters that measure the `strength' of pseudo boiling, and the likelihood of reaching pseudo boiling through heat addition.

\subsubsection{A supercritical boiling number}
The `strength' of the pseudo boiling effect can be quantified by comparing the amount of energy needed to heat the fluid $h_\mathrm{th}$ to the amount of energy used to expand the fluid and separate its particles, $h_\mathrm{st}$ - see Figure~\ref{fig:PB:cp-budget} and Eq.~(\ref{eq:PBh}). The boiling number $B$ \cite{Banuti2015} then assesses the respective contributions,

\begin{equation}
    B = \frac{h_\mathrm{st}}{h_\mathrm{th}} \; .
\end{equation}

For subcritical transitions with phase coexistence, the transition occurs at a single constant temperature, so that $h_\mathrm{th}=0$ and $B\rightarrow \infty$. For $\pr=3$, $B \approx 0.1$ and the effect becomes negligible. For even higher pressures, $h_\mathrm{th} \gg h_\mathrm{st}$ and $B\rightarrow 0$.  

This parameter thus gives us evidence about an upper pressure limit of $\pr \approx 3$ for pseudo boiling and associated phenomena.

\subsubsection{Heat flux vs mass flow}
In order for pseudo boiling to play a role in a process, the pseudo boiling state has to be reached from the inflow condition, subscript (in). The enthalpy required to do so for a given mass flow rate $\dot{m}$ is then \cite{BanutiPoF2016} 
\begin{equation}
\Delta \dot{H}_\mathrm{pb} = \dot{m}(h_\mathrm{pb} - h_\mathrm{in}) \; .
\end{equation}
For an isobaric heating process, we can write 
\begin{equation}
\Delta \dot{H}_\mathrm{pb} = 
\dot{m} \int_{T_\mathrm{in}}^{\Tpb}c_p(T)\dd T =
\dot{m} \left[h(\Tpb) - h(T_\mathrm{in})\right] \; .
\end{equation}

We can then define a parameter $\Lambda$ \cite{BanutiPoF2016} that relates the required energy addition $\Delta \dot{H}_\mathrm{pb}$ from inflow conditions to the supplied heat $\dot{q}$ 
\begin{equation}
    \Lambda = \frac{\Delta \dot{H}_\mathrm{pb}}{\dot{q}} \; .
    \label{eq:Lambda}
\end{equation}
Then, $\Lambda=1$ means that pseudo boiling conditions are exactly reached through added heat, and $\Lambda\gg1$ means that pseudo boiling is not reached.

Equation~(\ref{eq:Lambda}) under the condition $\Lambda=1$ can be rearranged to determine the limiting mass flow rate for pseudo boiling to be reached
\begin{equation}
    \dot{m}_\mathrm{pb} = \frac{\dot{q}}{(h_\mathrm{pb} - h_\mathrm{in})}
    \label{eq:mpb}
\end{equation}
Equation~(\ref{eq:mpb}) suggests that pseudo boiling can be avoided for 
\begin{equation}
    \dot{m} \gg \dot{m}_\mathrm{pb}\; .
    \label{eq:mnopb}
\end{equation}

The ill-defined `supercritical boiling number' SBO was later introduced that is reminiscent of the reciprocal of Eq.~(\ref{eq:Lambda}) \cite{Zhu2019}, $SBO = q_\mathrm{w}/ \dot{m} \, i_\mathrm{pc}$. However, here, $i_\mathrm{pc}$ denotes the specific enthalpy at the pseudo boiling temperature rather than an enthalpy difference. As the absolute value of the enthalpy is meaningless, and indeed arbitrary reference values can be chosen \cite{Elliott2012}, $i_\mathrm{pc}$ and thus SBO are likewise arbitrary and thus not physically meaningful.

\subsection{Conclusion}
It has become clear that distinct phases and phase transitions can be distinguished at supercritical pressures. Then, phase transition phenomena that are known at subcritical pressures, such as the boiling crisis, can equally be expected at supercritical pressures.  Pseudo boiling theory provides criteria for onset and upper pressures limits for these phase transitions. 

The absence of a phase equilibrium beyond the critical point is no contradiction: heat transfer is inherently a non-equilibrium process. The existence of different regions with distinctly liquid or gaseous character, such as a gaseous film layer embedded in liquid bulk flow, is caused by the existence of regions with different temperatures.

\newpage
\section{Fluid Mechanics and convective heat transfer}
\label{fluid}

The rapidly changing thermo-physical properties in the vicinity of the critical point can introduce significant complexity to even simple channels or pipes. However, measuring the internal flow field in the supercritical condition is an extremely challenging task due to the high pressure involved \citep{Licht2009,edlebeck2014measurements}. The feasibility of particle image velocimetry (PIV) measurement in a thermal convection with supercritical fuid was investigated by \citet{valori2018rayleigh, valori2019particle}. A preliminary background oriented schlieren (BOS) study confirmed that the tracer particles remained visible despite of signifcant local blurring. Main difficulty was a result of blurring and optical distortions in the boundary layer and thermal plumes regions.

Advances in computational power over the last decade have enabled the use of scale-resolving approaches such as DNS to study the fundamental fluid mechanics and convective heat transfer of supercritical fluids. Significant progress has been made in this area through the utilization of various high-fidelity numerical methods, which have contributed to the fundamental understanding of fluid mechanics at supercritical pressure.

Approximately 20 years ago, \citet{Bae2005,Bae2008} conducted a series of DNS on a spatially-developing long pipe filled with CO$_2$ at a pressure of 8 MPa. Wall heat flux caused the near-wall fluid temperature to rise above the pseudocritical temperature $T_\mathrm{pc}$, while the bulk fluid temperature remained below $\Tcr$. Heat transfer deterioration was observed in upward flow orientation, while heat transfer enhancement was seen in downward flow, highlighting the role of the buoyancy effect. Although a validation with the strongly heated airflow experiments \citep{shehata1998mean} was given by \citet{Bae2005,Bae2008}, a direct validation with the experiments of supercritical CO$_2$ was missing. Based on these results, \citet{Yoo2013} published a survey covering thermal-hydraulic considerations and fundamental fluid mechanics. DNS has become more affordable for turbulent flow at low Reynolds numbers in the last decade, making it a major contributor to new physical understanding and modeling. This section is aimed to summarize the relevant new findings.

\begin{figure}[hbt!]
\centering
\includegraphics[width=0.6\textwidth]{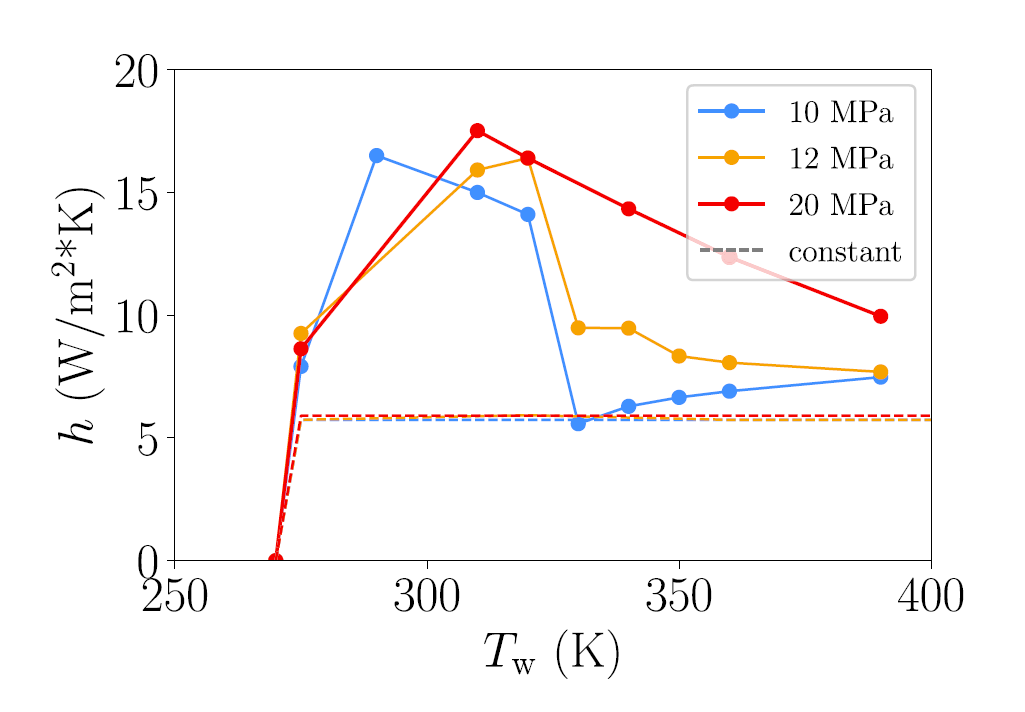}
\caption{Calculated heat transfer coefficient for CO$_2$. At the lower pressures an increase in the heat transfer coefficient is seen before the pseudo boiling temperature then a sharp decrease in the heat transfer coefficient occurs right after the pseudo boiling temperature. Reproduced from \citep{longmire2022onset} with permission given by Elsevier.}
\label{fig:laminar1}
\end{figure}

The occurrence of supercritical heat transfer deterioration (HTD) is commonly linked to mass flow rate, buoyancy, pressure, fluid properties, and temperatures \cite{Yoo2013,Pizzarelli2018}. Mostly, turbulence reduction or relaminarization is suggested as the underlying physical mechanism, as no true phase change -- like during subcritical film boiling and the associated boiling crisis -- occurs at supercritical conditions \cite{Yoo2013,Pizzarelli2018}.
However, the discussion of pseudo boiling above demonstrates that distinct liquid and gaseous states, as well as sharp interfaces between them, can exist in a flow field. Based on this, Longmire \& Banuti \cite{LongmireIJHMT2022} demonstrated that HTD does indeed occur in absence of turbulence and buoyancy in simulations of laminar forced convection over a flat plate. Their simulations showed that pseudo boiling alone is a sufficient physical mechanism to cause HTD, in the absence of turbulence or buoyancy, as seen in Figure \ref{fig:laminar1}. They observed the formation of a supercritical gas film at the wall, with low density, low thermal conductivity, a high compressibility factor, and low viscosity -- just like during subcritical film boiling and the boiling crisis.

Technical heat transfer systems, however, are typically operated in turbulent flow regimes. I.e., the interaction of pseudo boiling and turbulent flow governs heat transfer and supercritical phase transitions. This is the focus for the remainder of the paper.

\subsection{The scaling law of velocity and temperature in the vicinity of pseudo boiling point} \label{scaling law}

One of the most fundamental knowledge of the wall-bounded turbulent flow is the scaling of the mean velocity profile and the corresponding law of the wall. 
The law of the wall forms the cornerstone of modeling for wall-bounded turbulent flows, including techniques such as wall functions in RANS or wall-modeled LES. Yet, in conditions where strong thermo-physical property variations occur, conventional turbulence behavior is altered, causing scaling laws based on constant property flows to fail. In such cases, the van Driest velocity transformation \citep{vanDriest1951} can accommodate density variations provided that Morkovin’s hypothesis \citep{Morkovin1962} is met. However, when dealing with transcritical fluids, Morkovin's hypothesis is often violated, making the derivation of new scaling laws a priority.

A conventional turbulent channel flow (Figure \ref{fig:DNSchannel}) offers a suitable starting point to test these scaling laws and to investigate turbulence modulation in transcritical fluids. This configuration allows for the design of DNS with two homogeneous and statistically steady-state directions: streamwise and spanwise. The homogeneous setting of the streamwise direction disables the appearance of the heat transfer deterioration makes a simplification to the condition. With isothermal walls set at differing temperatures—one hot side $T_{\mathrm{hot}}>T_{\mathrm{pc}}$ and one cold sie $T_{\mathrm{cold}}<T_{\mathrm{pc}}$-the wall-normal direction encompasses a temperature range on either side of the pseudocritical point. This facilitates a detailed exploration of the effects of significant variations in thermo-physical properties on wall-bounded turbulent flow. Many recent DNS studies have employed this setup \citep{Patel2015, Patel2016, Patel2017,Pecnik2017, Peeters2016, Peeters2017, Ma2018, Guo2022, Wan2020,Foll2019}, despite differences in numerical methods, fluids, and temperature variations between the walls. 


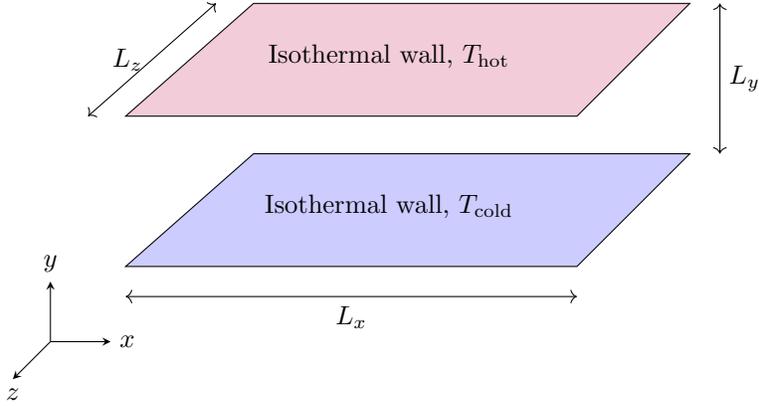
\begin{figure}[hbt!]
\centering
\begin{tikzpicture}[scale=1, transform shape]
    \coordinate (A) at (1,0);
    \coordinate (B) at (7,0);
    \coordinate (C) at (8.5,1.5);
    \coordinate (D) at (2.7,1.5);
    \coordinate (E) at (1,2);
    \coordinate (F) at (7,2);
    \coordinate (G) at (8.5,3.5);
    \coordinate (H) at (2.7,3.5);   
    \draw[fill=blue!20] (A) -- (B) -- (C) -- (D) -- cycle;
    \draw[fill=purple!20] (E) -- (F) -- (G) -- (H) -- cycle;
    \draw[<->] ($(A)+(0,-0.4)$) -- node[below] {$L_x$} ($(B)+(0,-0.4)$);
    \draw[<->] ($(C)+(0.4,0)$) -- node[right] {$L_y$} ($(G)+(0.4, 0)$);
    \draw[<->] ($(E)+(-0.5,0)$) -- node[left] {$L_z$} ($(H)+(-0.5, 0)$);
    \draw[-stealth] (0,-1) -- (0.8,-1) node[right] {$x$};
    \draw[-stealth] (0,-1) -- (0,-0.2) node[above] {$y$};
    \draw[-stealth] (0,-1) -- (-0.5,-1.5) node[below] {$z$};    
    \node at ($(A)!0.5!(B)+(0.5,0.8)$) {Isothermal wall, $T_{\mathrm{cold}} $};
    \node at ($(E)!0.5!(F)+(0.5,0.8)$) {Isothermal wall, $T_{\mathrm{hot}}$};

\label{fig:DNSchannel}  
\end{tikzpicture}
\caption{Channel flow setup in the DNS with two homogeneous (streamwise- and spanwise) directions, one side hot wall $T_{\mathrm{hot}}>T_{\mathrm{pc}}$ and one side cold wall $T_{\mathrm{cold}}<T_{\mathrm{pc}}$}
\label{fig:DNSchannel}
\end{figure}

In conventional wall-bounded turbulent flows with constant thermo-physical properties, the time-averaged streamwise velocity profile follows the law of the wall:

\begin{equation}
\begin{array}{l}
\displaystyle    \overline{u}^+ = y^+, \, \textrm{in the viscous sublayer,} \, y^+<5, \\[10pt]
\displaystyle    \overline{u}^+ = \frac{1}{\kappa}\log(y^+)+B, \, \textrm{in the logarithmic region,} \, y^+ \ge 30 ,
\end{array}
\end{equation}

where $u$ is the streamwise velocity, $y$ is the wall-normal coordinate, $^+$ indicates normalization by wall units, $\kappa\approx0.41$ is the von Kármán constant, $B\approx5.2$ is an additive constant and $\bar{\cdot}$ indicates an ensemble average. The wall unit for normalization is $u_\tau=\sqrt{\tau_\mathrm{w}/\overline{\rho}_\mathrm{w}}$ as friction velocity and $\delta_\nu=\overline{\mu}_\mathrm{w}/(\overline{\rho}_\mathrm{w} u_\tau)$ as the viscous length scale. Correspondingly, the dimensionless coordinate is $y^+=y/\delta_\nu$ and dimensionless velocity is $\overline{u}^+ = \overline{u}/u_\tau$. For variable-property flows such as compressible boundary layer flow and supercritical fluid flow, fluid density, dynamic viscosity and/or other flow properties are functions of temperature and pressure. The classical law of the wall could fail. About 60 years ago, \citet{vanDriest1951} derived a velocity transformation, commonly referred as van Driest transformation:

\begin{equation}
   \overline{u}^+_\mathrm{VD} =\int_0^{\overline{u}^+}(\frac{\overline{\rho}}{\overline{\rho}_\mathrm{w}})^{1/2}d\overline{u}^+ ,
\end{equation}

The performance of van Driest transformation has been assessed by various experiments and DNS of different conditions.
It works quite well for boundary conditions with adiabatic walls but less successful for flows with heat transfer involved. 
Therefore, the development of new velocity transformations is urgently required.


  

\citet{Patel2015,Patel2016} derived a semi-local transformation and tested it in their designed DNS. This semi-local transformation is derived by equating the turbulent momentum flux and the viscous stress to their incompressible counterparts. Interestingly, this transformation is the same as the one derived by \citet{Trettel2016} at almost the same time, although \citet{Trettel2016} intended for the compressible boundary-layer flow and used a slightly different approach. The validity of the semi-local velocity transformation was further proved using the setup as in Figure \ref{fig:DNSchannel} with synthetic variable thermo-physical properties. 
This semi-local scaling law reads as:

\begin{equation}
\begin{array}{l}
\displaystyle   y^\star = \frac{y Re^\star_\tau}{h} \\[6pt]
  
\displaystyle  \overline{u}^\star = \int_0^{\overline{u^{}}_\mathrm{VD}}(1+\frac{y}{Re_\tau^\star}\frac{dRe_\tau^\star}{dy})d\overline{u^{}}_\mathrm{VD} \\[6pt]

\displaystyle Re^\star_\tau =  \frac{Re_\tau \sqrt{\overline{\rho}/\overline{\rho}_\mathrm{w}}}{\overline{\mu}/\overline{\mu}_\mathrm{w}}
\end{array}
\end{equation}

where $y^\star$ is the semi-local wall-normal coordinate, $\overline{u}^\star$ is the transformed velocity and $Re^\star_\tau$ is semi-local friction Reynolds number. The semi-local scaling is proved to be quite useful on supercritical fluids. In the DNS test cases \citep{Patel2016}, it is able to collapse velocity profiles as a function of the semi-local wall coordinate despite of property gradients in the channel. Moreover, other turbulent characteristics such as streamwise velocity streaks and quasistreamwise vortices are also strongly governed by $Re^\star_\tau$ and their dependence on the density and viscosity profiles is minor. 

More recently, \citet{volpiani2020data} developed a data-driven velocity transformation with the assumption that the compressibility effect can be accounted for by the variation of mean density and mean viscosity. The exponent parameters are calibrated using DNS databases of adiabatic and diabatic boundary layers.

\begin{equation}
\begin{array}{l}
\displaystyle   u_\mathrm{V}^+ = \int_0^{\tilde{u}^+} \left( \frac{\bar{\rho}}{\bar{\rho}_\mathrm{w}}\right)^{\frac{1}{2}} \left( \frac{\bar{\mu}}{\bar{\mu}_\mathrm{w}}\right)^{-\frac{1}{2}} d \tilde{u}^+\\[6pt]
  
\displaystyle  y_\mathrm{V}^+ = \int_0^{\tilde{y}^+} \left( \frac{\bar{\rho}}{\bar{\rho}_\mathrm{w}}\right)^{\frac{1}{2}} \left( \frac{\bar{\mu}}{\bar{\mu}_\mathrm{w}}\right)^{-\frac{3}{2}} d \tilde{y}^+\\[6pt]
\end{array}
\end{equation}

In addition, \citet{Griffin2021} derived a total-stress-based velocity transformation. It is fundamentally distinct from previous methods, which have presumed that compressibility influences both the viscous stress and Reynolds shear stress through a single mechanism. In contrast, the present approach handles the viscous stress by considering mean property variations through the application of semi-local nondimensionalization. Meanwhile, it addresses the Reynolds shear stress in a manner that preserves the approximate equilibrium between turbulence production and dissipation.

\begin{equation}
\begin{array}{l}
\displaystyle u_\mathrm{G}^+ = \int_0^{y^\star} S_\mathrm{G}^+ d y^\star  \\[6pt]
 
\displaystyle  S_\mathrm{G}^+ =\frac{\tau^+ S_\mathrm{eq}^+}{\tau^+ + S_\mathrm{eq}^+-S_\mathrm{TL}^+} \\[6pt]

\displaystyle  S_\mathrm{eq}^+(y^\star)=\frac{1}{\bar{\mu}^+}\frac{\partial \tilde{\mu}^+}{\partial y^\star}=\frac{\partial u_\mathrm{eq}^+}{\partial y^\star} \\[6pt]

\displaystyle   S_\mathrm{TL}^+(y^\star)=\bar{\mu}^+\frac{\partial \tilde{\mu}^+}{\partial y^+}=\frac{\partial u^\star}{\partial y^\star} \\[6pt]
\end{array}
\end{equation}


\citet{Bai2022} conducted an evaluation of various velocity transformations for non-canonical wall-bounded turbulent flows, including those involving supercritical fluids. The assessment encompasses four databases from different groups (\citet{Wan2020}, \citet{Kawai2019}, \citet{Toki2020}, \citet{Ma2018}), which can be categorized into two groups: the low-density ratio group (with a typical density ratio of $\mathcal{O}(<10)$, from \citet{Wan2020} and \citet{Toki2020}) and the high-density ratio group (with a higher density ratio of $\mathcal{O}(20)$, from \citet{Ma2018} and \citet{Kawai2019}).
The transformations were successful in collapsing the mean velocity profiles in the low-density ratio group, with the best performance achieved by the transformation proposed by \citet{Patel2015}. However, in the case of the high-density ratio group, none of the examined transformation methods were able to collapse the profiles to match the incompressible reference.

\begin{figure}[hbt!]
\centering
\includegraphics[width=0.48\textwidth]{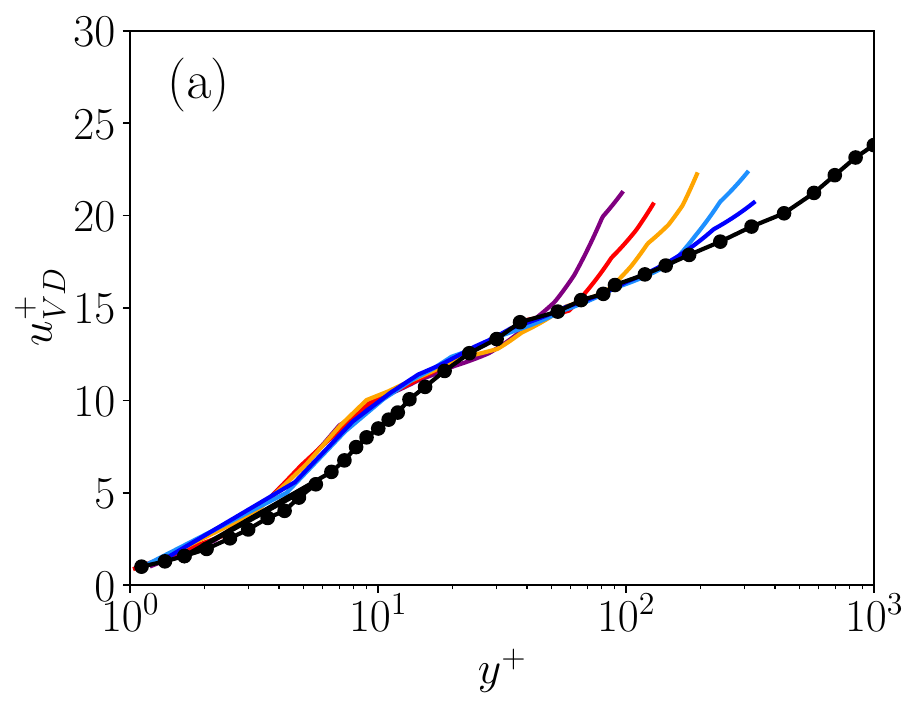}
\includegraphics[width=0.47\textwidth]{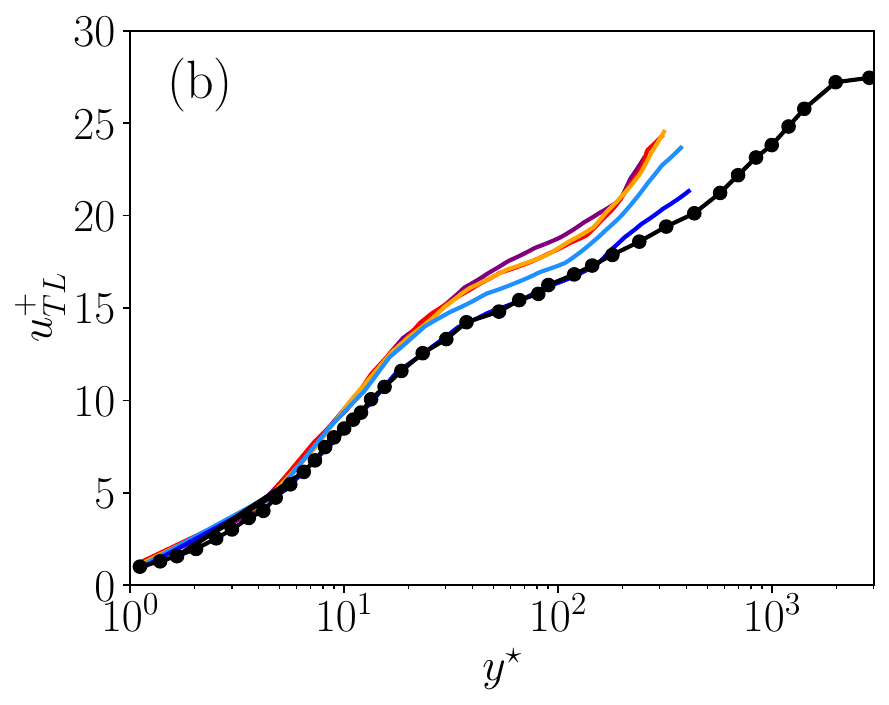}
\includegraphics[width=0.48\textwidth]{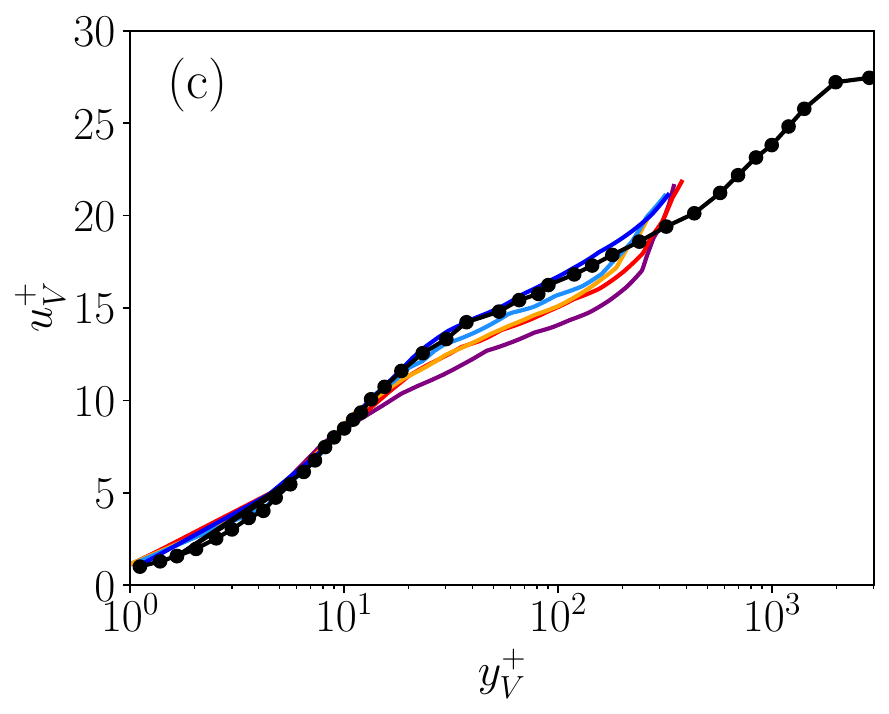}
\includegraphics[width=0.47\textwidth]{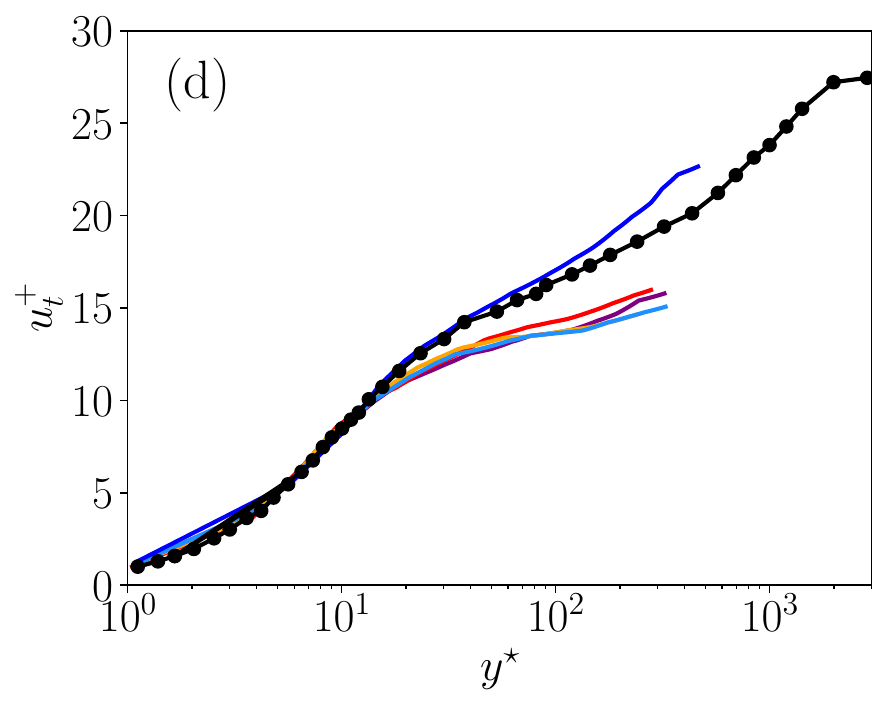}
\caption{Validation of velocity transformation laws on the DNS data (5 cases) of \citet{Kawai2019}: Van-Driest tranformation \citet{vanDriest1951}  in (a), the transformation from \citet{Patel2015} in (b), the transformation from \citet{volpiani2020data} in (c), the transformation from \citet{Griffin2021} in (d), and incompressible the zero-pressure-gradient boundary layer (black dotted line) by \citet{Sillero2013} as comparisons (reproduced from \citep{Bai2022} with permission given by AIAA)}
\label{fig:LinFu}
\end{figure}

Figure \ref{fig:LinFu} illustrates the results from one of the four assessments, with the data from \citet{Kawai2019}. While all four transformations demonstrate a qualitative agreement, the best performance is observed using the method of \citep{volpiani2020data}, though it still falls short of success.
The conditions under which these transformations are successful appear to be limited to cases with a density ratio of $\mathcal{O}(<10)$. In practical engineering applications involving transcritical fluids, large density ratios on the order of $\mathcal{O}(10\sim100)$ and significant density fluctuations are frequently encountered. Therefore, the development of advanced velocity transformations that are suitable to these realistic conditions is both necessary and urgent \citep{hasan2023incorporating}.
An enhanced velocity transformation has the potential to significantly refine the performance of turbulence modeling. This assertion has been demonstrated and supported by the work of \citet{Patel2018}.

Analog to the velocity transformation, the temperature profile is also expected to follow a universal scaling behavior. Firstly, we define $\overline{\theta}=|\overline{T}-T_w|$ where $T$ is the fluid temperature and $T_w$ donates the wall temperature. 
Normalization of $\theta$ reads:

\begin{equation}
   \overline{\theta}^+ =\frac{\overline{\theta}\rho_\mathrm{w} c_{p,\mathrm{w}} u_\tau}{q_\mathrm{w}}=\frac{\overline{\theta}}{T_\tau}=\frac{|\overline{T}-T_\mathrm{w}|}{T_\tau} ,
\end{equation}
with $\rho_w$, $c_{p,w}$ and $Pr$ as the density and special heat capacity on the wall as well as the Prandtl number. 
A general form of the temperature law of the wall \citep{Kader1981} can be expressed as:

\begin{equation}
\begin{array}{l}
\overline{\theta}^+=\overline{Pr}y^+ \textrm{, in the viscous sublayer,}\\[10pt]

\overline{\theta}^+=\frac{1}{\kappa_\mathrm{T}}\ln{y^+}+\beta (\overline{Pr}) \textrm{, in the logarithmic (log) region,}\\
\end{array}
\end{equation}

$q_\mathrm{w}=|\lambda \partial T/\partial y|_\mathrm{w}$ denotes the wall heat flux and $T_\tau=q_\mathrm{w}/(\rho_\mathrm{w} c_{p,\mathrm{w}} u_\tau)$ is the friction temperature. Similar to the van Driest transformation of velocity, the van Driest transformed temperature can be written as:

\begin{equation}
   \overline{\theta}^+_\mathrm{VD} =\int_0^{\overline{\theta}^+}(\frac{\overline{\rho}}{\overline{\rho}_\mathrm{w}})^{1/2}\frac{\overline{c_p}}{\overline{c_p}_\mathrm{w}}d \overline{\theta}^+,
\end{equation}


\citet{Patel2017} and \citet{Wan2020} evaluated this transformation using their own DNS studies, one employing synthetic properties and the other focusing on transcritical CO$_2$. Both studies found that the transformation did not provide a satisfactory collapse. They concurred that this failure could be attributed to the influence of near-wall property variations, which the transformation did not adequately account for. This evidence suggests that considering only the mean density ($\overline{\rho}$) and the mean specific heat at constant pressure ($\overline{c_p}$) is not enough to create an effective temperature transformation. An important progress is the extended van Driest transformation derived from the low-Mach energy equation by \citet{Patel2017} and later by \citet{Wan2020}. The extended van Driest transformation starts from the low-Mach-number Reynolds averaged energy equation, with a final expression as:

\begin{equation}
   \overline{\theta}^\star_\mathrm{eVD} =\int_0^{\tilde{y}^+} \frac{1}{\frac{\alpha_\mathrm{t}}{\overline{\mu}}+\frac{1}{Pr^\star}} d\varphi ,
\end{equation}
 
with the turbulent eddy conductivity $\alpha_\mathrm{t}=(-\overline{\rho}\tilde{\nu^{\prime\prime}\theta}^{\prime\prime})/(d\overline{\theta}/dy)$, the semi-local Prandtl number is defined as $Pr^\star=Pr_\mathrm{w}(\overline{c_p}/c_{p,\mathrm{w}})(\overline{\mu}/\mu_\mathrm{w})(\overline{\lambda/\lambda_\mathrm{w}})$.
$\overline{\theta}^\star_\mathrm{eVD}$ has been seen with a reasonable collapse in the less-challenging cases \citet{Patel2017, Wan2020} as well as an improvement in \citet{Guo2022}, which has a density difference of $\mathcal{O}(20)$. In all three tests, $\overline{\theta}^\star_\mathrm{eVD}$ exhibits a Prandtl-number-dependent shift and a similar slope in the log-law region. To eliminate the influence of local Prandtl number in the log regime, $\overline{\theta}^\star_\mathrm{eVD}$ can be decomposed into two parts: $\overline{\theta}^\star_\mathrm{\mathcal{T}, eVD}$ with constant local Prandtl number $Pr^\star=1$ and local-Prandtl-number depend part $\overline{\theta}^\star_\mathrm{\mathcal{P}, eVD}$
\begin{equation}
   \overline{\theta}^\star_\mathrm{eVD} =\overline{\theta}^\star_\mathrm{\mathcal{T}, eVD}+\overline{\theta}^\star_\mathrm{\mathcal{P}, eVD}, \textrm{with }  \overline{\theta}^\star_\mathrm{\mathcal{T}, eVD}=\int_0^{\tilde{y}^+}\frac{1}{(\frac{\alpha_\mathrm{t}}{\overline{\mu}}+1)}d\varphi
\end{equation}

By using the constant local Prandtl number part $\overline{\theta}^\star_\mathrm{\mathcal{T}, eVD}$, a quite successful temperature scaling has been proved by \citet{Patel2017, Wan2020} and also a significant improvement in the full compressible DNS of \citet{Guo2022}.

The above temperature scaling laws are all derived from the energy equation with the priori-assumed low-Mach condition, its validity could be questionable since the compressibility of transcritical fluids even in low bulk Mach number setup is not a clear issue. Therefore, \citet{Guo2022} took a step forward and developed a new temperature transformation within the same framework as above. \citet{Guo2022} derived the scaling law from the full compressible Navier-Stokes equations rather than the low-Mach-number assumption as other scaling laws. By analyzing the averaged energy flux equation obtained from DNS, it was found that the pressure work $\overline{u_i \partial p/ \partial x_i}$ is small enough to be neglected but the contribution of viscous dissipation of kinetic energy $\overline{\tau_{ij}\partial u_i / x_j}$ is significant. 
In addition, the hypotheses of linear heat flux is found to restrict the performance of the transformation and has also been replaced. 
This version of transformation can be considered as the state-of-art as it considers all possible physical processes allowed by the energy equation:

\begin{equation}
   \overline{\theta}^\star =\int_0^{\overline{y}^\star} \frac{\frac{1}{\overline{c_p}}\frac{d\overline{h}}{d\overline{\theta}}}{\frac{\alpha_\mathrm{t}}{\overline{\mu}}+\frac{1}{Pr^\star}} d\varphi , \textrm{with } \overline{\theta}^\star_\mathrm{\mathcal{T}} =\int_0^{\overline{y}^\star} \frac{\frac{1}{\overline{c_p}}\frac{d\overline{h}}{d\overline{\theta}}}{\frac{\alpha_\mathrm{t}}{\overline{\mu}}+1} d\varphi
\end{equation}

In the validation, the constant-Prandtl number part $\overline{\theta}^\star_\mathrm{\mathcal{T}}$ shows a nearly-universal scaling in all five cases. Further efforts are required to prove the validity of this temperature scaling in other conditions, such as an engineering-realistic density difference of $\mathcal{O}(100)$.

\subsection{Insights into convective heat transfer via DNS}
\subsubsection{Heat transfer deterioration in the vertical pipe flow}


\begin{figure}[hbt!]
\centering
\begin{tikzpicture}[
    arrowFilling/.style={single arrow, draw, minimum height=2.5cm, minimum width=1cm, single arrow head extend=.2cm, single arrow tip angle=150},
    heatArrow/.style={-{Triangle[length=3mm,width=2mm]},red, line width=1mm},
    ]
    \def\pipeLength{7cm}
    \def\pipeRadius{1cm}
    \def\D{2cm}
    \draw (0,0) -- node[fill=white,rotate=90,inner sep=-2.75pt,outer sep=0,anchor=center]{$\approx$} (8,0);
    \draw (0,2) -- node[fill=white,rotate=90,inner sep=-2.75pt,outer sep=0,anchor=center]{$\approx$} (8,2);
    \draw[dashdotted] (-1,\pipeRadius) -- (9,\pipeRadius);
    \draw (8,\pipeRadius) ellipse(\pipeRadius/2 and \pipeRadius);
    \draw[dashed] (1,\pipeRadius) ellipse(\pipeRadius/2 and \pipeRadius);
    \draw (0,2) arc(90:270:\pipeRadius/2 and \pipeRadius);
    \draw[|<->|] (0,-0.5) --++(\pipeRadius,0) node[midway,below=0] {Inflow generator};
    \draw[|<->|] (\pipeRadius,-0.5) --++(\pipeLength+0.4,0) node[midway,below=0] {Heated domain};

    \foreach \x in {0.3, 0.4, 0.5, 0.6, 0.7, 0.8, 0.9}{
        \draw[heatArrow] (\pipeLength*\x, \pipeRadius*3) -- ++(0,-1);
    }
    \node[above,red] at (\pipeLength*0.99, \pipeRadius*2 + 1) {$q_w$};

    \draw[->,very thick,green!45!black,line cap=round] (\pipeLength-1cm,\pipeRadius) -- ++(\pipeRadius,0) node[below left] {Flow direction};

    \draw[->,very thick,gray,line cap=round] (\pipeLength-3cm,\pipeRadius) -- ++(-\pipeRadius,0) node[below left] {$g$};

\end{tikzpicture}

\caption{Spatial developing pipe flow setup}
\label{fig:DNSpipe}
\end{figure}
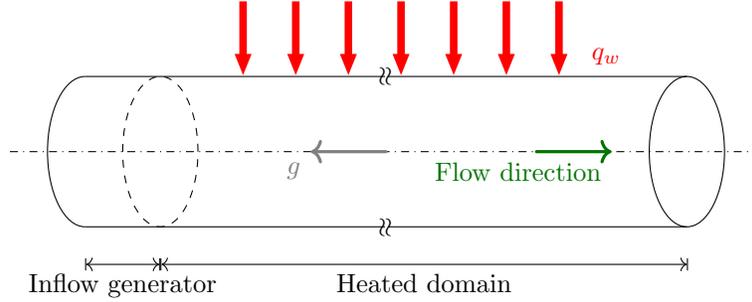


Over the past two decades, numerous attempts have been made to predict abnormal heat transfer through turbulence modeling \cite{He2008}. Unfortunately, most of these attempts have been successful only under specific conditions, and they have failed to accurately reproduce the trends observed in heat transfer. DNS, on the other hand, has shown the capacity to provide valuable insights into the physics of flows within confined ranges of flow and heat transfer conditions \cite{Yoo2013}.
As the pioneer in this area, \citet{Bae2005} designed and performed a series of DNS for turbulent heat transfer to sCO$_2$ and elucidated the effects of buoyancy on turbulent heat transfer in a vertical tube as shown in Figure \ref{fig:DNSpipe}. In their DNS, the wall temperatures for downward flow showed lower values than for upward flow and a local peak in temperature was observed in upward flow, a characteristic of deteriorated heat transfer. 
\citet{Bae2005} explained the heat transfer deterioration with the very strong buoyancy forces due to density variations. The velocity profile flattens out during the relaminarization. The flow can recover with a new kind of turbulence with a further increase of the buoyancy force, and the mean velocity profile acquires then an $\mathcal{M}$-shape. 

This setup has been inspiring many researchers and significant progress has been achieved since then. 
From today's perspective, the spatial resolution from \citet{Bae2005} ($0.18<\Delta r^+<5.34$, $\Delta (R\theta)^+\approx 9.41$, $\Delta z^+ \approx 14.55$ based on the inlet velocity and properties) is insufficient. After the inlet, the bulk Reynolds number goes higher since the bulk temperature rises due to the heat input. Moreover, the thermal scales are expected to be much smaller than the Kolmogorov length scale since the Prandtl number is $Pr_0=3.19$ at the inlet and reach a maximum value of $Pr=14$ for CO$_2$ at the pseudo-critical temperature. The smallest thermal scale, i.e. the Batchelor scale has a relationship of $\eta_\theta=\eta/\sqrt{Pr}$ \citep{Tennekes1972}. Thus, the mesh resolution in each direction to resolve the temperature field should be approximately twice to four times as fine as required to resolve the velocity field. 
For instance, in the DNS of \citet{Wan2020} and \citet{Nemati2016}, both showed the mesh resolution in each direction compared to the Batchelor scale $\Delta/\eta_\theta$, which could exceed $\Delta/\eta_\theta>10$ in the position where pseudo-critical temperature appears. This indicates a slight under-resolution. 
However, this has only minimal influence on the first- and second-order statistics such as mean wall temperature, as pointed out by the grid independence studies by \citet{Nemati2015,Cao2021}.


Given the rapid advancements in HPC resources, recent simulations can now be performed with high-order methods or higher spatial resolution. As such, a convincing validation usually serves as the initial step before delving into scientific analysis. Although DNS is often regarded as the ground truth or a numerical experiment, noticeable differences may exist between numerical methods, spatial and temporal resolution, thermo-physical libraries, and other factors. DNS of canonical turbulence cases, such as channel flow \citep{Lee2015,lozano2014effect}, pipe flow \citep{wu2008direct}, and turbulent boundary layers, generally exhibit a cohesive agreement across different studies. However, the discrepancy between simulations tends to increase with the complexity of the flow. In more intricate scenarios like multi-phase turbulent flows \citep{nemati2021direct,chu2020turbulence,liu2023large,liu2023}, reaching mutual agreement among different simulations can be particularly challenging. Most existing DNS studies of supercritical fluids have been validated using well-recognized experiments of strongly heated air flow \citep{Shehata1998,Mceligot2020,mceligot2018internal, Chu2016c}, which also exhibit slight heat transfer deterioration. However, these are typically considered less challenging cases. A direct validation between experiments and DNS of supercritical fluids remains rare, as the Reynolds number that DNS can cover is generally 1-2 orders of magnitude lower than that in experimental settings. 


\begin{table}[h!]
\centering
\begin{tabular}{|c|cccc|}
\hline
\makecell[c]{Case name in \\ \citet{Bae2005}}& \makecell[c]{Pipe diameter \\ (mm)} & \makecell[c]{heat flux \\(kW/m$^2$)}  & Flow direction &convection type\\
\hline
A   &1  & 61.74 & upward &forced\\
B   &1  & 61.74 & upward &mixed\\
C   &2  & 30.87 & upward &mixed\\
E   &1  & 30.87 & downward &mixed\\
F   &2  & 30.87 & downward &mixed\\
H   &2  & 61.74 & upward &mixed\\
\hline
\end{tabular}
\caption{Validation cases, $p_0=8$ MPa, $T_0$=301.15 K, Re$_0$=5400}
\label{tab:2}
\end{table}

\begin{figure}[hbt!]
\centering
\includegraphics[width=0.48\textwidth]{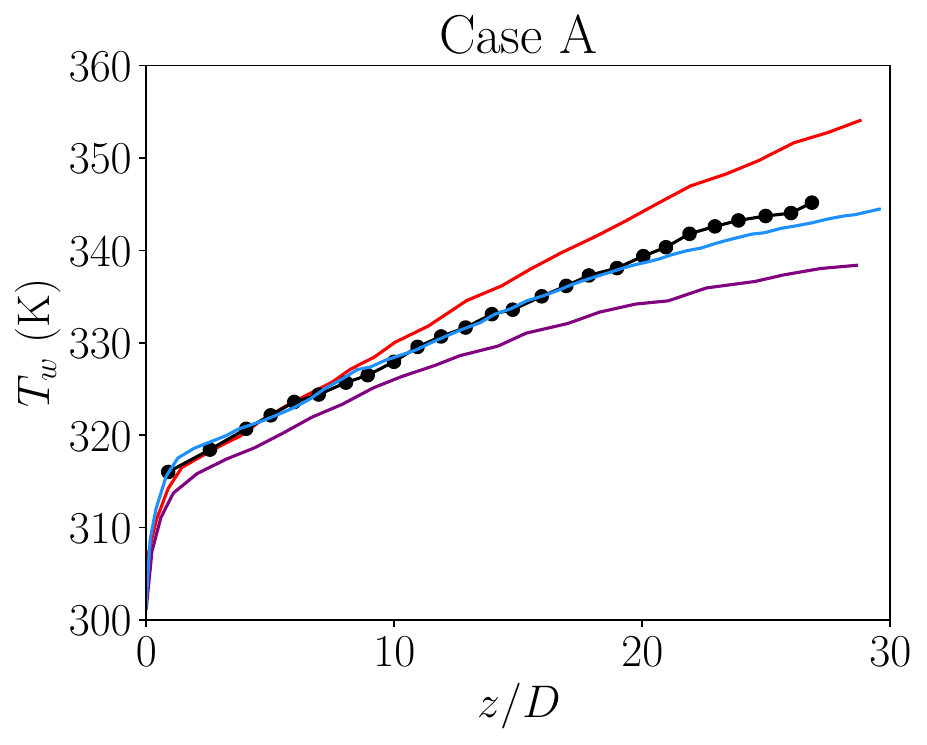}
\includegraphics[width=0.48\textwidth]{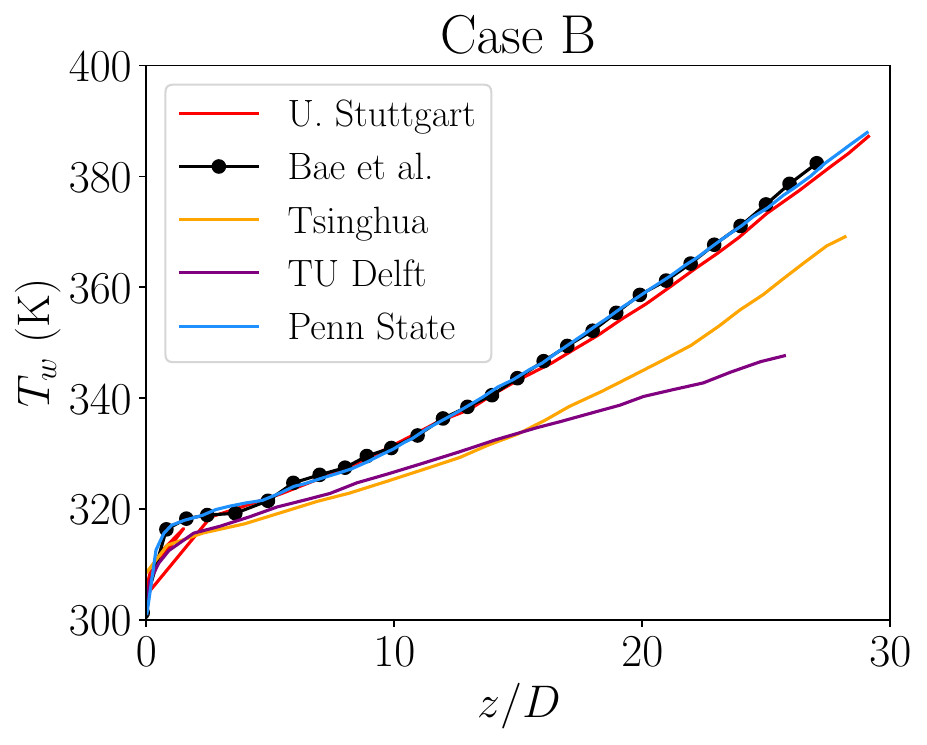}
\includegraphics[width=0.48\textwidth]{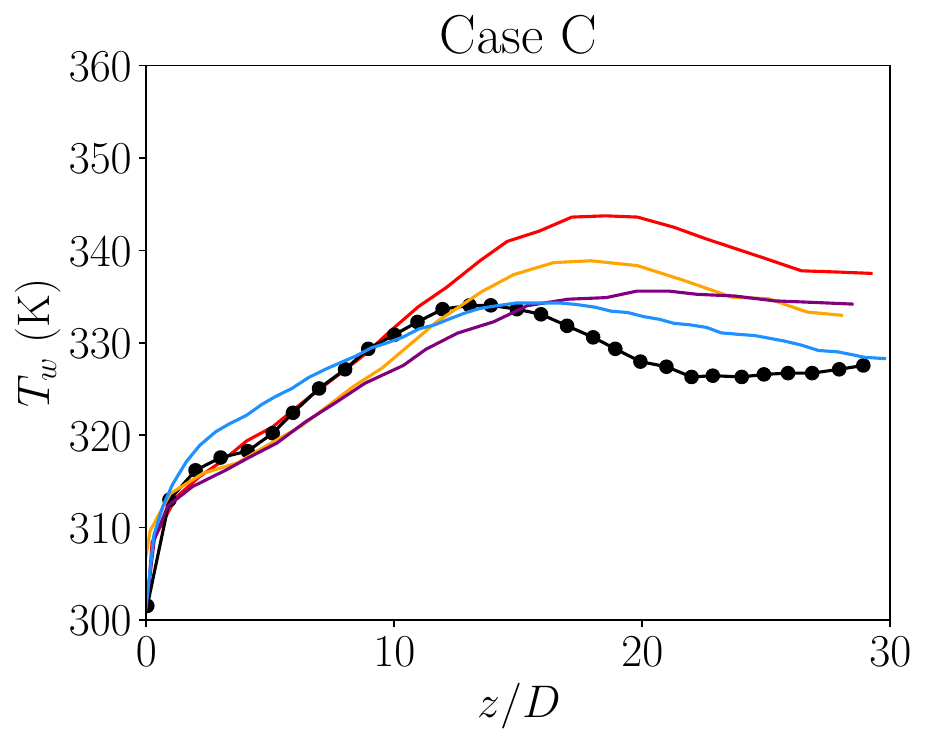}
\includegraphics[width=0.48\textwidth]{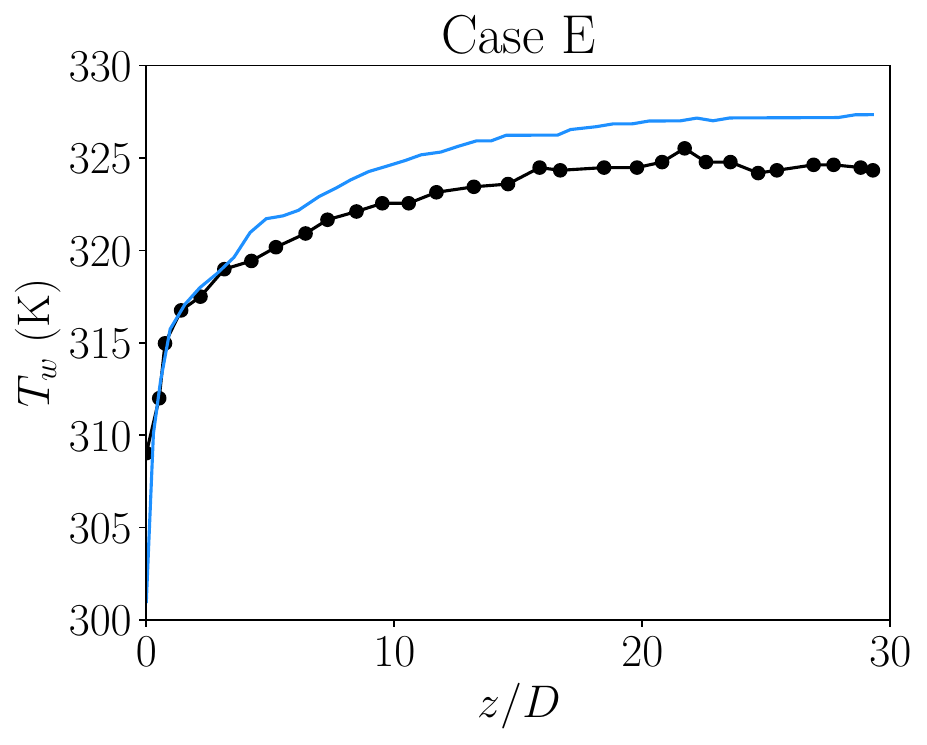}
\includegraphics[width=0.48\textwidth]{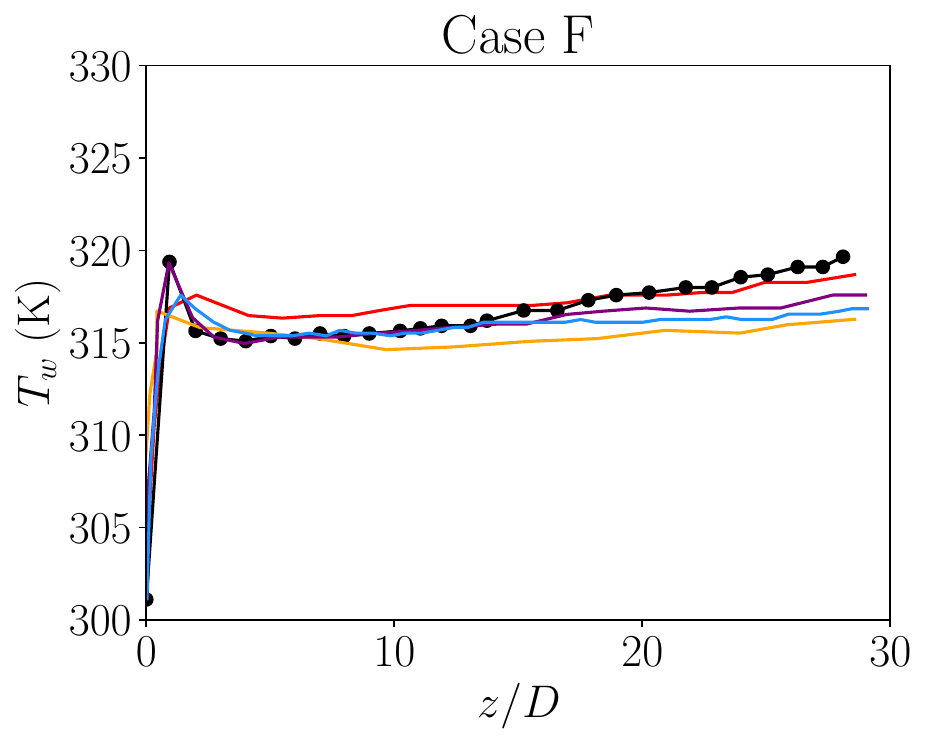}
\includegraphics[width=0.48\textwidth]{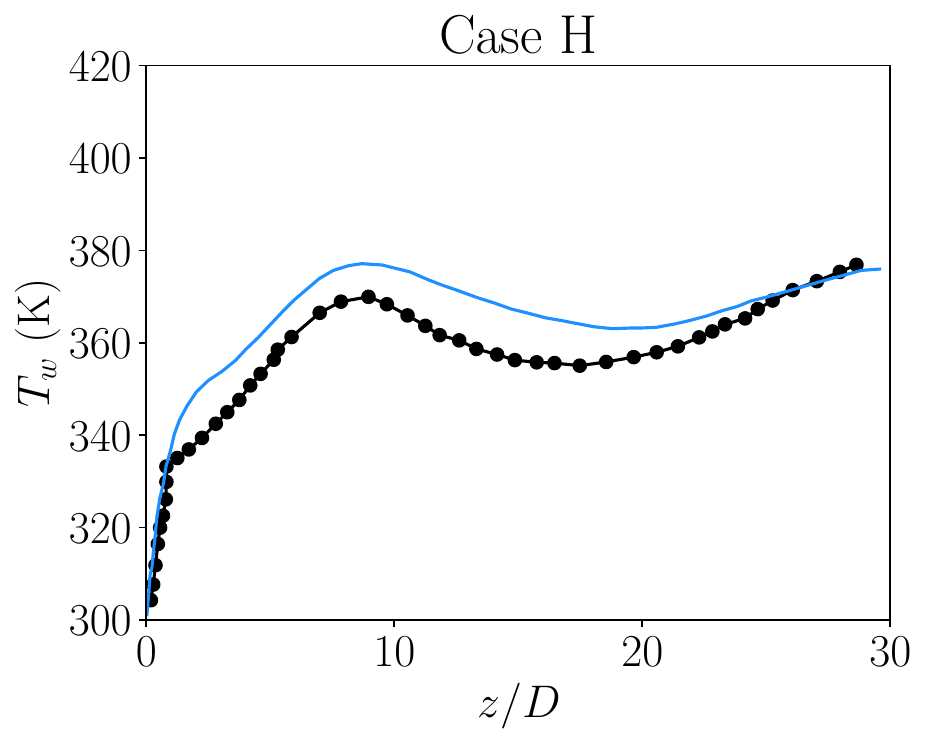}
\caption{A comparison of wall temperature between DNSs from five groups (\citet{Bae2005,Cao2021,Nemati2015,Nguyen2022, Chu2016})}
 \label{fig:DNSvalidation}
\end{figure}

Bringing together all the available DNS for a comprehensive comparison would indeed be a valuable contribution to the field, as such a detailed examination has not been conducted elsewhere. Such a unified reference could become an essential tool for researchers, guiding future simulations, assisting in the validation of new models, and fostering collaboration across the field.
Table \ref{tab:2} details the simulation conditions for all six validation cases originally designed by \citet{Bae2005}, and the reproduced results from different authors are depicted in Figure \ref{fig:DNSvalidation}. Among them, one case pertains to forced convection, while the other five cases involve mixed convection, including three upward flowing instances and two downward cases. These simulations were later independently performed by several research groups: \citet{Cao2021,Nemati2015,Chu2016,Nguyen2022}, each using different numerical methods, as outlined in Table \ref{tab:3}.
All of the numerical solvers leveraged thermo-physical properties extracted from NIST/PROPATH and implemented them as functions of temperature. Generally, four of the solvers maintained second-order accuracy in both space and time, whereas \citet{Nguyen2022} utilized a high-order spectral-element method known as NekRS \citep{Fischer2021}, employing a polynomial order of 7. To ensure a consistent foundation, all the DNS incorporated fully developed turbulent flow at the inlet, validated against the standard metrics for mean and fluctuation statistics.

\begin{table}[h!]
\centering
\begin{tabular}{|c|ccc|}
\hline
Authors& Solver & \makecell[c]{Mesh resolution\\ $r\times \theta \times z$} & Thermo library\\
\hline
\citet{Bae2005} & In-house & $69\times129\times769$ & PROPATH \\
\citet{Cao2021} & CHAPSim & $64\times128\times768$ & NIST \\
\citet{Chu2016} & OpenFOAM \citep{Jasak2007}& $168\times172\times2400$ & NIST\\
\citet{Nemati2015} & In-house & $128\times288\times1728$ & REFPROP \\
\citet{Nguyen2022} & NekRS \citep{Fischer2021} & 33 Mio. & PROPATH \\
\hline
\end{tabular}
\caption{Numerical details (partially) of five DNSs}
\label{tab:3}
\end{table}

It is notable that all simulations performed admirably in the validation against the heated air flow simulation of \citep{Shehata1998}. However, the agreement with supercritical CO$_2$ flow, as depicted in Figure \ref{fig:DNSvalidation}, is less successful.
For the cases exhibiting a monotonic trend of wall temperature (case A, B, E, F), the wall temperature $T_w$ in those simulations demonstrates a similar slope with only minor differences. The best agreement across all the DNS is observed in downward flowing cases E and F. On the other hand, the largest discrepancies are found in upward flowing cases C and H, characterized by a non-monotonic trend of wall temperature, i.e., the phase involving deteriorated heat transfer and subsequent recovery.
This side-by-side comparison highlights that the heat transfer of supercritical fluids is highly sensitive to the choice of discretization methods and simulation details. This sensitivity makes it a far more complex and demanding task than constant-property simulations or heat transfer with mild property variation, such as with air. A blind benchmark with experimental measurements could be extremely valuable in establishing a uniform standard for the community, guiding future research, and ensuring more consistent and reliable results.


\begin{figure}[hbt!]
\centering
\includegraphics[width=0.7\textwidth]{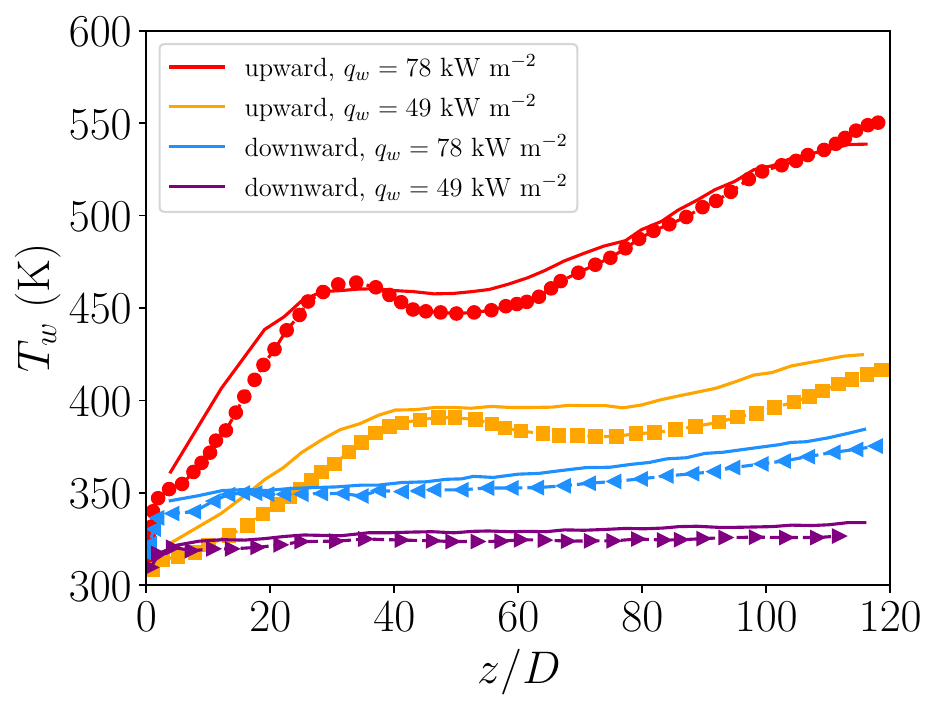}
\caption{Validation between DNS (\citet{Cao2021}, permission given by Camebridge University Press) in lines and experiments (\citet{Wang2019}) with markers}
\label{fig:Jiang1}
\end{figure}

Recently, \citet{Cao2021} conducted an insightful, non-blind validation alongside experiments from \citet{Wang2019} at a low Reynolds number. The DNS in their study was configured with parameters matching those in the experiments, encompassing both upward and downward arrangements. Figure \ref{fig:Jiang1} illustrates a validation of the averaged wall temperature. Within the experiment, the tube's inside diameter measures 1 mm, and its outside diameter is 2 mm. Thermocouples measured the temperatures of the tube's outer surface, and the inner tube surface's local temperatures were subsequently estimated. This validation approach serves as a notable example of how experimental data and computational simulations can be cohesively aligned, thereby enhancing our understanding of complex thermal and flow phenomena.
In general, the DNS results show a commendable agreement with the experimental data across all four considered cases. The trend of heat transfer deterioration, evident in cases A and B, has been accurately captured. Interestingly, case A demonstrates a better alignment with experimental results than case B, even though the wall heat flux in case A is higher. This successful validation is quite encouraging, indicating the potential for further exploration and refinement of DNS in understanding complex flow dynamics. The availability of this validation set, which includes both experimental measurements and DNS, represents a valuable asset for researchers and engineers in the field.



\begin{figure}[hbt!]
\centering
\includegraphics[width=0.4\textwidth]{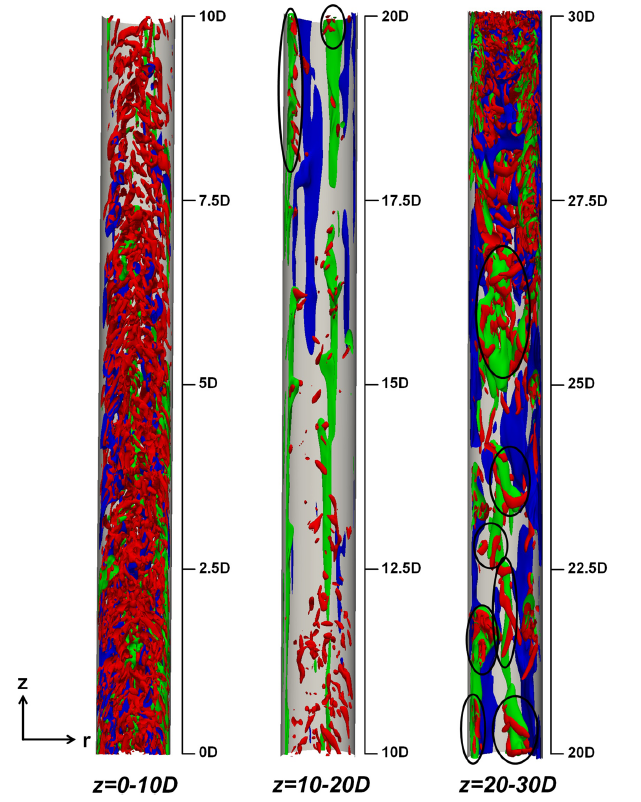}
\caption{Iso-surfaces of streaks and vortex structures along the pipe of cooling; blue: low-speed streaks; green: high-speed streaks. Reproduced from \citep{Pandey2018} with permission given by American Institute of Physics}
\label{fig:DNSPandey}
\end{figure}

Until now, there is still debate about the causality of the heat transfer deterioration. \citet{longmire2022onset} examined forced laminar convection of supercritical CO$_2$ over a flat plate, excluding factors like buoyancy or turbulence. The paper demonstrated pseudoboiling as a sufficient physical mechanism to cause heat transfer deterioration, even without turbulence or buoyancy. The study also observes the formation of a supercritical gas film at the wall, characterized by low density, low thermal conductivity, high compressibility factor, and low viscosity. These findings suggest that pseudoboiling is a sufficient mechanism for heat transfer deterioration.
In the turbulent regime, one theory may present a plausible explanation. In a heated upward flow or cooled downward flow, a flow relaminarization can appear which leads to the deteriorated heat transfer. This observation has been confirmed by many DNS as well as experiments. The flow development and heat transfer in a spatial-developing pipe can be divided into several stages:


\uppercase\expandafter{\romannumeral1}: In the vicinity of the wall, the fluid temperature rises past the pseudo-critical point, while the bulk temperature remains below this threshold. This vast discrepancy in thermo-physical properties exerts a profound influence on the turbulent flow field. Specifically, the near-wall region's density is much lower than in the bulk region, resulting in a non-uniform body force that flattens the mean velocity profile meaning a low velocity gradient and diminishes the shear production from the mean flow. This effect can be seen in the shear production term  in the turbulent kinetic energy budget. Consequently, the flow tends to become laminar, though turbulence may still be generated by buoyancy production. Near the wall, elongated and stable streaks are formed, largely driven by buoyancy production without any vorticity generation (see Figure \ref{fig:DNSPandey}). These streaks maintain the streamwise fluctuation, but the radial and circumferential fluctuations are notably weak. This rod-like turbulence, characterized by one-dimensional fluctuation, has been vividly depicted through the Anisotropy invariant map (AIM) \citep{Pandey2018,pandey2018development}. These elongated streaks stay stable and therefore the generic Self-Sustaining Process in Shear Flows is broken. The ability of a non-uniform body force to relaminarize turbulent flow has been confirmed in various studies \citep{He2016,Kuhnen2018b,Pandey2020}.



\uppercase\expandafter{\romannumeral2}: As the flow progresses further downstream, it becomes distorted, forming a $\mathcal{M}$-shape velocity profile indicating a recovered velocity gradient and therefore shear production. In this region, turbulence begins to recover. The elongated streaks generated in the relaminarization section turn unstable, leading to the creation of turbulence spots \citep{He2021}. Both buoyancy production and shear production act as significant sources of turbulence generation in this complex environment. Contrary to the monotonic velocity gradient typically seen in constant-property turbulence, the $\mathcal{M}$-shape  velocity profile possesses both positive and negative velocity gradients, with an inflection point in between. This means that both the shear production and turbulent fluctuation could exhibit two peaks in the radial direction. The recovered turbulence, characterized by an $\mathcal{M}$-shape velocity profile, exhibits significant differences from the turbulence at the inlet. According to quadrant analysis, this unique form of turbulence is dominated by Q1/Q3 events \citep{Pandey2018}, in contrast to the traditional sweep and ejection events (Q2/Q4) commonly observed in constant-property turbulence. 


\uppercase\expandafter{\romannumeral3}: As the bulk fluid temperature crosses the pseudo-critical point due to the constant input of heat flux, the variations in properties across the radial direction become less pronounced, and the non-uniformity of the body force diminishes. Consequently, the $\mathcal{M}$-shape velocity profile begins to flatten again, leading to the possibility of secondary relaminarization and secondary heat transfer deterioration. In most existing DNS, the pipe flow is not long enough to observe this phenomenon. However, in some DNS simulations with extreme pipe lengths, such as a pipe of $75D$ as studied in \citep{Zhao2021}, the secondary relaminarization and heat transfer deterioration can be observed. Interestingly, \citet{Zhao2021} also noted that the second instance of heat transfer deterioration is not as pronounced as the first one.


\uppercase\expandafter{\romannumeral4}: As the thermo-physical properties in the bulk of the flow approach those in the near-wall region, the velocity profile, initially flattened for a second time, will gradually return to a parabolic shape, typical of constant-property flow. Consequently, the phenomena of secondary relaminarization and heat transfer deterioration will dissipate. This evolution in flow characteristics underscores the complex interplay between thermo-physical properties and flow behavior in supercritical fluid systems.

The cause for the relaminarization and heat transfer deterioration could be even more complicated. 
Among many, buoyancy is considered as the dominant effect to alternate the heat transfer behavior. 
In the forced convection condition where the buoyancy is absent, strongly-relaminarized flow and the transition from the $\mathcal{M}$-shape velocity profile have not been observed as the mixed convection cases \citep{He2020, Chu2016c, Pandey2018}.
Instead, the wall temperature shows a monotonic trend.
One of the advantages of simulations is to artificially design and control the thermo-physical properties.
\citet{He2020} designed a series of DNS with one or more thermo-physical properties artificially frozen. 
The DNS solely with the real density profile and constant other properties is still able to show the $\mathcal{M}$-shape velocity profile and therefore the flow relaminarization and recovery.
On the other hand, keeping the density constant and remain the other real properties can not do the same.
This confirmed the dominant role of the buoyancy effect brought by the density variance.
Even though the heat transfer deterioration is more frequently found in the mixed convection under the strong influence of buoyancy, it should be also noted that sometimes heat transfer deterioration occurs in the forced convection with a negligible buoyancy effect \citep{Shiralkar1970}.

In addition to the local buoyancy effect, the global thermal acceleration can attenuate the heat transfer as well. The thermal acceleration is known as the global flow acceleration driven by the temperature gradient in the axial direction \citep{Mceligot2004}. Its interaction with the buoyancy effect belongs to the cornerstone of the supercritical heat transfer.
However, its impact is much weaker than the buoyancy effect.
Apart from density, viscosity $\mu$ also plays a direct role in the momentum equation. The rapid decrease in viscosity near the wall leads to a reduction in skin friction, consequently accelerating the flow in this region. In the DNS study with artificially combined thermo-physical properties, \citet{He2020} discovered that the effect of variable viscosity alone can lead to a reduction in turbulence by flattening the velocity profile. However, this effect does not suffice to transform the velocity profile into an $\mathcal{M}$-shape; this transformation can only be achieved through the influence of buoyancy.

Until now, there is no clear evidence of the exact cause of the deteriorated heat transfer. The variance of density, viscosity, heat capacity, pseudo-boiling, thermal acceleration, buoyancy and others are among the possible reasons. A potential way to quantify this systematically is to use the causality analysis in the framework of information theory.
Traditionally, causal inference has been reduced to analyzing the time correlation between pairs of signals. However, this approach is flawed, as correlation does not possess the necessary directionality and asymmetry to definitively establish causation. Granger's 1969 test \citep{granger1969investigating} sought to address this by assessing causality through the statistical utility of one time signal to predict another. Despite this advancement, concerns remain about assumptions related to the joint statistical distribution of data and the applicability of Granger causality to highly nonlinear systems. Recent efforts to mitigate these issues have shifted focus to information-theoretic measures of causality.
The concept of transfer entropy was proposed by \citet{Schreiber2000} as an effective approach to evaluate the directional information transfer between a source signal and a target signal. The transfer entropy measures the information contained in the source about the next state of the target that was not already contained in the target's past. This allows one to differentiate the direction of the information flux, and can therefore be used to quantify causal interactions. Recently, \citet{lozano2020causality,lozano2020cause,wang2021information, wang2022spatial,liu2023,liu2023large} highlighted the importance of causal inference in fluid mechanics and proposed leveraging information-theoretic metrics to explore causality in various scenarios. 

\subsubsection{Flow and heat transfer in complex geometries and boundary conditions}


Certainly, a vertical circular pipe with uniform heat flux is the most representative configuration for studying the heat transfer behavior of supercritical fluids. However, in practical applications such as printed circular heat exchangers (PCHE), additional geometrical complexities such as non-circular channels, non-uniform heat flux, and horizontal alignment often arise.
It is not difficult to imagine that the introduction of these complex boundary conditions will significantly alter the flow and heat transfer characteristics, especially since fluids near the critical point are highly sensitive to even small changes. Over the last decade, substantial progress has been made in understanding these complex geometries and boundary conditions. Many intriguing findings have been reported, providing valuable insights for engineering design and further expanding the scope of our understanding of supercritical fluid dynamics. However, the introductions here mainly focused on the high-fidelity simulations with DNS.

\begin{figure}[hbt!]
\centering
\includegraphics[width=1\textwidth]{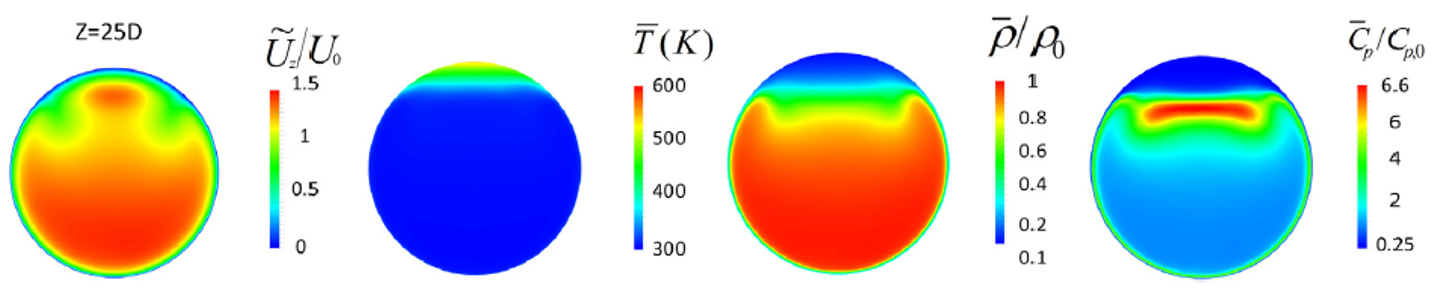}
\caption{DNS of a heated horizontal pipe flow, reproduced from \citep{Chu2016b} with permission given by Elsevier}
\label{fig:horizontal}
\end{figure}

\citet{Nemati2016} conducted a comparison between using constant heat flux as a boundary condition and employing averaged constant wall temperature as the boundary condition under forced convection. They found that suppressing temperature fluctuations at the wall led to a $7\%$ reduction in the Nusselt number. Conversely, allowing temperature fluctuations resulted in large property variations near the wall, triggering an increase in wall-normal velocity fluctuations and thereby enhancing overall heat flux and skin friction.
\citet{He2022effect} investigated the effect of conjugate heat transfer demonstrating that the inclusion of solid pipe wall conduction in the computational model dampened enthalpy fluctuations near the wall and caused uneven distribution of wall heat flux. These changes had minimal impact on flow development and turbulence but led to differences in heat transfer at the flow's early stage. When considering the thermal conduction of the solid wall, the Nusselt number exhibited a stronger laminar contribution and a weaker turbulent contribution. This resulted in a larger $Nu$ at an early stage and a lower $Nu$ at a later stage compared to the flow case without a solid wall.
\citet{pucciarelli2018effect} presented the findings from LES analyses regarding the influence of conjugate heat transfer. Specifically, the presence of a wall with realistic properties strongly dampens the large temperature and fluid property fluctuations that occur when imposing constant heat flux. Therefore, contrary to fluids in standard conditions, heat transfer to supercritical fluids appears to depend on the actual coupling between the fluid and the wall, adding another layer of complexity to an already challenging subject. 
The design of applications such as PCHE may require various arrangements of pipes or channels, including horizontal flows and inclined angles. While buoyancy plays a significant role in vertical flow, it is also expected to greatly influence horizontal flows and flows at other incline angles. Building on previous DNS of vertical flows \citep{Chu2016}, \citet{Chu2016b} unveiled new DNS results concerning horizontal pipe arrangements. Here, the density difference across the pseudo-critical point gives rise to strong flow stratification (as illustrated in Figure \ref{fig:horizontal}), with low-density hot fluid located in the upper layer and high-density cold fluid in the lower layer, separated by a layer with high specific heat capacity. Such stratification is driven by buoyancy-induced secondary flow from Dean vortices, resulting in a temperature difference between the upper and lower parts that is significantly larger than in vertical flows with the same setup. This information is invaluable for heat exchanger design.
Furthermore, \citet{wang2023direct} conducted DNS to investigate the underlying characteristics of turbulence and heat transfer in supercritical fluid flowing between two differentially heated parallel plates. By examining both forced and mixed convection, they sought to understand the effects of buoyancy and variable properties on stably-stratified horizontal flow. Their findings revealed that the variable property effect contributed to turbulence deterioration near the cold wall, but increased turbulence stress near the hot wall. While direct buoyancy effects tended to dampen turbulence throughout the channel, indirect buoyancy effects fostered flow laminarization near the cold wall and enhanced turbulence near the hot wall. Interestingly, the net buoyancy effect near the hot wall was found to be negligible, as competing direct and indirect effects canceled each other out. Overall, the turbulence behavior near the hot wall was largely governed by the variable property effect, while the wall heat flux was significantly reduced by buoyancy but less influenced by property variation.

\begin{figure}[hbt!]
\centering
\includegraphics[width=1\textwidth]{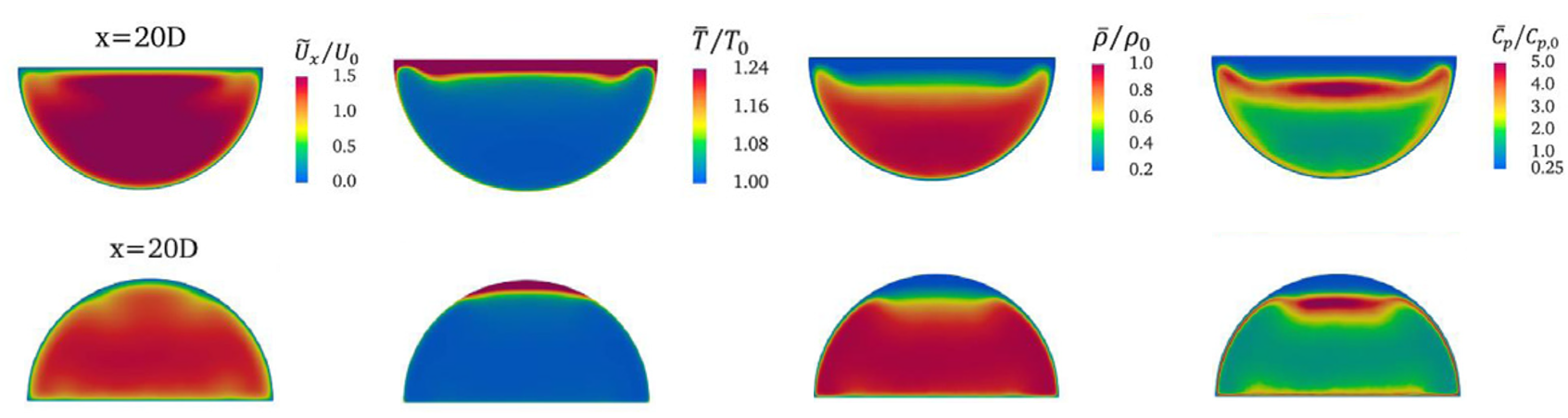}
\caption{DNS of a heated horizontal semicircular pipe flow, reproduced from \citep{Wang2021} with permission given by Elsevier}
\label{fig:semicircular}
\end{figure}

Due to manufacturing reasons, the channels in PCHE are usually semicircular rather than circular. The semicircular cross-section is featured with two additional corners which are similar to the duct flow.  \citet{Wang2021} performed DNS of a horizontal semicircular duct with a diameter of $D=2$ mm corresponding to a hydraulic diameter of $D=1.22$ mm, which is close to that of a typical PCHE channel. The flat wall and curved wall are both imposed with a constant heat flux. 
In the typical semicircular configuration (top flat-wall), the wall temperature was highest at the corner, which is known as the “corner effect” (Figure \ref{fig:semicircular}).
The second highest temperature is observed at the top wall owing to the buoyancy effect.
The secondary flow as in the horizontal circular pipe \citep{Chu2016b} was also seen here.
In the top curved-wall configuration (Figure \ref{fig:semicircular}), the temperature at the top flat wall was the highest.
In both horizontal configurations, the thermal stratification and secondary flow are obvious.

The effect of non-uniformity of wall heat flux has been studied by \citet{Pandey2021}. They performed DNS of vertical circular pipe flow with both axial and circumferential varying heat flux. It has been found that the circumferential heat flux induces the thermal stratification which resulted in a higher wall temperature at one side where heat flux was maximum.




\subsubsection{Flow relaminarization and stablization brought by non-uniform body force}

\begin{figure}[hbt!]
\centering
\includegraphics[width=0.48\textwidth]{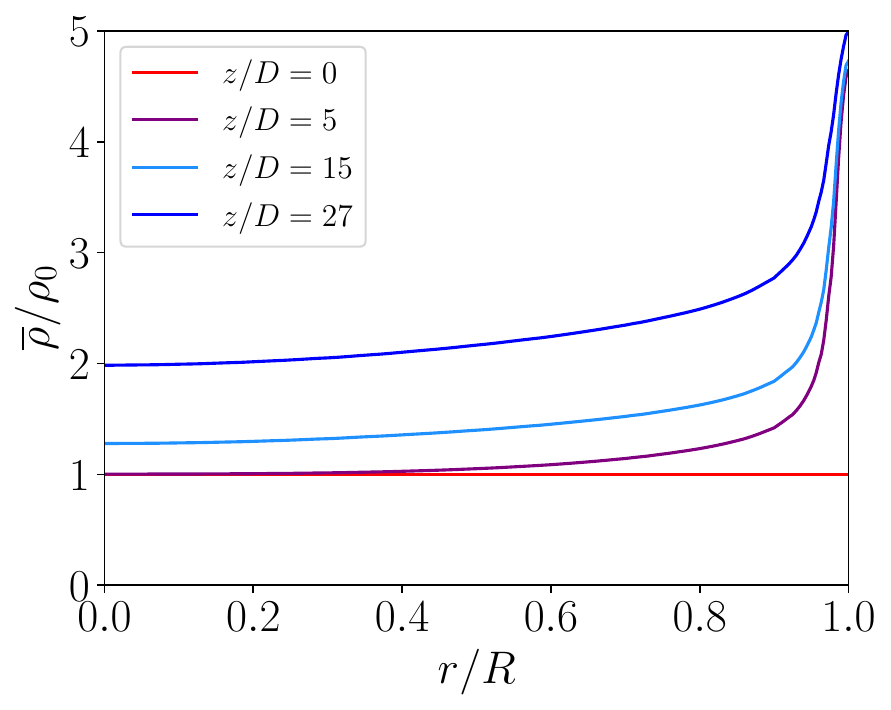}
\includegraphics[width=0.48\textwidth]{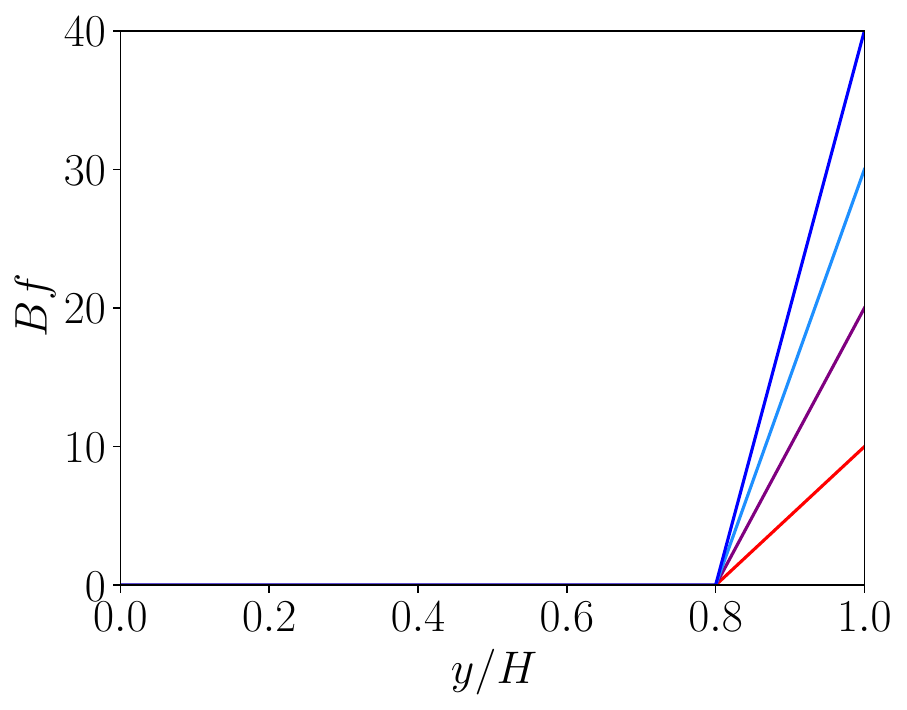}
\caption{(a):Density variation observed in DNS of \citet{Pandey2018}, (b):designed non-uniform body force by \citet{Pandey2020}. Reproduced with permission given by American Physical Society.}
\label{fig:bf}
\end{figure}

The numerous DNS results in the previous section provided evidence that the buoyancy-induced non-uniform body force flow can flatten the velocity profile leading to relaminarization and subsequently to heat transfer deterioration. This complex setup can be simplified using constant-property incompressible Navier-Stokes equations without the energy equation, and by including an additional body force term in the momentum equation to simulate the non-uniform body force caused by buoyancy, as illustrated by the linear change of body force in Figure \ref{fig:bf} \citep{He2016,Pandey2020}. In fact, \citet{He2016} designed four different types of body force, including stepwise and linear changes, all of which were able to reduce turbulence and cause the flow to relaminarize.

Applying non-uniform body force in a fully-developed wall-bounded turbulent flow can reproduce the fluid phenomena in the heated supercritical fluids. Various intensity of the non-uniform body force leads to flattened mean velocity profile or $\mathcal{M}$-shape velocity profile corresponding to partially relaminarize, fully relaminarize or even recover the turbulence respectively. 
This again confirms the dominant role of the buoyancy in the heat transfer of supercritical fluid.
Furthermore, \citet{He2016} introduced their new insights into the influence of non-uniform body force on turbulence and named with author's initials as "HHS". 
The idea of "HHS" is based on the body-force influenced flow behaves in the same way as the the equal pressure gradient flow and relaminarisation occurs when the apparent Reynolds number $Re_{\tau,\mathrm{HHS}}=u_{\tau,\mathrm{HHS}} R/\nu$ drops below a certain threshold where turbulence cannot be sustained any more. 
The new definition of the apparent friction velocity $u_{\tau,\mathrm{HHS}}=\sqrt{\tau_{wp}/\rho}$ relies on the wall friction $\tau_{wp}$ from the equal pressure gradient reference from the bulk instead of the measured $\tau_w$ on the wall.

\begin{figure}[hbt!]
\centering
\raisebox{0.5cm}{\includegraphics[width=0.49\textwidth]{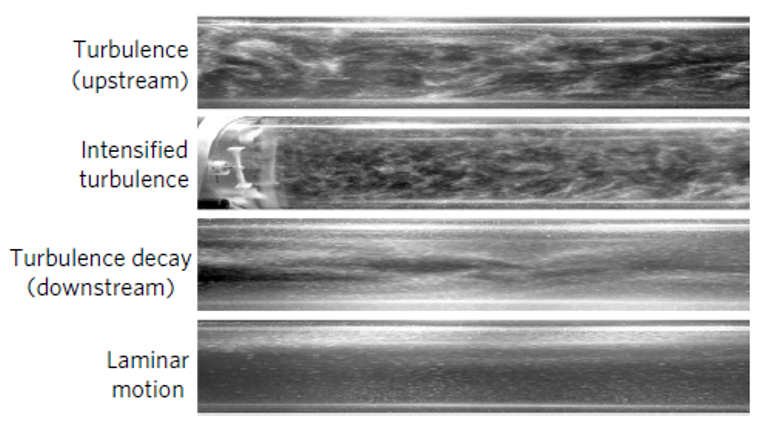}}
\includegraphics[width=0.46\textwidth]{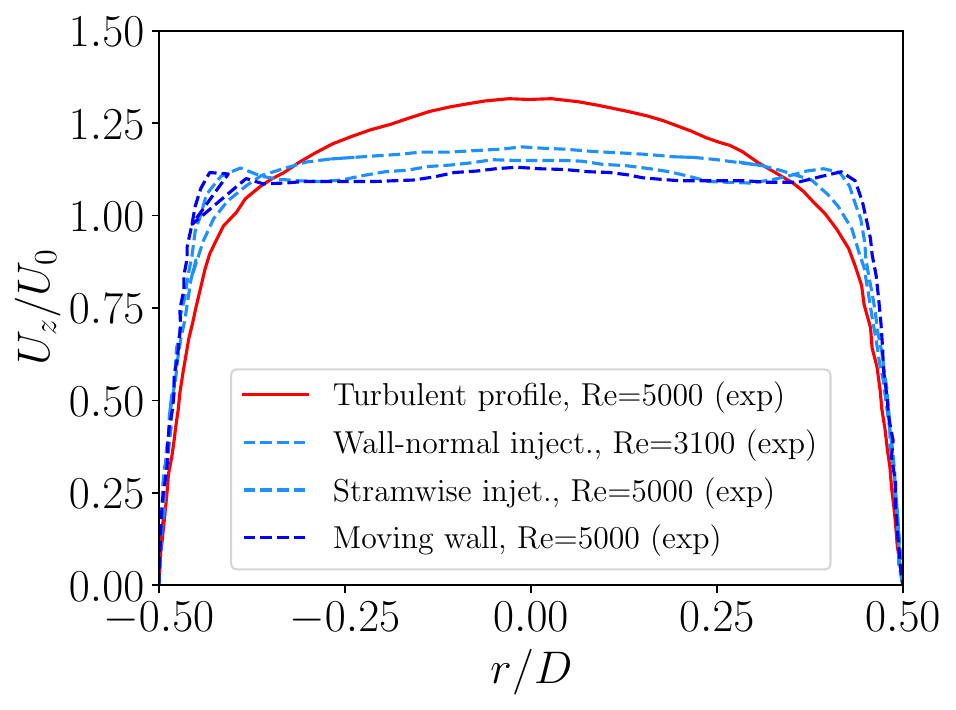}
\caption{Experimental measurement of relaminarization by the non-uniform body force \citet{Kuhnen2018b}, reproduced with permission given by Springer Nature}
\label{fig:naturephy}
\end{figure}

The interest to the relaminarized turbulent flow by the flattened velocity profile attracts broader interests than the supercritical fluids community. \citet{Kuhnen2018b} from Institute of Science and Technology Austria designed the pipe flow facility and the corresponding DNS to investigate the spatial development of the pipe turbulence under the influence of non-uniform body force. One of the approach to introduce non-uniform body force is the axial injection at a certain location. They observed a stabilization of the turbulent field when the velocity profile becomes flat (Figure \ref{fig:naturephy}). 

In wall-bounded turbulent flow, continuous energy transfer from the mean shear to smaller eddy motions is necessary to maintain turbulence. A critical aspect of this process is the interaction between streamwise vortices and streaks. The role of streamwise vortices is to elevate low-velocity fluid from the wall and transport it towards the center, a phenomenon referred to as the lift-up mechanism. This mechanism, which is rooted in the non-normality of the linear Navier–Stokes operator \citep{Brandt2014}, is a primary contributor to energy growth in shear flows. Its significance is often quantified by transient growth calculations.
In the study by \citet{Kuhnen2018b}, transient growth was computed for flattened flow profiles in DNS, revealing that transient growth monotonically declines as the velocity profile becomes flatter, reaching its minimum value just before turbulence collapses. In general, the flatter the velocity profile, the more the interaction between streaks and vortices is suppressed, thereby reducing the capability of the flow to sustain turbulence.


\section{Flow instability and laminar-turbulent transition}
\label{stability}
\begin{figure}[hbt!]
\centering
\includegraphics[width=0.95\textwidth]{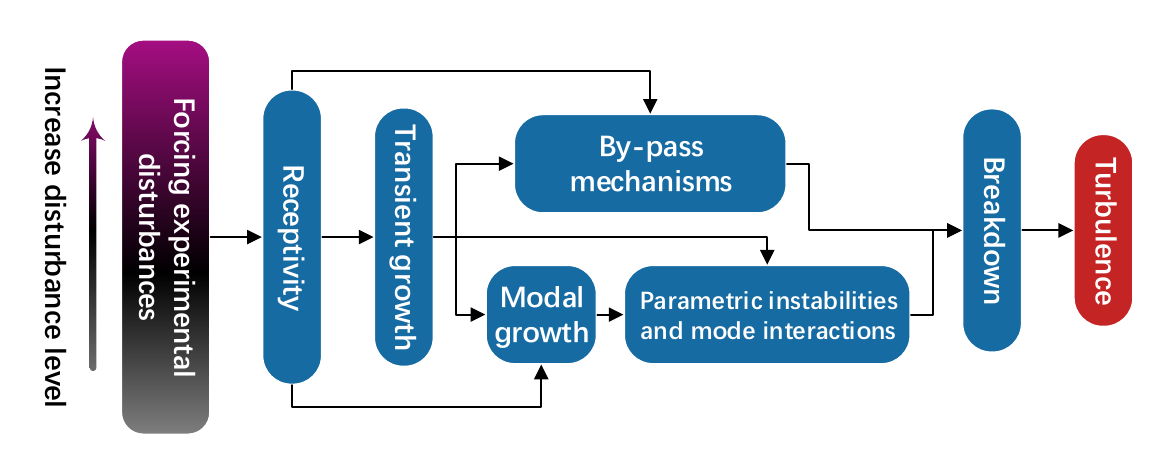}
\caption{The route leading to turbulence. Reproduced from \citep{morkovin1994transition} with permission given by American Physical Society.}
\label{fig:trans}
\end{figure}

Emerging applications involve flows that are not fully turbulent. In such considerations, one must take the laminar-turbulent transition into account. The flow stability analysis thus helps to understand the mechanisms and further build transition prediction models whose objective is to minimize uncertainties in designing purposes. A comparatively accepted transition route is shown in Figure \ref{fig:trans}. As can be seen, transition is associated with various mechanisms (receptivity, eigenmode growth, transient growth, nonlinear interactions and bypass mechanisms), whose route depends on environmental disturbance levels. Due to its high complexity and sensitivity to various parameters, flow instability and transition have remained an enduring research topic even in the framework of ideal-gas assumption. Readers are referred to contemporary reviews covering viscosity-stratification \citep{govindarajan2014instabilities}, bypass transition \& modeling \citep{jovanovic2021bypass},  nonlinear theories \citep{kerswell2018nonlinear,wu2019nonlinear}, chemical reactions \citep{de2020chemo} and hypersonic flows \citep{lee2019recent}. 

\subsection{Flow stability theory and its extension to non-ideal fluids}
Modern hydrodynamic stability theory commonly dates back to the setup of O. Reynolds in 1883 \citep{reynolds1883}, where laminar-turbulent transitions in a pipe flow were experimentally investigated. 
From a contemporary perspective, the initial step toward turbulence begins with the receptivity stage \citep{saric2002boundary}: the penetration of perturbations from the outside or border of the boundary layers. 
Once excited, perturbations in the boundary layer may experience transient growth mathematically due to the non-normality of the stability operator \citep{schmid2007nonmodal}. The more vigorous growth amounts to the eigenmode growth \citep{reed1996linear}, which usually causes a significant accumulation of the growth rate and disturbance energy, based on which the N-factor is obtained, enabling engineering transition prediction \citep{arnal1994boundary,fu2013rans}. Beyond the linear eigenmode or transient growth, the larger perturbation scales break down and build a typical turbulent energy cascade following nonlinear interaction or bypass mechanisms. In boundary layer flows of ideal gases, the unstable eigenmode (the Tollmien–Schlichting wave) is responsible for the transition in the subsonic or modest supersonic regimes. In the hypersonic regime, there are additionally multiple modes involved, commonly referred to as Mack's second mode. The reason is due to the metamorphosis of governing equation's nature as shown by \citet{Mack1969}. Mack's second mode plays a decisive role in the laminar-to-turbulent transition of an insulating boundary layer for Mach numbers greater than 4. For flows of higher Mach numbers, the real gas effects become influential, therefore, received increasing engagement since the 1990s \citep{Malik1991, Zhong2012}, addressing the physics of high-enthalpy flow encountered \citep{chen_xi_ren_fu_2022}, for example, during high-speed re-entry. However, for non-ideal gases such as supercritical fluid flow, the knowledge about the mechanism of laminar-turbulent transition is still minimal. Only until quite recently have some researchers started to reach this area and revealed  interesting new physics.

Naturally, the stability equations are derived by subtracting governing equations for the perturbed and laminar base flow (both satisfy the N-S equations). Note that the framework remains analogous to ideal fluids, but the coupling with thermodynamics shall be adequately accounted for. Without loss of generality, the stability equation reads
\begin{multline}\label{eq_stab}
\left(-i\omega\mathsfbi{L_t}+\mathsfbi{L_y}\mathsfbi{D}+i\beta\mathsfbi{L_z}+\mathsfbi{L_q}+\mathsfbi{V_{yy}}\mathsfbi{D}^{2}+i\beta\mathsfbi{V_{yz}}\mathsfbi{D}-\beta^{2}\mathsfbi{V_{zz}}\right)\hat{\boldsymbol{q}}\\ =\alpha\left(\beta\mathsfbi{V_{xz}}-i\mathsfbi{V_{xy}}\mathsfbi{D}-i\mathsfbi{L_x}\right)\hat{\boldsymbol{q}}+\alpha^{2}\mathsfbi{V_{xx}}\hat{\boldsymbol{q}},
\end{multline}
where the perturbation vector $\boldsymbol{q} = (\rho^\prime, u^\prime, v^\prime, w^\prime, T^\prime)^T$. In addition to the velocity perturbation, one may choose any two non-trivial thermodynamic variables (e.g., pressure, density, temperature, internal energy, to name a few)  depicting the thermodynamic state  to define the perturbation vector $\boldsymbol{q}$. In addition, the analytical equation-of-state and viscosity/thermal conductivity law may not be present. In such circumstances, it calls for implementing lookup tables based on a trusted database. According to the state postulate for a simple compressible system, the dependence of thermodynamic and transport properties (e.g., viscosity) and their perturbations are functions of two primary thermodynamic variables (e.g., density and temperature). As a result, the non-ideality of the stability problem is delivered in the stability matrix $\mathsfbi{L}$, $\mathsfbi{V}$ (see \citet{Ren2019a}) and nonlinear terms $\mathsfbi{N}$. 

\subsection{Results for canonical flows}
\begin{figure}[hbt!]
\centering
\includegraphics[width=1.0\textwidth]{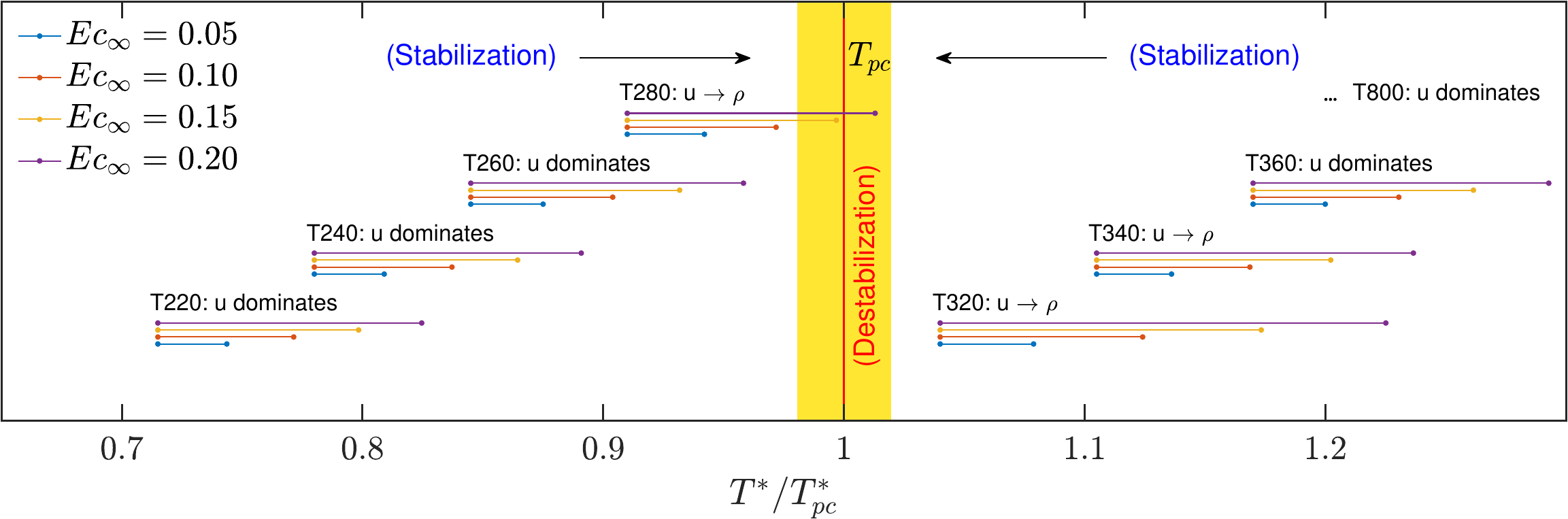}
\caption{Summary of stability analysis results from \citet{Ren2019b}. Horizontal line segments stand for the temperature range of the base flow for each case. The red vertical line shows the pseudo-critical temperature $T_{pc}=307.7$ K. For instance, the case name T220 denotes a freestrem temperature of $T_\infty=220$ K. Four line colors indicates four different Eckert number (viscous heating). Reproduced from \citep{Ren2019b} with permission given by Cambridge University Press.}
\label{fig:Ren1}
\end{figure}

In early 2019, the first knowledge of Poiseuille flows' linear stabilities with highly non-ideal fluids (supercritical CO$_2$ at 80 bar) was brought by \citet{Ren2019a}. A strong influence of fluid properties on the linear stability of these flows was demonstrated. The results show that the flow is more unstable in the subcritical (relative to the Widom line) regime compared to an ideal gas in the same conditions but less unstable in the supercritical regime. In the transcritical regime, the flow turns  inviscid unstable. Additionally, the authors showed that except for the transcritical regime, the linear stability analysis with simple cubic EoS (The van der Waals, Redlich–Kwong and Peng–Robinson) gives qualitatively similar results to using the more accurate multi-parameter EoS implemented in the REFPROP library. In the following study, \citet{Ren2019b} have extended these results for the first time to boundary layer flows near the pseudo-critical point. Depending on the free-stream temperature $T_\infty$ (with adiabatic wall) and the Eckert number $Ec_{\infty}=u_{\infty}^2/(c_{p,\infty}T_{\infty})$ that determines the viscous heating, the boundary-layer temperature profile was divided into sub-, trans- or supercritical regimes (Figure \ref{fig:Ren1}). The stability analysis showed that the non-ideal-gas exhibit a stabilization influence in the subcritical or supercritical regime, as shown in Figure \ref{fig:Ren1}.

\begin{figure}[hbt!]
\centering
\includegraphics[width=0.65\textwidth]{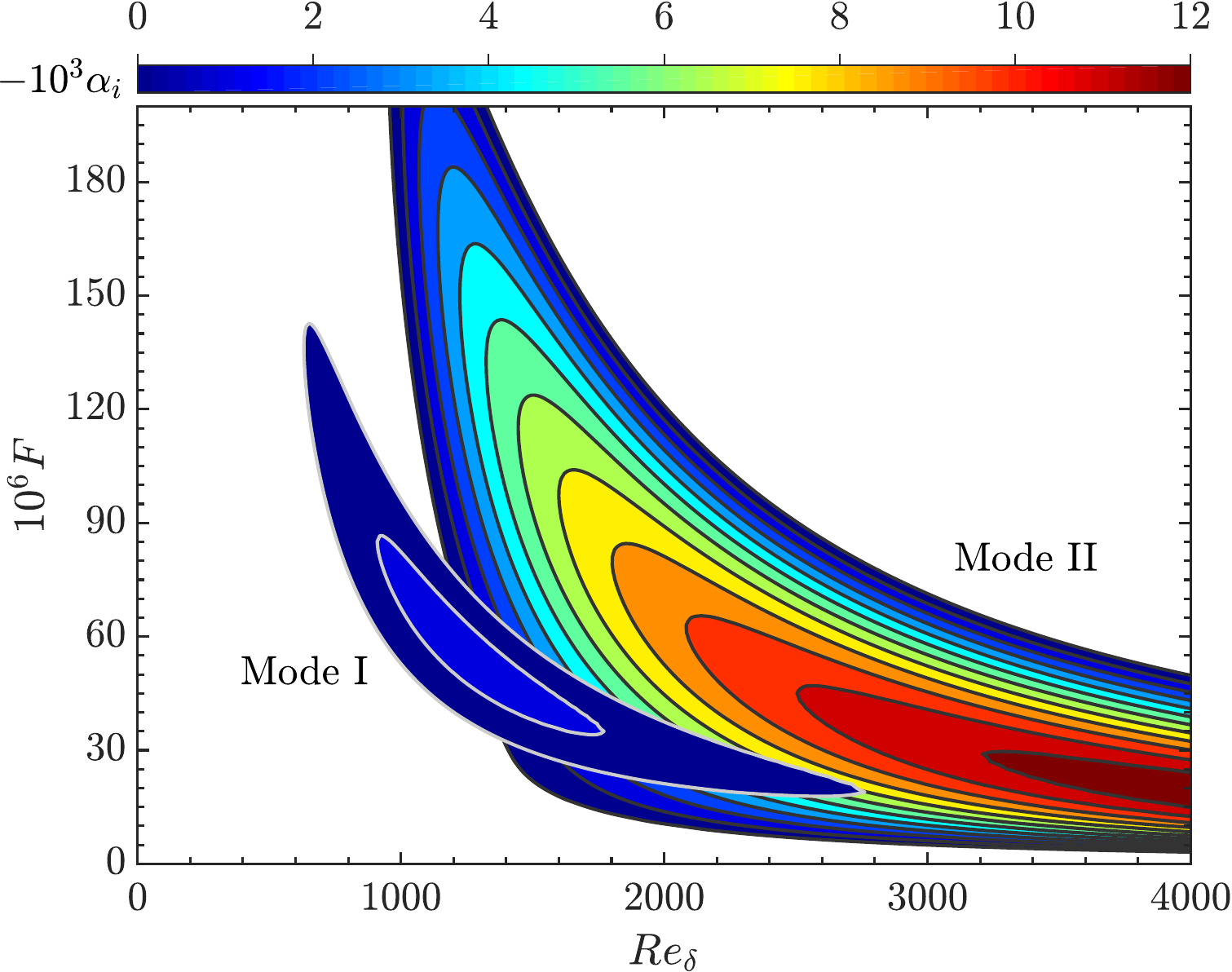}
\caption{The onset of Mode II in a boundary layer flow with transcritical CO$_2$ at $p=8$ MPa. Growth rates (in color) of perturbations in the $F–Re_\delta$ stability diagram with subcritical free-stream temperatures of $T_\infty^\star=280$ K, the horizontal axis is the boundary layer Reynolds number based on boundary  $Re_\delta=\rho_\infty u_\infty \delta/\mu_\infty$ with boundary layer thickness $\delta$. Reproduced from \citep{Ren2019b} with permission given by Cambridge University Press.}
\label{fig:Ren2}
\end{figure}

An exciting discovery was made in the transcritical regime. The flow is extremely destabilized due to the co-existence of dual unstable modes: Mode I, related to viscous Tollmien–Schlichting instability and Mode II, a new inviscid mode with a tremendous growth rate  (Figure \ref{fig:Ren2}). Mode II does not exist in the sub- and supercritical regimes. Only Mode I drives the instabilities: viscous and two-dimensional for the subcritical regime and inflectional and three-dimensional for the supercritical regime \citep{robinet2019instabilities}. In addition, \citet{Ren2019b} showed that Mode II is not connected to Mack's second mode \citep{Mack1969} since the Mach number relative to perturbations' phase velocity is not supersonic in the transcritical regime, which is a necessary condition to change the nature of the stability equation thus supporting acoustic modes. The work of \citet{Ren2019a,Ren2019b} presents intriguing perspectives on the study of transcritical fluids. In particular, the dual-mode instability exhibits dramatic instability and may act as a catalyst for the transition to turbulent flows.

More recently, \citet{ly2022destabilization} found that the coupling of Widom-line transition and binary mixing-layer instabilities (e.g., in fuel injection systems) leads to a destabilised novel thermodynamical instability that bears striking analogies with the dual-mode instability behaviour in boundary-layer flows \citep{Ren2019b}. An intuitive observation is a competition between a backward- and a forward-travelling eigenmode, where the former represents the thermodynamically induced instability. In addition to Widom line transition, the binary mixing flows gives contradictory results compared to supercritical single-species mixing layers when examining the influence of reduced pressure. The study provides essential theoretical foundations for applications with binary mixtures, commonly encountered in fuel injection systems.

In addition, \citet{Rinaldi2017} studied linear stability of buffer layer streaks in turbulent channels with synthetic properties: variable density and viscosity as in the DNS of \citet{Patel2015, Patel2016}. The mean velocity profile from this DNS is taken as the base flow for the linear stability analysis. The study aims to investigate the linear stability of buffer layer streaks under the condition of variable properties rather than laminar-turbulent transition. The semi-local scaling developed by  \citet{Patel2015, Patel2016} provides an effective parametrization of the effect of variable properties on optimal streaks and their critical mode of secondary instability calculated using the linear theory.

New knowledge also extends to three-dimensional boundary layer flows \citep{Ren2022CF,ren2022non}. Towards the goal of a cleaner sky, crossflow (CF) instability that rises on wing surfaces of aircraft has attracted persisting research forces since the 1990s. Understanding and controlling CF instability are essential to achieving laminar flow control and drag reduction. Physically, the instability is caused by a secondary inflectional flow profile perpendicular to the potential-flow direction, that balances the pressure gradient and centrifugal force inside the boundary layer. CF instability typically happens in e.g. a swept flow in the favourable pressure gradient region of a wing surface. Three dimensional boundary layers were investigated with ideal gas assumption due to its early focus on aeroplanes. Recent study indicates that a strong coupling between pseudo-boiling and the linear instability can completely change its nature of instability from cross-flow mode to inviscid T-S mode. 

\begin{figure}[hbt!]
\centering
\includegraphics[width=0.4\textwidth]{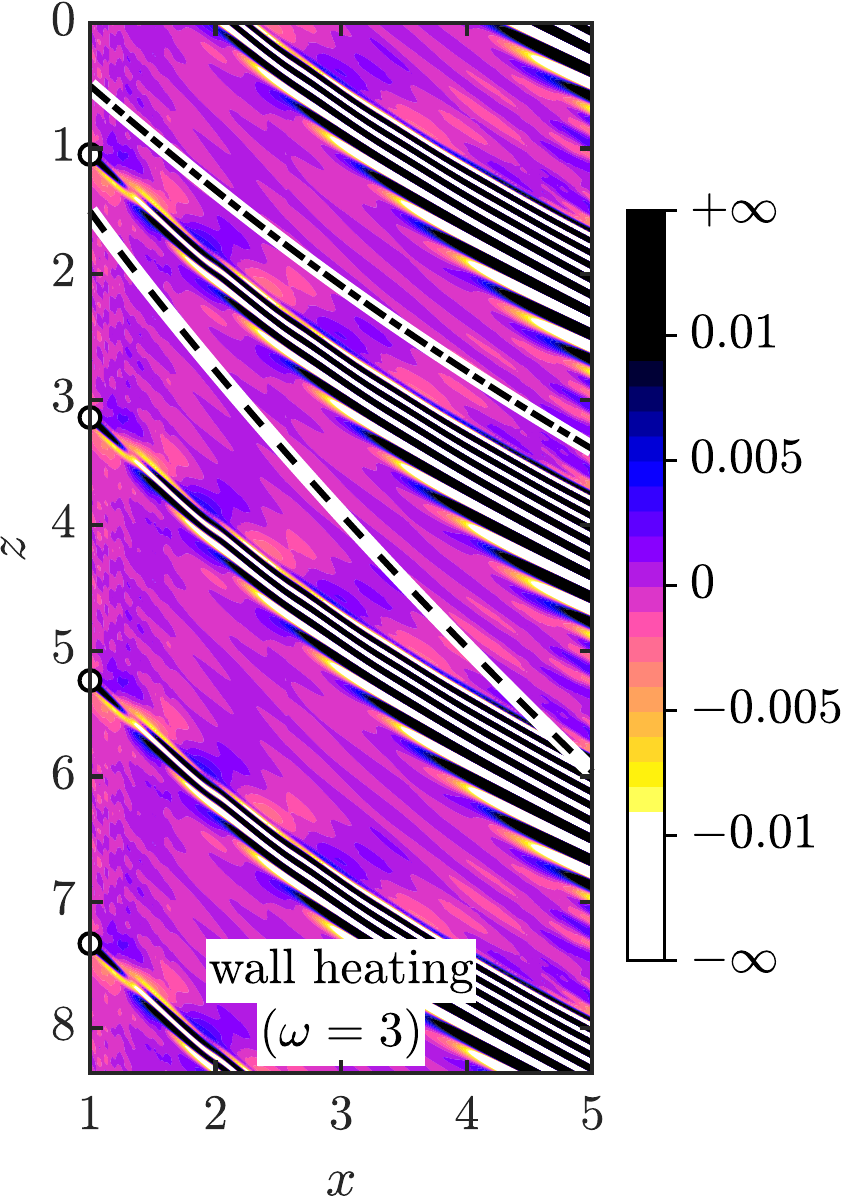}
\includegraphics[width=0.4\textwidth]{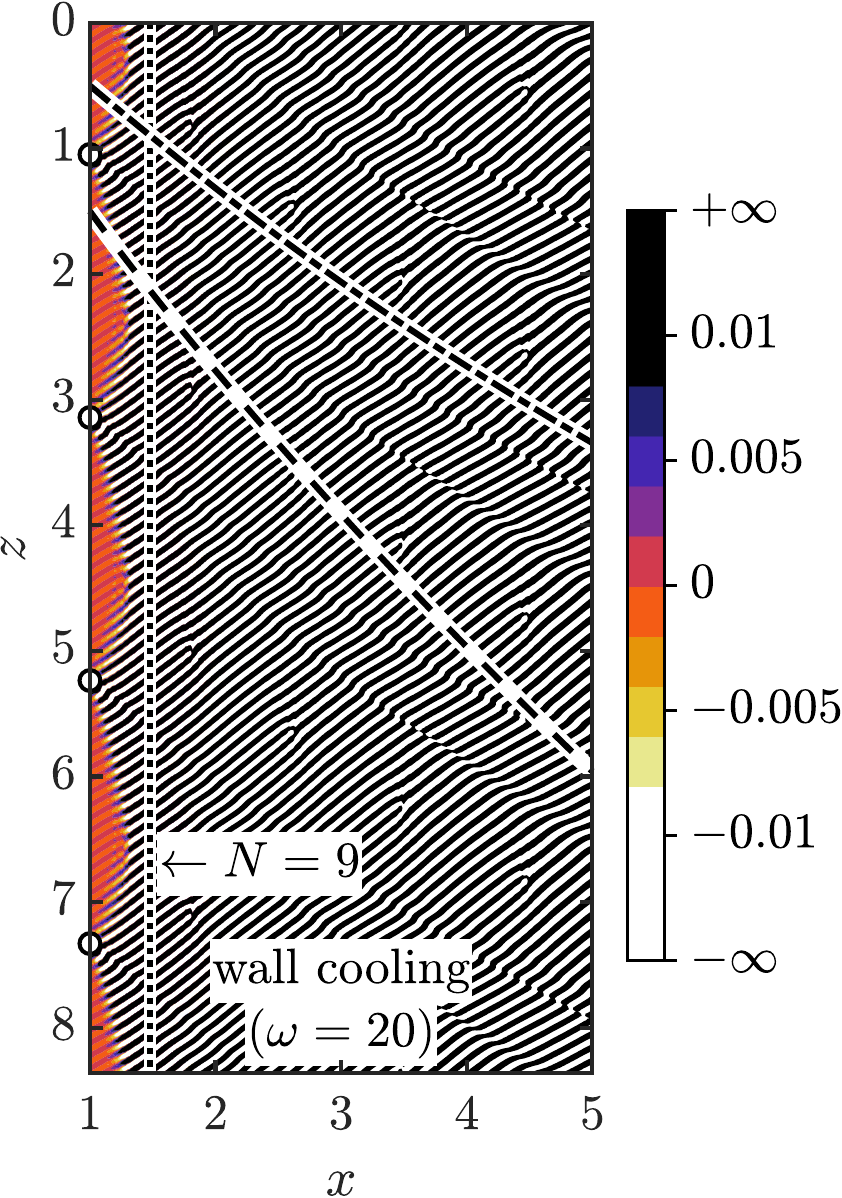}
\caption{Stability sceanario shown with physical perturbation $\left.\partial u^{\prime}/\partial y\right|_{y\rightarrow0}\left(x,z,t\right)$. The dashed and dash-dotted line correspond to the potental and wall streamlines.  The colormap displays perturbation whose amplitude is large than a threshold with black-and-white contours. The point sources (shown with blue circles) are located at $x=1$ and $z=$ 1.05, 3.14, 5.24 and 7.33. Reproduced from \citep{ren2022instabilities} with permission given by Cambridge University Press.}
\label{fig:CF2}
\end{figure}

As shown in figure~\ref{fig:CF2}, in the transcritical regime a changeover of the dominating instability mechanism from CF to inviscid TS type appears here for the wall cooling case. The maximum growth rate of the inviscid TS instability, raised by an inflection point in the streamwise-velocity profile, is one order of magnitude larger than the CF mode rate despite the flow acceleration. The strong instability will lead to a rapid flow transition and must be avoided if possibly long laminar flow is sought. The scenario resembles somewhat the situation with an ideal fluid with a CF boundary layer and sudden strong flow deceleration by an adverse pressure gradient, leading to flow separation. With the non-ideal fluid considered here, just a slight wall cooling in the transcritical regime does it.

The new physics observed in recent studies has spurred efforts to elucidate the underlying mechanisms. The work of \citet{bugeat2022instability} represents an  advancement in this domain, revealing that within stratified plane Couette flow, the presence of a minimum kinematic viscosity in the baseflow profile induces a generalized inflection point, satisfying Fjørtoft's generalized inviscid instability criterion.

\citet{boldini2024transient} investigated the non-modal growth of perturbations. By eliminating the pressure-work term, they defined a new energy norm under which the growth of perturbation energy is considered. In non-transcritical regimes, the optimal perturbations are identified as streamwise-elongated streaks resulting from the lift-up mechanism, similar to what is observed in an ideal gas. When the temperature becomes transcritical under wall heating, the optimal energy growth arises from an interplay between the lift-up and Orr mechanisms due to the additional presence of the inviscid Mode II. The most significant change occurs with wall cooling across the Widom line, where energy amplification is significantly enhanced by streamwise-independent optimal disturbance structures.

From the above research, it is evident that new instability mechanisms emerge when the temperature crosses the Widom line, particularly with wall cooling (if a wall is present). Significant portions of flow theory, such as linear modal/non-modal and inviscid theories, have been extended to non-ideal fluids. The results obtained thus far primarily pertain to canonical flows and remain within the linear framework. Future work is necessary to explore the nonlinear regime, model the transition, and apply these findings to complex geometries. 

\section{Modeling of turbulence and heat transfer and the involvement of data-driven approaches}
\label{model}

\subsection{Progress in RANS modeling}

Even though computational power has been increasing for a long time, making DNS more accessible, turbulence modeling remains the most suitable approach for engineering applications. However, there is a common consensus within the community that the existing Reynolds-averaged Navier–Stokes equations (RANS) models are not reliable in predicting the heat transfer of supercritical fluids \citep{He2008, Yoo2013,rohde2016blind,pecnik2021characterizing}. 
This is certainly not a surprise, considering the observable discrepancy published in simulations without any models (DNS) among different groups. Even in fluids with constant thermo-physical properties, standard RANS models tend to fail under conditions of transition and flow separation. The emergence of flow relaminarization and transition poses further challenges, in addition to dealing with complex supercritical properties.
In 2008, \citet{He2008} conducted a comprehensive assessment of the performance of RANS models, utilizing the very first DNS data from \citet{Bae2005}. They concluded that none of the tested models were capable of accurately predicting the results derived from DNS. This conclusion was highlighted in a well-known review by \citet{Yoo2013}, published approximately a decade ago. Since that time, there has been a recognition of the need for more DNS data to better understand the flow physics and to support improvements in modeling. 
In 2016, \citet{rohde2016blind} organized a blind numerical benchmark focusing on supercritical water heat transfer experiments in a 7-Rod Bundle. This initiative, undertaken by the Gen-IV International Forum, aimed to demonstrate the predictive capability of the currently available models in real-life applications. The findings concluded that none of the numerical simulations were able to accurately predict the wall temperature in the test case where a deterioration of heat transfer was observed.

Modeling the heat transfer of supercritical fluids continues to be a challenging task, yet some progress has been made recently. \citet{Bae2016} modeled the turbulent Prandtl number, $Pr_t$, as a function of thermo-physical properties. Utilizing the mixing length theory, he derived the Reynolds stress and turbulent heat flux, taking into account density variation, which led to a modified $Pr_t$. 
In subsequent research, \citet{Bae2016b} explained that during the relaminarization-transition stage, the effective viscous sublayer thickness is not constant but varies as a function of buoyancy effects. Consequently, a new function was derived for the viscous sublayer thickness, taking into account the deformation of the turbulent boundary layer. \citet{Tang2016} employed a similar approach aimed at developing an alternative expression for the turbulent Prandtl number, $Pr_t$. However, the modeling of $Pr_t$ tends more towards empirical methods rather than physical ones. 

An alternative method to enhance modeling performance involves advanced modeling of turbulent heat flux. Many studies have utilized the Algebraic Heat Flux Model (AHFM) proposed by \citet{launder1988computation}, which serves as a replacement for both the Simple Gradient Diffusion Hypothesis (SGDH) and the Generalized Gradient Diffusion Hypothesis (GGDH) models. The AHFM is derived from a detailed analysis of the exact transport equation for turbulent heat flux, $\overline{u^\prime_i T^\prime}$:
\begin{equation}
    \overline{u^\prime_i T^\prime}= - C_t \frac{k}{\varepsilon} \left[C_{t1} \overline{u^\prime_i u^\prime_j}\frac{\partial T}{\partial x_j}+(1-C_{t2})\overline{u^\prime_j T^\prime}\frac{\partial \overline{u}_i}{\partial x_j}+(1-C_{t3})\beta g_i \overline{T^{\prime 2}}\right]
\end{equation}
This model simplifies the balance equation of turbulent heat flux into a steady-state algebraic form, thereby obviating the need for a turbulent Prandtl number. This approach necessitates the definition of four coefficients and requires the computation of the temperature variance distribution. Consequently, it increases computational demands by requiring two additional partial differential equations to calculate this variable and its dissipation rate.

The first term on the right-hand side is identical to that in the GGDH model, whereas the second term adjusts the ratio of the flux in different directions, and the third term models the buoyancy effect. The four constant parameters are defined differently by various authors. Several studies have confirmed the enhanced performance offered by the AHFM. \citet{Zhang2012} demonstrated that the AHFM model is superior in calculating the buoyancy term compared to the SGDH and GGDH models, which consistently underestimate this term. \citet{Xiong2014} reported even further improvements with the use of the Elliptic Blending-Algebraic Flux Model.
\citet{pucciarelli2016prediction} implemented the AHFM in commercial code and found that the calculated wall temperature values are generally well reproduced. \citet{pucciarelli2018use} further tested the AHFM across a wide range of experimental cases involving supercritical carbon dioxide, with operating conditions that included both high and low mass flux values and ranged from relatively low inlet temperatures to values exceeding the pseudocritical threshold. The model successfully replicated several intriguing aspects observed in the experimental data, accurately forecasting the trends of wall temperature both qualitatively and quantitatively.

In addition to the studies based on AHFM, \citet{Li2020} developed a novel non-linear eddy viscosity turbulence model, deriving from the AKN $k-\varepsilon$ low Reynolds-number model. Unlike the linear Boussinesq approximation for eddy viscosity, this model represents the Reynolds stress tensor as a quadratic function of a set of integrity bases derived from the strain rate tensor and the rotation tensor \citep{Pope2000}. Furthermore, this non-linear eddy viscosity model incorporates a new pressure strain model, which includes both rapid and slow components, aligning with the principles of Reynolds stress transport modeling to account for the non-equilibrium effects of Reynolds stress transport. By integrating the pressure strain model with the non-linear Reynolds stress constitutive equation, modified values for the eddy viscosity, $\mu_t$, and the turbulent Prandtl number, $Pr_t$, are derived. Ultimately, the modified model was tested across 10 experimental and DNS data cases, demonstrating significant improvements over the baseline model in terms of wall temperature predictions and other turbulent statistics.

\citet{Pecnik2017} made a fundamental and systematic contribution to the modeling. They utilized the key finding from their DNS \citep{Patel2015,Patel2016}: a semi-local scaling equation system for turbulence modeling. They derived an alternative form of the TKE equation based on semi-local scaling and demonstrated that if a turbulence model is solved in the semi-local scaling form, instead of its conventional form, there is an excellent agreement with the DNS data \citep{Patel2015,Patel2016}. In the subsequent study, \citet{Patel2018} tested the semi-local scaling approach on five different turbulence models: the eddy viscosity correlation of Cess, the one-equation model of Spalart-Allmaras (SA), the $k-\varepsilon$ model of Myong and Kasagi (MK), Menter’s shear stress transport model (SST), and the four-equation $v^2-f$ model (V2F). The agreement with results obtained by DNS shows that the modification significantly improves the results of eddy viscosity models for fluids with variable transport properties. It should be noted, however, that these validations were based on channel flow with synthetic thermo-physical properties, avoiding the complexities of deteriorated heat transfer, which is less challenging than real properties. 

\begin{sidewaystable}
\centering
\begin{tabular}{|c|c|c|c|}
\hline
Authors & Model type & Major improvements  \\
\hline
\citet{Bae2016b} &  Low-Re Myong–Kasagi k-$\varepsilon$ model  & Variable turbulent Prandtl number \\
\citet{Li2020}  & non-linear eddy viscosity model & modeling of rapid and slow pressure terms  \\
\citet{Patel2018}  & various models & Based on new semi-local scaling \\
\citet{pucciarelli2018use}& Lien Low-Re k-$\varepsilon$ model & Algebraic Heat Flux Model \\
\citet{cao2023physics} & Low-Re k-$\varepsilon$ model & data-driven approach \\

\hline
\end{tabular}
\caption{A summary of recent modifications of turbulence models}
\label{tab:4}
\end{sidewaystable}

In recent years, machine learning has been successfully applied in the fields of fluid mechanics and heat transfer, thanks to its inherent ability to learn from complex data \citep{Duraisamy2019,brunton2020machine,chung2022interpretable,chu2020turbulence}. This approach can be customized for various applications to optimize computational resource usage and minimize the risk of overfitting. Currently, different types of neural networks have led to state-of-the-art accuracy in challenging domains such as computer vision and natural language processing. Over the past decade, a key factor behind the success of deep learning has been the ability of neural networks to represent general non-linear functions, while providing multiple alternatives for fine-tuning their design. Notable works in applying deep learning to CFD include the study of \citet{Ling2016}, who developed deep neural networks to model turbulence with embedded Galilean invariance. Despite the wealth of recent research, substantial investigation is still required regarding the implementation of machine learning in the realm of CFD. Furthermore, there is a need for the development of rich datasets to explore turbulence under complex conditions. 
\citet{sanhueza2023machine} introduced a field inversion machine learning (FIML) approach \citep{parish2016paradigm} to improve predictions of RANS turbulence models in channel flows with varying thermophysical properties. The approach includes efficient optimization routines for field inversion within CFD and a unique neural network architecture with logarithmic and hyperbolic tangent neurons for numerical stability. A weighted relaxation factor methodology refines predictions, and L2 regularization minimizes overfitting. Utilizing K-fold cross-validation, the model significantly reduced the L-infinity modeling error on the velocity profile from 23.4\% to 4.0\% in successful cases. The study concludes that this method offers a valid alternative for enhancing RANS models without introducing prior assumptions.
\citet{cao2023physics} proposed an innovative approach to modeling through a combination of direct turbulent production models and indirect transport closures utilizing deep neural networks (DNN). An iterative framework involving DNS-DNN-RANS is introduced, incorporating prior physics knowledge and feature engineering strategies. By employing machine learning algorithms to modify the low Reynolds number $k-\varepsilon$ model, the study successfully validates the new model against DNS and experimental data for upward pipe flows.

\subsection{LES approach}

Due to the low fidelity of RANS, LES is considered as the future turbulence modeling approach \citep{slotnick2014cfd}, though at the expense of increased computational demand. Specifically, wall-modeled LES, combined with modern heterogeneous HPC systems, is becoming increasingly viable for industrial applications, such as aerodynamic design \citep{goc2021large}. 
For a considerable period, LES has been regarded as having the potential to supplant RANS in industrial applications, contingent upon the continued growth of computational power. LES strikes a balance between accuracy and computational cost by not resolving the smallest scales of turbulence, thereby offering a practical compromise for complex flow simulations.
In our domain of application, efforts have been made to leverage LES alongside the widespread implementation of RANS modeling. \citet{Niceno2013} pioneered the use of LES for turbulent heat transfer in water at supercritical pressures, employing a standard Smagorinsky sub-grid model. In both upward and downward flows, the wall temperatures demonstrated qualitative agreement with experimental measurements.
\citet{Pucciarelli2018} conducted LES analyses incorporating conjugate heat transfer. The commercial solver STAR-CCM+ \citep{Cd2012}, in conjunction with the Wall-Adapting Local Eddy-viscosity (WALE) subgrid model \citep{Nicoud1999}, was employed. The analyses revealed a more pronounced heat transfer deterioration phenomenon in the case that accounted for conjugate heat transfer.
\citet{Nabil2019} conducted a LES on supercritical CO$_2$  convection within microchannels, focusing on nonuniform heat fluxes. The investigation employed OpenFOAM \citep{Jasak2007} as the solver, with the WALE subgrid model being utilized.
The geometry of microchannels closely mirrors that of a typical heat exchanger, thereby underscoring the importance of utilizing high-fidelity LES in this context. Such application of LES offers a compelling demonstration of its utility in practical engineering scenarios.
In subsequent research, \citet{Alkandari2022} employed the same numerical method to simulate an impinging jet of supercritical CO$_2$. They found that heat transfer deterioration was most pronounced in the stagnation zone, with its relative magnitude decreasing more sharply downstream in turbulent flows than in laminar flows. Intriguingly, an enhancement in heat transfer was observed when the temperatures at both the jet inlet and the impingement plate were situated at opposite ends of the pseudo-critical temperature range.
\citet{Xie2022} assessed heat transfer deterioration to supercritical CO$_2$ via LES. The wall temperature in the LES shows excellent consistency with the DNS data.
\citet{Wang2021b} introduced their LES solver framework with the WALE subgrid model and used it to simulate a vertical pipe with supercritical water flow. Later, they \citep{Wang2022} used this numerical method on a square subchannel.
\citet{indelicato2023dataset} presented wall-resolved LES of cryogenic hydrogen at supercritical pressure with varying wall heat fluxes. 
This dataset is poised to serve as a pivotal reference for the development of wall functions under trans- and supercritical conditions, which are prevalent in liquid rocket engine applications. The study encompasses a comprehensive validation and grid-convergence analysis, alongside an in-depth examination of axial, radial, and azimuthal refinements. Additionally, it conducts a parametric analysis with a focus on turbulent pseudoboiling. Notable outcomes highlight a significant increase in wall temperature correlating with heat flux, a phenomenon attributed to a pseudochange in the core flow phase. Furthermore, the research provides valuable insights into the interplay between turbulent pseudoboiling and near-wall gradients.
Moreover, \citet{indelicato2022assessment} explored the application of an algebraic equilibrium wall function to supercritical conditions, highlighting its potential to significantly reduce computational costs compared to wall-resolved LES. The research primarily assesses the efficacy of existing algebraic wall functions through a systematic analysis, leveraging a wall-resolved LES database \citep{indelicato2023dataset} of cryogenic para-hydrogen flow in a heated pipe under supercritical pressure. The findings indicate that the model tends to overestimate wall temperatures and slightly underestimate skin friction velocity as heat flux increases. The investigation delves into the underlying causes of these discrepancies, scrutinizing the equilibrium boundary layer hypothesis, the applicability of the Van Driest transformation for stratified supercritical flows, and the implications of the ideal-gas assumption in the original model formulation. To enhance the model's adaptability across different Equations of State (EOS), a thermodynamic correction is proposed. Although this adjustment leads to improved predictions, notable deviations persist. These discrepancies are primarily attributed to the equilibrium assumption and the limitations of the Van Driest transformation, particularly under conditions of increasing stratification, which introduce errors that vary by specific percentages depending on the case. Further analysis into recent transformations for variable property flows also uncovers similar constraints.


Compared to the situation a decade ago \citep{Yoo2013}, notable progress has been achieved in enhancing RANS and LES modeling. At that time \citep{Yoo2013}, the limitations of existing RANS models were well-recognized, yet the path to improvement remained uncertain. Since then, the adoption of AFHM \citep{pucciarelli2018use} and the non-linear eddy viscosity model \citep{Li2020} has demonstrated significant benefits over preceding models. Furthermore, the integration of physics-informed machine learning presents promising opportunities for future advancements. Despite these innovations, a universally accepted model that achieves comprehensive success across all applications has yet to emerge.
An updated blind test between various models such as the one from \citet{rohde2016blind} could be extremely helpful.
Moreover, the validation of these RANS models is often based on experimental or DNS data from vertical, circular, and straight pipes, 
not reflecting their application to more complex engineering geometries. 
As the cost of wall-resolved/modelled LES becomes increasingly manageable, its capacity for scale-resolving simulations presents substantial potential for scenarios that require high precision. However, the impact of subgrid scale modeling on simulation outcomes remains an area requiring clearer understanding. Establishing a benchmark for these models could offer crucial insights, guiding future research endeavors. The significant decrease in LES computational expenses opens up new possibilities for its application to more intricate geometries or segments of heat exchangers. Drawing on the findings of \citet{Nabil2019}, it is projected that simulating a section of a heat exchanger with LES would necessitate between $\mathcal{O}(10^8)$ and $\mathcal{O}(10^9)$ mesh elements, a requirement that is well within the reach of current HPC resources. This advancement signals a promising direction for the application of simulation methods in addressing complex engineering challenges.

\subsection{Machine learning based heat transfer model}
\label{ML}


Despite a long history of developing empirical correlations or physics-based models, there is still no satisfactory approach to addressing the growing complexity of new heat transfer correlations/models \citep{jackson2017models,pandey2017modified}, which is prone to the issue of overfitting. 
Inspired by recent advancements in artificial intelligence across multiple disciplines, a data-driven approach utilizing neural networks has been proposed to swiftly predict heat transfer characteristics. This innovative methodology harnesses the power of machine learning, enabling efficient and precise analyses. Such an approach underscores the burgeoning trend of incorporating AI technologies into engineering practices, reflecting a shift towards more intelligent and adaptable solutions in the field.

\begin{figure}[ht!]
\centering
\begin{tikzpicture}[node distance=2cm]
\foreach \i in {1,...,5} {
    \node[fill=yellow!20, draw=black!60, circle] (input\i) at (0,-\i/2) {};
}
\foreach \i in {1,...,10} {
    \node[fill=blue!15, draw=blue!60, circle] (hidden1\i) at (2,-\i/2+1.25) {};
}
\foreach \i in {1,...,10} {
    \node[fill=blue!15, draw=blue!60, circle] (hidden2\i) at (4,-\i/2+1.25) {};
}
\foreach \i in {1} {
    \node[draw=red!60, fill=red!10, circle] (output\i) at (6,-3/2) {};
}
\foreach \i in {1,...,5} {
    \foreach \j in {1,...,10} {
        \draw (input\i) -- (hidden1\j);
    }
}
\foreach \i in {1,...,10} {
    \foreach \j in {1,...,10} {
        \draw (hidden1\i) -- (hidden2\j);
    }
}
\foreach \i in {1,...,10} {
    \foreach \j in {1} {
        \draw (hidden2\i) -- (output\j);
    }
}
\node[above of=input1, node distance=2cm] {Input};

\node[above of=output1, node distance=3cm] {Output};
\node[above of=hidden11, node distance=0.75cm] {Hidden layer};
\node[above of=hidden21, node distance=0.75cm] {Hidden layer};
\node[left of=input1, node distance=1cm] {$p$};
\node[left of=input2, node distance=1cm] {$G$};
\node[left of=input3, node distance=1cm] {$q$};
\node[left of=input4, node distance=1cm] {$d_{in}$};
\node[left of=input5, node distance=1cm] {$h_b$};
\node[right of=output1, node distance=1cm] {$T_w$};

\end{tikzpicture}
\caption{A three-layer multi-layer-perceptron (MLP) neural network structure used by \citet{ANN_sun2021}. Reproduced with permission given by Elsevier}
\label{sun2021NN}
\end{figure}
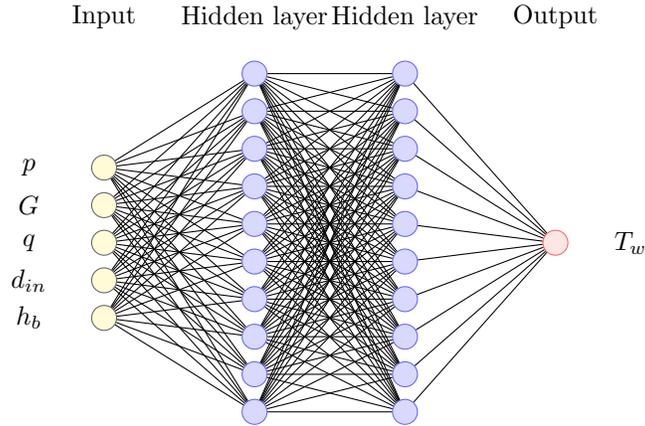

\begin{figure}[hbt!]
\centering
\includegraphics[width=0.48\textwidth]{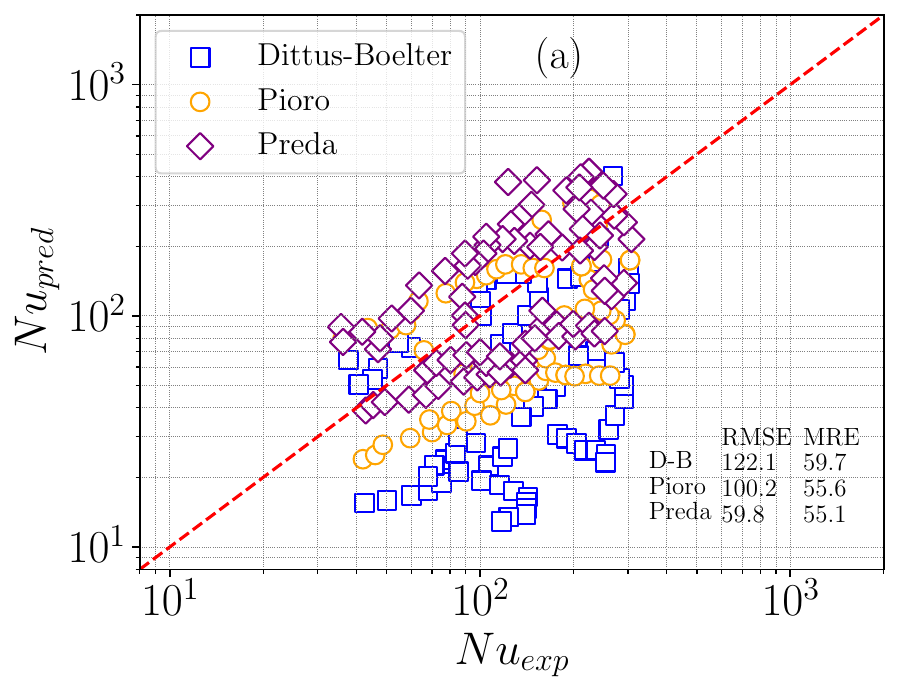}
\includegraphics[width=0.48\textwidth]{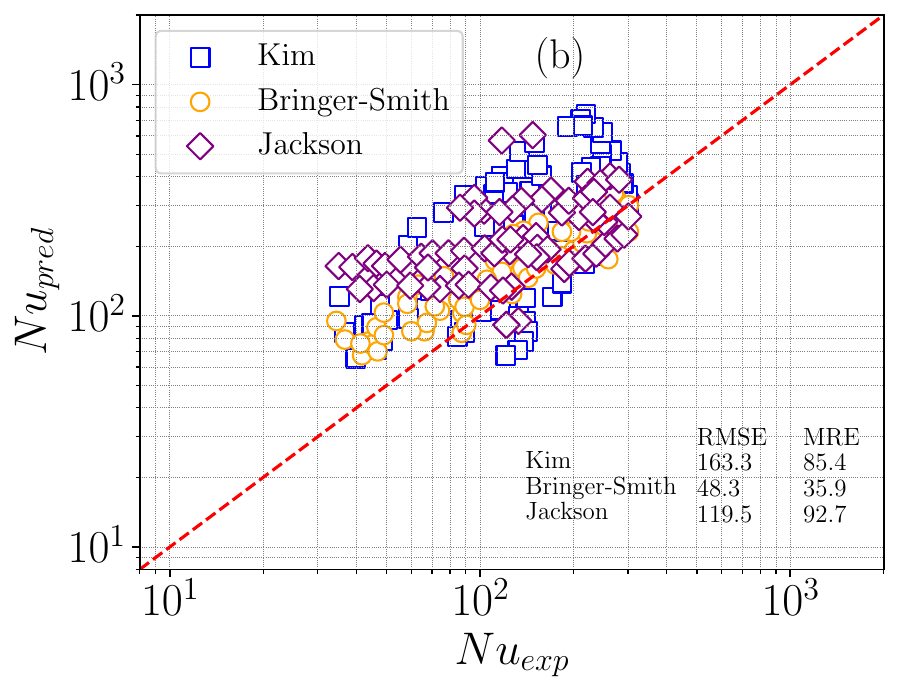}
\includegraphics[width=0.48\textwidth]{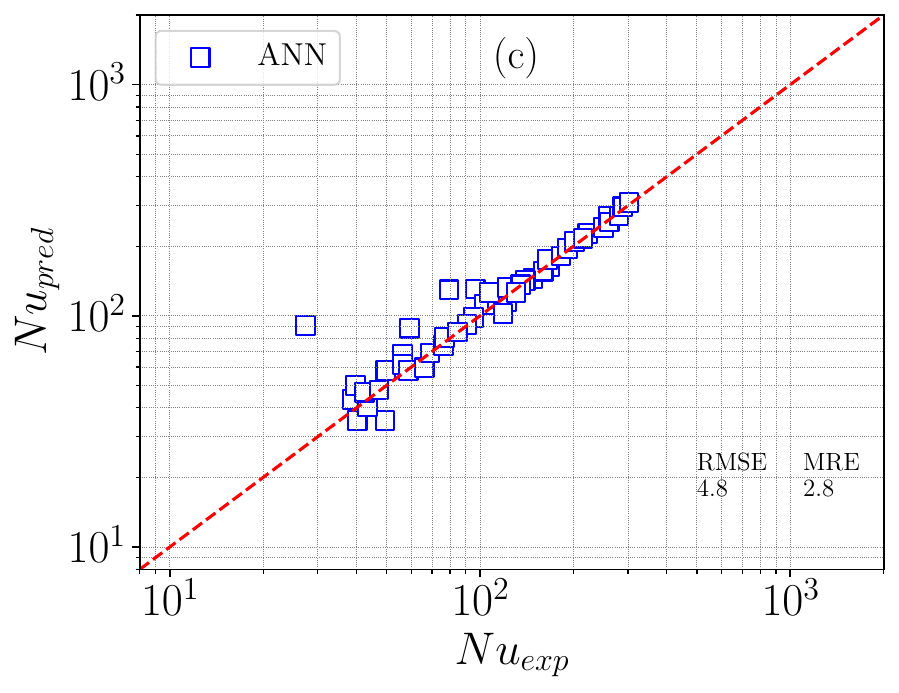}
\caption{Nusselt number predictions compared with correlations (Dittus-Boelter, Pioro, Kim, Jackson, Preda, Bringer-Smith), reproduced from \citep{ANN_sun2021} with permission given by Elsevier}
\label{sun2021}
\end{figure}

\begin{figure}[hbt!]
\centering
\includegraphics[width=0.7\textwidth]{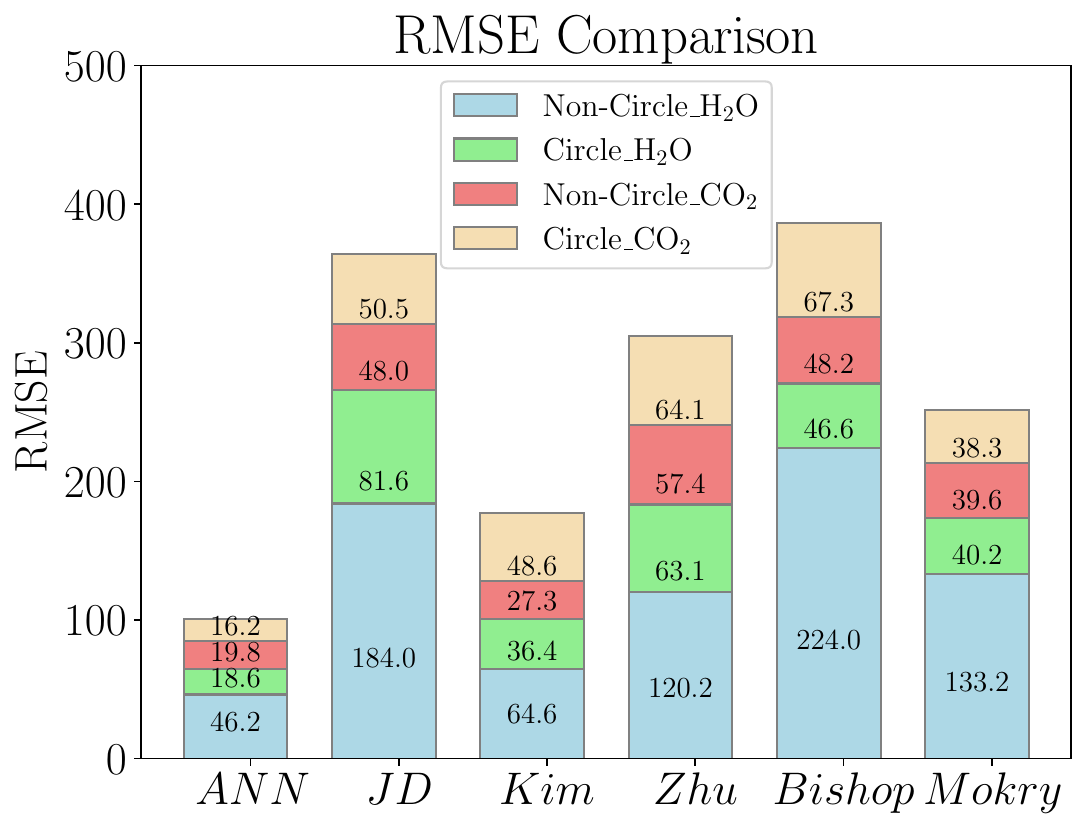}
\caption{Root mean square error (RMSE) of Nusselt number predictions of circle and non-circle channels. Fluids are H$_2$O and CO$_2$. Reproduced from \citep{zhang2023ann} with permission given by Elsevier}
\label{zhang2023fig}
\end{figure}

\citet{ANN_chang2018} introduced a shallow artificial neural network (ANN) model, trained using experimental data, to predict heat transfer behavior in supercritical water. The model inputs include mass flux, heat flux, pressure, tube diameter, and bulk specific enthalpy, with the tube wall temperature as the output. The training dataset comprised 5280 data points, all derived from published experimental findings. A rigorous validation process, distinct from the training phase, demonstrated mean error percentages and standard deviations below 0.5\%. In addition, the ANN's performance was evaluated under operational conditions not covered by the training or validation datasets, comparing its predictions against those from four well-established correlations. The results indicated a significant improvement in prediction accuracy by the ANN. Remarkably, the ANN training process requires less than an hour on a standard computing setup, and it can generate predictions within milliseconds.
Subsequently, \citet{ANN_sun2021} created a forecasting model utilizing ANN  (Figure \ref{sun2021NN}) that was trained using 5780 sets of experimental wall temperature data collected from upward flows. The model yielded a remarkably low root mean square error (RMSE) and mean relative error, demonstrating its superior predictive capabilities compared to conventional empirical correlations 
(Figure \ref{sun2021}).
The superiority of using a data-driven approach such as ANN has been demonstrated by independent research groups \citep{ANN_waxenegger2020,ye2019modeling, zhang2023ann, ANN_chu2018,scalabrin2003analysis,zhu2022study}.
In a landmark development, \citet{zhang2023ann} have developed an advanced ANN model that is capable of simultaneously predicting both the wall temperature and the Nusselt number for supercritical H$_2$O and CO$_2$ fluids (Figure \ref{zhang2023fig}). This model represents a pioneering achievement as it is the first to offer predictive capabilities for multiple fluids. Remarkably, although the model was initially trained on data from circular tubes, it demonstrated exceptional accuracy when applied to data from non-circular tubes. This indicates a significant level of generalization, suggesting that the model's predictive accuracy may not be limited by the geometry of the channel.
Recently, \citet{shi2024prediction} introduced a transfer learning model that leverages multi-fidelity data to enhance prediction accuracy and generalization capabilities. Their findings underscore the significant impact of the training dataset's data distribution on both the accuracy and the generalization ability of deep learning models. Notably, the transfer learning model, benefiting from the assimilation of prior knowledge during pre-training, exhibited superior performance on both high-fidelity and validation datasets. This achievement demonstrates that the generalization ability of the transfer learning model significantly surpasses that of models trained exclusively on sparse high-fidelity data.
Departing from conventional 1-dimensional (streamwise direction) data-driven models, \citet{shi2023rapid} have innovated a rapid prediction model that captures both streamwise and circumferential dimensions for supercritical CO$_2$ heat transfer in inclined tubes. This deep learning-based model is capable of predicting the 2D temperature distribution for such cases, achieving a maximum absolute relative error of 16.60\%. The introduction of a 2-dimensional perspective is particularly significant, addressing the frequent occurrence of circumferential non-uniformity in heat exchangers and representing a valuable advancement in the field.

The research mentioned above demonstrates the remarkable superiority of machine-learning algorithms over traditional heat transfer correlations, with significant potential for accurately predicting results despite unclear underlying mathematical equations. 
However, this emerging field still presents unanswered questions and challenges that necessitate further investigation and exploration.

Generalizability and Blind Tests:
One of the paramount advantages of ANNs lies in their capacity for generalization, which enables them to make accurate predictions for data points not encountered during training. To validate the robustness of an ANN, employing blind tests is essential. These tests involve comparing the model's predictions against actual outcomes, thereby providing a measure of the model's accuracy. Conducting such assessments ensures that the ANN can reliably generalize beyond the specific instances it has learned from, affirming its utility in practical applications.

Various fluid types:
The investigation into heat transfer mechanisms within supercritical fluids extends beyond merely H$_2$O and CO$_2$, encompassing a variety of other fluids such as helium, hydrogen, and nitrogen. Preliminary findings suggest the feasibility of employing ANN to predict heat transfer behaviors across these diverse fluids as well \citep{zhang2023ann}. To ensure the ANNs' capacity for accurate generalization, it is critical that the training datasets comprehensively encapsulate a wide spectrum of fluid properties, including specific heat, pressure, and temperature. Such diversity in the training data is imperative for the ANN to achieve precise and reliable generalizations across different fluid types.

Complex geometry:
ANNs hold the potential to accurately predict heat transfer phenomena in complex geometrical configurations, such as horizontal flows, rod bundles, and non-circular pipes. Nevertheless, it necessitates a comprehensive and diverse dataset that encompasses a wide range of geometries, flow rates, and other pertinent variables. Additionally, selecting the input features for the ANN, such as surface roughness, Prandtl number, and Reynolds number, is crucial to ensure precise predictions. This aspect is undeniably relevant to the generalizability of the trained models.

Small Data Training:
In contexts where data availability is constrained, several strategies can be employed to train an ANN efficiently. Transfer learning is a notable technique in this regard, which involves refining a pre-trained neural network on a smaller dataset to adapt it to specific needs. To enhance generalization and prevent overfitting, regularization techniques such as dropout and weight decay can be utilized. Moreover, it might become necessary to enhance the available data by creating artificial data points using methods like data interpolation or data augmentation.
  
In summary, ANN is a potent instrument that can predict heat transfer in supercritical fluids in various situations. However, precise attention to the training data, input features, and model architecture is crucial to producing accurate and generalizable predictions.


\section{Conclusion and perspectives}


Supercritical fluids, marked by their existence above the liquid-vapor critical point, offer a rich and fascinating area of study that spans both scientific and industrial realms. From the intriguing properties like zero surface tension at the critical point and the gradual vanishing of the latent heat of vaporization, to their diffusive and solvent characteristics, supercritical fluids present a myriad of potentials. Their applications  extend to analytical chemistry, material processing, and environmental science, as well as powering cutting-edge energy generation technologies. The present review navigated the complex landscape of supercritical fluids' thermodynamics, fluid dynamics, instability, and heat transfer. Recent advances in machine learning for enhanced turbulence modeling and heat transfer predictions were also discussed, showing a promising pathway for efficient modeling. Our synthesis of the current state of knowledge emphasizes the critical role supercritical fluids play in aligning with global climate commitments, technological innovation, and the underlying physics that govern their behavior. This review, therefore, stands as a comprehensive testament to the ongoing progress in understanding and leveraging supercritical fluids.

The review delves into various facets of supercritical fluids, from microscopic physics to macroscopic heat transfer. In the section \ref{thermodynamics}, it explores the near-critical thermophysical properties, phase transitions, and fluid dynamics of supercritical fluids, grounding the investigation in cutting-edge findings. A highlight is the discovery of a "pseudo boiling" process, a novel concept extending classical phase transitions and revealing intricate mechanisms affecting heat transfer in supercritical states. 


The section \ref{fluid} starts with a summary of the new velocity and temperature scaling laws and then focuses on the summary and discussion of the recent findings on the turbulent heat transfer. A comprehensive comparison of DNS results from various research groups sheds light on discrepancies in wall temperature predictions, accentuating the uncertainties inherent in numerical solvers. The section also includes innovative findings on horizontal flows, non-circular pipes, and non-uniform heat fluxes, significantly broadening the understanding of heat transfer phenomena in supercritical fluids. In addition, we link the appeared non-variable density field to the non-uniform body-force, which is able to relaminarize the flow in the experiments and DNS.

Section \ref{stability} elucidates the forefront of advancements unraveling the complexities inherent in the laminar-turbulent transition, with a particular emphasis on the linear stage of flow instabilities. This examination encompasses an array of flow configurations, including Poiseuille and Couette flows, mixing layers, and both two-dimensional and three-dimensional boundary layer flows. A pivotal discovery within this domain is the identification of a novel inviscid mode, which is markedly influenced by the significant gradients in thermodynamic and transport properties proximal to the Widom line. Furthermore, we underline that the entirety of the transition process is intrinsically characterized by profound nonlinear phenomena. It articulates that there is a substantial avenue for future research endeavors to demystify the underlying principles of this transition process comprehensively. Such explorations are imperative for the advancement of precise control mechanisms and the refinement of predictive models, thereby contributing to a holistic understanding and manipulation of transitional flows. 

Finally, section \ref{model} introduces encouraging progress in turbulence modeling based on AFHM and non-linear eddy viscosity models. In contrast to a decade ago when existing RANS models were tested, the recent decade has witnessed substantial advancements in enhancing and refining these models. Non-blind tests have yielded promising outcomes regarding heat transfer behavior, yet these findings require validation through impartial blind tests. Subsequently, validated models can then be applied to intricate real-world geometries and heat exchangers for further testing.
Except RANS modeling, the newly developed data-driven 1-dimensional heat transfer models have demonstrated superior performance over traditional heat transfer correlations while maintaining comparable prediction speeds. Consequently, these trained models are well-suited for engineering prototyping, though their capacity for extrapolation merits additional tests.
Overall, the review offers a multifaceted view of the progress of single-phase supercritical heat transfer in the last decade, unearthing new scientific understandings and engineering-orientated modeling, bridging fundamental physics with innovative methodologies, and paving the way for future research and applications.

The exploration of heat transfer of supercritical fluids is far from complete; it continues to open new horizons in both theoretical understanding and practical application. On the scientific front, the uncharted territories of thermodynamics and fluid dynamics at the supercritical phase offer opportunities for further research, particularly in understanding the nuanced behavior at the molecular level and the pseudo boiling process. 
The recently employed S-CO$_2$ pilot loop indicates a urgent need for optimized designs in turbomachines, crucial for reaching the system's thermal efficiency to the target level. This advancement requires a deeper comprehension of supercritical fluids in the compressible regime and the impact of non-ideal gas effects on the turbomachines.
In the industrial domain, the applications of supercritical fluids in energy generation, including nuclear, fossil fuel, waste heat, and renewable energy sources, are paving the way for cleaner and more efficient power systems. The continuous refinement of these applications, underlined by fiscal and environmental considerations, is crucial to meeting the ever-growing global energy demands and climate goals.

The supercritical phase represents an extraordinary convergence of physics, chemistry, and engineering. The intersection of these domains provides fertile ground for innovation, driving technological advancement and fostering a deeper understanding of the fundamental principles of matter. The research community is poised to continue unraveling the multifaceted potentials of supercritical fluids, ensuring that their exploration remains at the forefront of scientific inquiry and industrial application. The journey has only just begun, and the coming years promise further exciting developments in the study and utilization of supercritical fluids.

\section*{CRediT authorship contribution statement}

\textbf{Zhouhang Li}: Conceptualization, Writing - Original Draft, Funding acquisition.
\textbf{Daniel Banuti}: Conceptualization, Writing - Original Draft, Funding acquisition.
\textbf{Jie Ren}: Conceptualization, Writing - Original Draft, Funding acquisition.
\textbf{Junfu Lyu}: Writing - Review \& Editing, Project administration.
\textbf{Hua Wang}: Writing - Review \& Editing, Project administration.
\textbf{Xu Chu}: Conceptualization, Writing - Original Draft, Funding acquisition

\section*{Declaration of competing interest }

The authors declare that they have no known competing financial interests or personal relationships that could have appeared to influence the work reported in this paper. 

\section*{Data availability}

Data will be made available on request. 

\section*{Acknowledgments}

The authors appreciates the funding support from Deutsche Forschungsgemeinschaft (DFG, German Research Foundation) under Germany’s Excellence Strategy-EXC2075-390740016, the National Natural Science Foundation of China (Grant No. 12372215, 52176073). The authors gratefully appreciate the access to the high performance computing facility Hawk at HLRS, Stuttgart of Germany.



\bibliography{lit2,PECS}

\newpage

\begin{figure}[hbt!]
\centering
\includegraphics[width=0.3\textwidth]{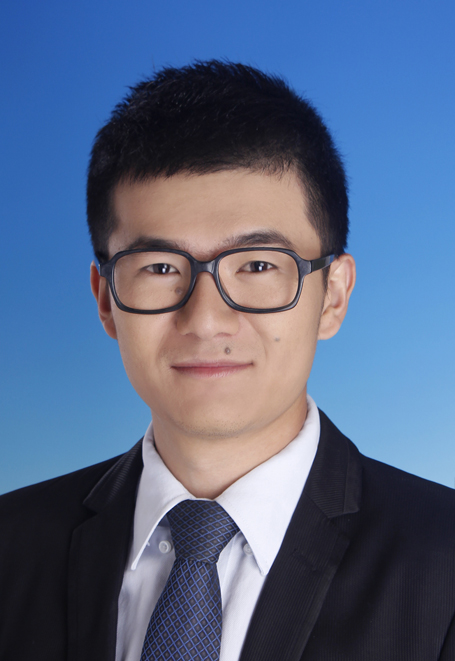}
\end{figure}

Prof. Zhouhang Li is currently a Professor in the School of Metallurgical and Energy Engineering at Kunming University of Science and Technology (KUST, China). Prof. Li previously held research positions at Tsinghua University and University of Stuttgart. He received his B.S. degree (2010) and Ph.D. (2015) from Tsinghua University, where he won the Excellent Doctoral Dissertation Award. His research covers several topics in the field of multiphase flow and heat transfer, particularly in fluid dynamics and heat transfer near the critical point. He has published over 30 articles in internationally recognized journals, edited two books, and has been involved in about ten research projects funded by the Chinese government, the Alexander von Humboldt Foundation (Germany), and industries.\\

\begin{figure}[hbt!]
\centering
\includegraphics[width=0.3\textwidth]{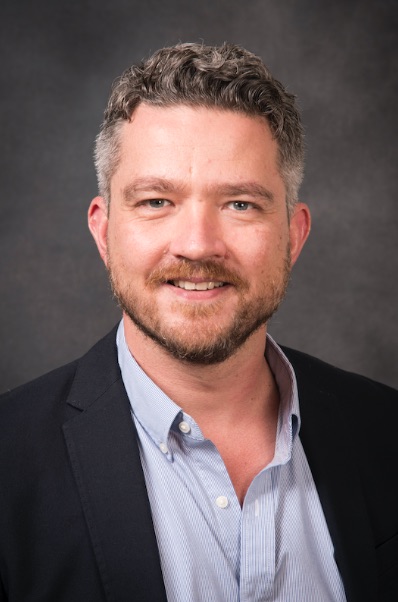}
\end{figure}

Prof. Dr.-Ing. Daniel T. Banuti holds the professorship (full) for “Hydrogen-based energy systems” at the Karlsruhe Institute of Technology (KIT) and is the head of the Institute for Thermal Energy Technology and Safety (ITES). Prof. Banuti previously held research positions at the German Aerospace Center (DLR) in Göttingen, Stanford University, Caltech, NASA JPL, and the University of New Mexico. He holds a MSc in Mechanical Engineering from RWTH Aachen University and a PhD in Aerospace Engineering from the University of Stuttgart, both in Germany. With an original research background in supercritical fluids modeling for rocket engines, his research focus has shifted to carbon-neutral thermal energy technologies, including propulsion and power, as well as fundamental behavior of fluids at supercritical conditions. Prof. Banuti is a a member of the directorate of the German Research Association Renewable Energy (FVEE) and the editorial board of the Journal of Supercritical Fluids.\\

\begin{figure}[hbt!]
\centering
\includegraphics[width=0.3\textwidth]{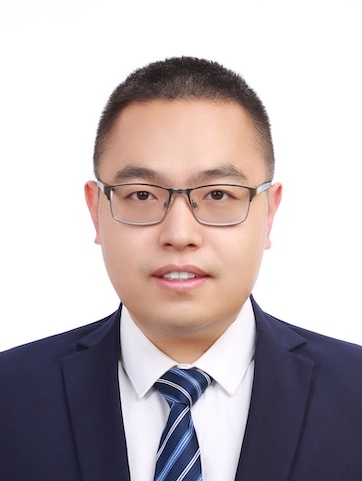}
\end{figure}

Prof. Jie Ren obtained his B.S. and Ph.D. from Tsinghua University. Before accepting a professorship at the Beijing Institute of Technology, he worked as a postdoc researcher (TU Delft, University of Nottingham) and  Alexander von Humboldt Research Fellow (University of Stuttgart). His research is dedicated to theoretical and numerical advancement toward understanding laminar-turbulent transition in the supercritical and hypersonic regimes. He won the Excellent Doctoral Dissertation of Tsinghua University and the Springer Thesis Prize. His discovery of the dual-mode instability in boundary layer flows of supercritical fluids is highlighted in Focus on Fluids of the Journal of Fluid Mechanics.   \\

\begin{figure}[hbt!]
\centering
\includegraphics[width=0.3\textwidth]{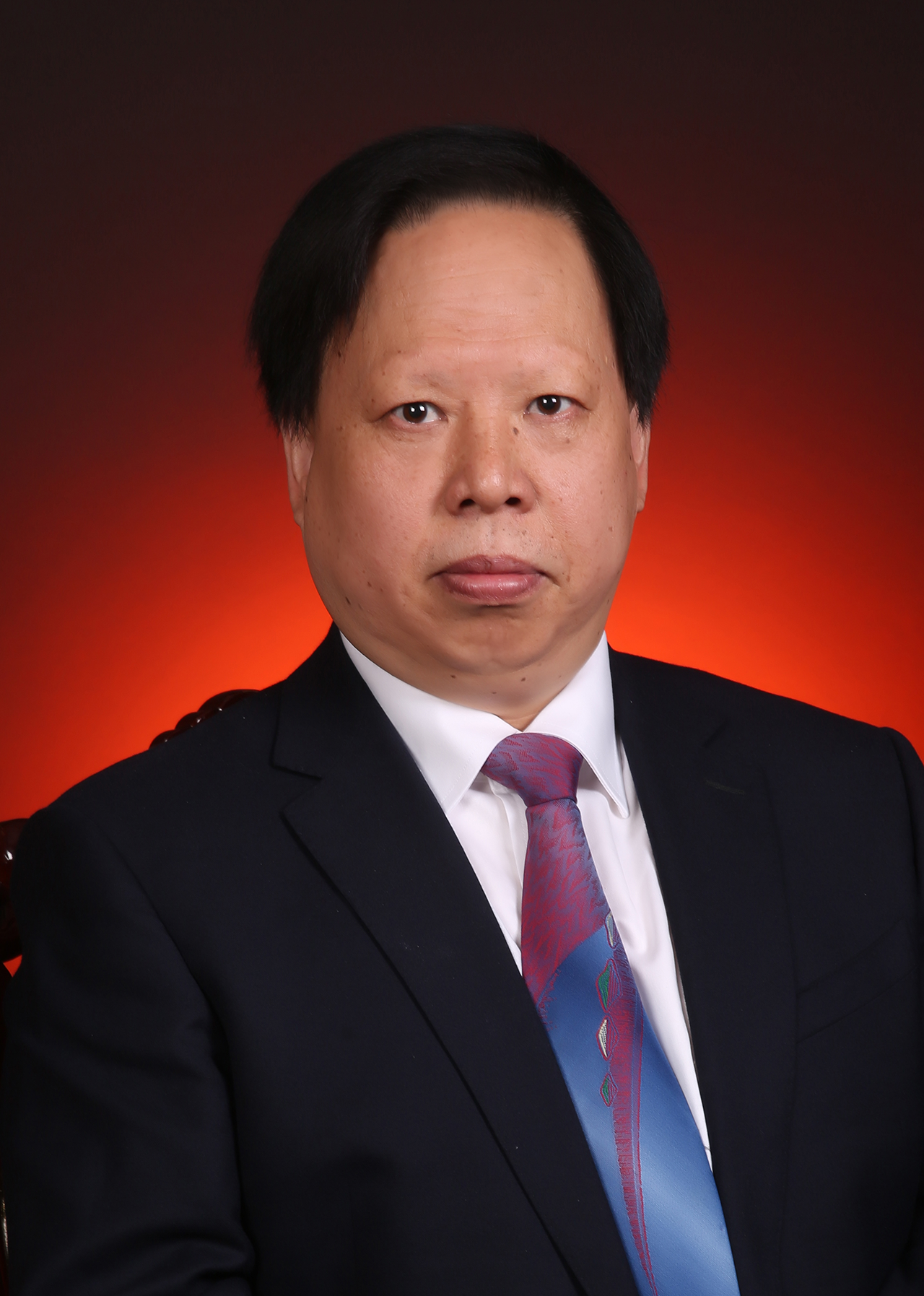}
\end{figure}

Prof. Junfu Lyu received his B.S. (1991), M.S. (1996) degrees and Ph.D. (2004) from Tsinghua University. In 1996 he joined the Department of Thermal Engineering at Tsinghua University and was a Professor since 2005. Dr. Lyu combines his expertise in research and technological innovation with a commitment to creating clean and efficient energy utilization. He has made significant contributions in the areas of gas-liquid two-phase flow, circulating fluidized bed combustion, boiler hydrodynamics, and coal gasification. He has published over 300 papers in various refereed journals, in addition to 5 edited books, 70 patents, and many keynote lectures at international conferences. Dr. Lyu served as the Chairman of the Executive Committee of IEA-FBC (International Energy Agency - Fluidized Bed Conversion) from 2020 to 2023, and is an editorial board member of several journals. In recognition of his research achievements, Prof. Lyu has received many awards in recent years, notable examples are the State Science and Technology Progress Award (State Council of the People's Republic of China, 2017, first prize), and the Guanghua Engineering Science and Technology Award (Chinese Academy of Engineering, 2020). He is a member of the Chinese Academy of Engineering (2023).\\

\begin{figure}[hbt!]
\centering
\includegraphics[width=0.3\textwidth]{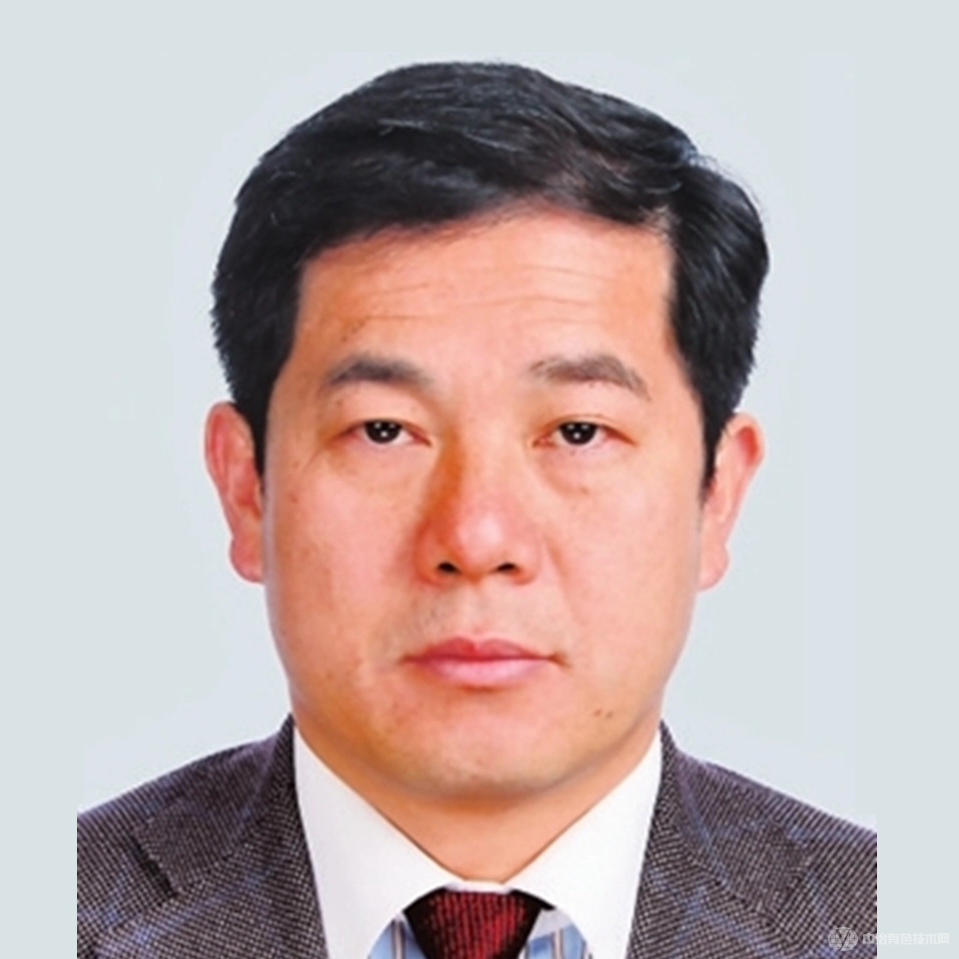}
\end{figure}

Prof. Hua Wang received his B.S. (1987) from the Northeastern University and both M.S. (1990) and Ph.D. (1996) from Kunming University of Science and Technology. Then he worked in Kyoto University as a postdoctoral researcher from 1998 to 2000. Since then he has been a Professor in Kunming University of Science and Technology. His research interest covers thermophysics in multiphase flow, chemical looping, and catalytic chemistry. He is author/co-author of more than 300 publications, 7 technical books, 216 papers in refereed journals, 113 papers in refereed proceedings of international and national conferences and symposiums, and 79 patents. Dr. Wang serves as a member of the international advisory committee of ICOPE (International Conference on Power Engineering) since 2009. He has received many awards including the State Science and Technology Progress Award (State Council of the People's Republic of China, first prize in 2012 and second prize in 2020), the Innovation Award (State Council of the People's Republic of China, 2019), and the Research Excellence Award of China (2014).\\

\begin{figure}[hbt!]
\centering
\includegraphics[width=0.3\textwidth]{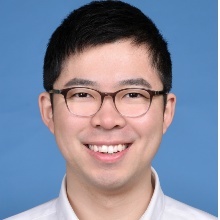}
\end{figure}

Dr. Xu Chu received his B.S. from Tongji University in China and both Dipl.-Ing. and Dr.-Ing. from University of Stuttgart in Germany. After his Ph.D. from Institute of Nuclear Technology and Energy Systems, University of Stuttgart, he led a research group in Institute of Aerospace Thermodynamics, University of Stuttgart. His research focuses on numerical and experimental multi-X thermofluids combined with data science. He has published more than 40 journal papers and book chapters on prestigious journals including Journal of Fluid Mechanics, Physical Review Fluids, Physics of Fluids and others. Dr. Chu serves as a guest editor for Physics of Fluids and the International Journal of Heat and Mass Transfer. Since 2024, he is the senior lecturer of Data-Centric Engineering in University of Exeter after submitted his Habilitation thesis jointly to the Faculty of Aerospace and Cluster of Excellence SimTech in the University of Stuttgart. \\

\end{document}